\documentclass[a4paper,11pt,bookmarks=true]{article}

\usepackage{jheppub} % for details on the use of the package, please
\usepackage{tocloft}

\usepackage[T1]{fontenc} % if needed
\usepackage{bm}

\def\beq{\begin{equation}}
\def\eeq{\end{equation}}
\def\bsp#1\esp{\begin{split}#1\end{split}}
\usepackage{multirow}
\newcommand{\bfp}{{\mathbf{p}}}

\newcommand{\cP}{{\mathcal{P}}}
\newcommand{\cPFS}{P}%{{\mathcal{P}_{\text{FS}}}}
\newcommand{\bfdelta}{{\bm{\delta}}}
\newcommand{\bfSigma}{{\bm{\Sigma}}}

\newcommand{\mMS}{\overline{m}}

\newcounter{RSQ}

\title{\boldmath An analysis of Bayesian estimates for missing higher orders in perturbative calculations}

\preprint{CERN-TH-2021-058}

\author[a]{Claude Duhr,}
\author[a]{Alexander Huss,}
\author[a]{Aleksas Mazeliauskas,}
\author[b]{Robert Szafron}
\emailAdd{claude.duhr@cern.ch}
\emailAdd{alexander.huss@cern.ch}
\emailAdd{aleksas.mazeliauskas@cern.ch}
\emailAdd{rszafron@bnl.gov}
\affiliation[a]{Theoretical Physics Department, CERN, CH-1211 Geneva 23, Switzerland}
\affiliation[b]{Department of Physics, Brookhaven National Laboratory, Upton, N.Y., 11973, U.S.A}

\abstract{With current high precision collider data, the reliable estimation of theoretical uncertainties due to missing higher orders (MHOs) in perturbation theory has become a pressing issue for collider phenomenology.
Traditionally, the size of the MHOs is estimated through scale variation, a simple but ad hoc method without probabilistic interpretation. Bayesian approaches provide a compelling alternative to estimate the size of the MHOs, but it is not clear how to interpret the perturbative scales, like the factorisation and renormalisation scales, in a Bayesian framework. Recently, it was proposed that the scales can be incorporated as hidden parameters into a Bayesian model.
In this paper, we thoroughly scrutinise Bayesian approaches to MHO estimation and systematically study the performance of different models on an extensive set of high-order calculations.
We extend the framework in two significant ways. First, we define  a new model that allows for asymmetric probability distributions. Second, we introduce a prescription to incorporate information on perturbative scales  without interpreting them as hidden model parameters. We clarify how the two scale prescriptions bias the result towards specific scale choice, and we discuss and compare different Bayesian MHO estimates among themselves and to the traditional scale variation approach.
Finally, we provide a practical prescription of how existing perturbative results at the standard scale variation points can be converted to 68\%/95\% credibility intervals in the Bayesian approach using the new public code \texttt{MiHO}.
}

\begin{document}
\maketitle

\section{Introduction}
\label{sec:intro}

Perturbative calculations are currently the only way to obtain Standard Model (SM) predictions for collider experiments. The backbone of almost all calculations for hadron collider observables is the QCD factorisation theorem, which relates an observable $\Sigma$ in proton-proton collisions (e.g., a total cross-section or differential distribution) to the corresponding partonic quantity,
\beq\label{eq:qcd_factorisation}
\Sigma= \sum_{i,j}\int_0^1 dx_1\,dx_2\,f_i(x_1,\mu_F)\,f_j(x_2,\mu_F)\,\hat{\Sigma}_{ij}(x_1,x_2,\mu_F) + \mathcal{O}\Big(\frac{\Lambda_\text{QCD}}{Q}\Big)\,,
\eeq
where $f_i(x,\mu_F)$ denotes the parton distribution function (PDF) for parton $i$ and $\hat{\Sigma}_{ij}$ denotes the observable computed on a collision of the partons $i$ and $j$. Here $\mu_F$ is the factorisation scale, $Q$ denotes a characteristic hard scale of the process, and $\Lambda_{\text{QCD}}\lesssim 1\text{ GeV}$ is the scale at which QCD becomes strongly coupled. 

The partonic coefficients $\hat{\Sigma}_{ij}$ are computable in perturbation theory, %i.e., they can be approximated by an expansion in the  coupling constants of the theory. For simplicity, we only discuss here the case of the strong coupling constant:
\beq\label{eq:dsigma_perturbative_expansion}
\hat{\Sigma}_{ij}(x_1,x_2,\mu_F) \simeq \sum_{k=0}^\infty \alpha_s(\mu_R)^{k_0+k}\,\hat{\Sigma}_{ij}^{(k)}(x_1,x_2,\mu_F,\mu_R)\,,
\eeq
where $k_0\ge0$ is a non-negative integer, $\alpha_s(\mu_R)$ denotes the renormalised strong coupling constant evaluated at the renormalisation scale $\mu_R$, and $\hat{\Sigma}_{ij}^{(k)}$ is the (next-to-)$^k$-leading order (N$^k$LO) contribution to the partonic coefficient.\footnote{The `$\simeq$'-sign in eq.~\eqref{eq:dsigma_perturbative_expansion} indicates that the perturbative expansion on the right-hand side is in general an asymptotic series with zero radius of convergence. For most phenomenological predictions only very few terms in the perturbative expansion are known (typically not more than the first 3 or 4 terms), and one does not have to worry about the non-convergent nature of the series.}
One can combine the QCD factorisation theorem in eq.~\eqref{eq:qcd_factorisation} with the perturbative expansion of the partonic coefficients in eq.~\eqref{eq:dsigma_perturbative_expansion} to obtain the N$^n$LO approximation for the observable $\Sigma$ within (fixed-order) perturbation theory
\beq\label{eq:Sigma_perturbative_expansion}
\Sigma \simeq \Sigma_n(\mu_F,\mu_R) := \sum_{k=0}^n\Sigma^{(k)}(\mu_F,\mu_R)\,.
\eeq
Computations are typically performed in the $\overline{\rm MS}$ renormalisation scheme.  Although $\Sigma$ is formally independent of the choice of the factorisation and renormalisation scales, the truncation of the perturbative series introduces a residual scale dependence into $\Sigma_n$.

Since a perturbative result is always an approximation of the exact value $\Sigma$, it is important to have robust methods to assess the quality of the approximation and to quantify the uncertainty attached to it. Three of the main sources of uncertainty are coming from the values for the strong coupling constant and other input parameters (we collectively refer to them as the {parametric uncertainty}), the PDFs\footnote{Note that the PDF uncertainty can itself be seen as a parametric uncertainty, just like the coupling constants or the masses. 
Indeed, there is no strong conceptual difference between the PDF and, e.g., $\alpha_s$ uncertainties, and often they are treated on the same footing.
For our discussion in later sections, we find it convenient to think of these two sources separately. This does, however, not impact any of our results, as our primary focus are MHO uncertainties.} and the truncation of the perturbative series.  
The parametric and PDFs uncertainties are related to their extraction from experimental data and the fitting methodology used. Consequently, the PDF and parametric uncertainties are not directly related to the observable $\Sigma$. They are computed using observable-independent recipes, cf.,~e.g.,~refs.~\cite{Butterworth:2015oua,Alekhin:2011sk,Botje:2011sn} (though care is needed as there may be correlations between $\Sigma$ and the data used to extract the PDFs and $\alpha_s$). The uncertainty resulting from the truncation of the perturbative series quantifies the effect of the missing higher orders (MHOs) in perturbation theory. Its estimation is one of the main topics of this paper. Under the assumption that the PDF, parametric, and MHO uncertainties are uncorrelated (see \cite{Harland-Lang:2018bxd,Ball:2018twp,AbdulKhalek:2019ihb,Ball:2021icz} for discussion of correlation between PDF and MHO uncertainties), they can be added in quadrature to obtain the uncertainty on the N$^n$LO prediction.\footnote{For a given observable there may be other uncertainties one needs to take into account, e.g., the choice of the renormalisation schemes used for the input parameters, the truncation of the heavy mass expansion etc. How to estimate these uncertainties and combine them is often a matter of debate for specific processes and observables. We do not discuss these issues in this paper as we intend to keep the discussion general.}

Currently, there is no reliable way to estimate the size of the MHOs (other than computing the next order in perturbation theory).
Since the effect of varying the scales at N$^n$LO is an N$^{n+1}$LO effect, one conventionally estimates the size of the MHO terms by studying the variation of $\Sigma_n(\mu_F,\mu_R)$ as the factorisation and renormalisation scales are varied in a certain range around a central scale $\mu_{F/R,\text{cent}}$ corresponding to a characteristic hard scale of the process.
A popular choice is the so-called \emph{7-point variation}, where the MHO uncertainty at N$^n$LO is estimated by the interval 
\begin{equation}\label{eq:7pointinterval}
    \Big[\min_{(k_F,k_R)\in S_7}\!\!\!\!\!\Sigma_n(k_F\,\mu_{F,\text{cent}},k_R\,\mu_{R,\text{cent}}),\max_{(k_F,k_R)\in S_7}\!\!\!\!\!\Sigma_n(k_F\,\mu_{F,\text{cent}},k_R\,\mu_{R,\text{cent}}) \Big]\,.
\end{equation}
$(k_F,k_R)$ take values in the following set:
\beq
S_7=\{(1/2,1/2), (1/2,1), (1,1/2),(1,1),(1,2),(2,1),(2,2)\}\,.
\label{eq:7points}
\eeq
This differs from the \emph{9-point variation} where $k_F,k_R\in\{1/2,1,2\}$ by the exclusion of large ratios of the scales, i.e., $k_F/k_R=4$ or $1/4$.
We stress that the procedure of estimating the MHO uncertainty from scale variation is completely ad hoc. There is no reason other than experience from NLO and NNLO calculations why the obtained uncertainty interval at N$^n$LO should capture the actual size of the N$^{n+1}$LO corrections $\Sigma^{(n+1)}$ or the true value of $\Sigma$. 
Indeed, the higher-order terms probed by varying the scales are universal and predictable from renormalisation group methods. They do not take into account the size of the genuine higher-order corrections, which are process-dependent and independent of the perturbative scales. With advances in precision calculations over the last 5 years, NNLO results are becoming routinely available for many LHC processes, and even N$^3$LO corrections to colourless final states have been recently computed. 
It is, therefore, time to critically assess the MHO uncertainties and to look for alternative prescriptions.
A number of attempts have been made in the past to estimate or predict the truncation errors and MHOs, see, e.g., refs~\cite{Ellis:1995jv,Gardi:1996iq,Ellis:1996zn,Weniger:1997zz,Ellis:1997sb,Karliner:1998ge,Caprini:1998wg,Cvetic:1998zm,Elias:2000iw,Caprini:2000js,Jentschura:2000iw,Beneke:2008ad,David:2013gaa,Furnstahl:2015rha,Boito:2016pwf,Jamin:2016ihy,Melendez:2017phj,Caprini:2019kwp,Melendez:2019izc,Costin:2020hwg,Boito:2020hvu,Drischler:2020hwi}.
Various prescriptions for reducing or eliminating scale ambiguity have also been discussed broadly in the literature, e.g., refs.~\cite{Politzer:1981vc,Pennington:1981sq,Stevenson:1982wn,Grunberg:1982fw,Brodsky:1982gc,Cvetic:1997ca,Maxwell:2000mm,Brodsky:2011ig,Kataev:2014jba,Kataev:2016aib,Czakon:2016dgf,Garkusha:2018mua,Wu:2019mky,Boito:2018rwt,Brodsky:2019tmf,Chawdhry:2019uuv,Chishtie:2020cen,DiGiustino:2020fbk}. 

Here we focus on a promising, widely applicable approach  to estimate the MHO contributions using Bayesian inference on the known fixed-order computations.  
This idea  was pioneered by Cacciari and Houdeau (CH) in ref.~\cite{Cacciari:2011ze}, and developed further in refs.~\cite{Forte:2013mda,Bagnaschi:2014wea,Bonvini:2020xeo}.
Bayesian inference is a powerful method to construct probability distributions in which Bayes' theorem is used to update the probability as new information becomes available iteratively. The procedure can be summarised as follows: one starts from some model and the prior distribution of the model parameters. These inputs are purely subjective. The next step is a construction of a posterior distribution for model parameters based on available data. This is the inference step, where one updates prior belief as new data become available. Finally, one constructs the posterior distribution for the next data point by marginalising over the posterior distribution for the model parameters. 
The main idea of the CH-method it to use Bayesian inference on the known perturbative orders to model the size of the MHOs. 
The original CH-model can be applied only after one specifies the choices for the perturbative scales. In this paper, we follow up on the recent work in ref.~\cite{Bonvini:2020xeo}, where the CH-approach was extended by the postulate that unphysical scales should be treated on par with other hidden model parameters. According to the lore of Bayesian inference, one  expects that the posterior distribution for the perturbative scales peaks at the `optimal value' for the scale choice, and the dependence on the scales is eliminated by marginalisation. However, this is only one example of possible prescriptions, as there is no reason, physical or mathematical, to interpret the perturbative scales as hidden model parameters. One of the main objectives of this paper is to thoroughly scrutinise the consequences of this postulate and to explore alternative approaches.

This paper is organised as follows: In section~\ref{sec:scale_marginalisation} we discuss two prescriptions of how to incorporate scale dependence into a model for Bayesian inference of fixed-order calculations, and we investigate implicit biases inherent to them. 
In section~\ref{sec:no_scale} we recap the mathematical formalism behind Bayesian inference at fixed scale and introduce a model to account for asymmetric probability distributions. In section~\ref{sec:qft_examples}
we study extensively the model performance for representative perturbative calculations without scale dependence or at fixed scale choice. Finally, in section~\ref{sec:hadronic_examples} we study the performance of the two prescriptions to incorporate scale information for an extensive set of collider processes known at NNLO and/or N$^{3}$LO. We conclude with a discussion in section~\ref{sec:discussion}.

\section{Bayesian methods for observables in quantum field theory}
\label{sec:scale_marginalisation}

\subsection{Incorporating scale information into Bayesian inference}

In this section we present a way to estimate the MHO uncertainty for an observable $\Sigma$ via Bayesian methods. To be concrete, let us assume that we have computed the first $n+1$ orders $\Sigma^{(k)}$, $0\le k\le n$, in the perturbative expansion of $\Sigma$. Our goal is to construct a probability distribution $\cP(\Sigma| \bfSigma_n)$ for $\Sigma$ given the sequence of perturbative coefficients $\bfSigma_n := (\Sigma^{(0)}, \Sigma^{(1)}, \ldots, \Sigma^{(n)})$. This distribution can be used to construct credibility intervals (CIs) for $\Sigma$, which may serve as the uncertainty estimate. In particular, we define the $x$\% CI as
$\text{CI}_x =[\Sigma_{x}^\text{low},\Sigma_{x}^\text{up}]$ where
\begin{equation}\label{eq:CIdef}
    \int_{-\infty}^{\Sigma_{x}^\text{low}}\! d\Sigma\, \cP(\Sigma| \bfSigma_n) =    \int^{\infty}_{\Sigma_{x}^\text{up}}\! d\Sigma\, \cP(\Sigma| \bfSigma_n) = \frac{1-\frac{x}{100}}{2}\,.
\end{equation}
We will mainly consider $x\%\approx 68\%$ and $x\%\approx 95\%$ CIs.

The original approach of ref.~\cite{Cacciari:2011ze}, known as the \emph{Cacciari-Houdeau} (CH) method, 
allows one to construct a probability $\cP(\Sigma| \bfSigma_n)$ in the case where the $\bfSigma_n$ are \emph{numbers}. In perturbation theory, however, the perturbative coefficients $\bfSigma_n$ are \emph{functions} of the perturbative scales, which are arbitrary quantities introduced by perturbation theory without any physical interpretation. Clearly, $\cP(\Sigma| \bfSigma_n)$ should be independent of the scale choice, but the value of the N$^n$LO prediction may depend strongly on this choice, and there is no reason to prefer one scale over another (as long as the scale does not differ too much from the hard scale of the process). 

Reference~\cite{Bonvini:2020xeo} introduced a possible way of incorporating perturbative scales into a Bayesian approach to estimating MHOs (see also ref.~\cite{Bagnaschi:2014wea} for an alternative approach). 
The approach of ref.~\cite{Bonvini:2020xeo}\footnote{Reference~\cite{Bonvini:2020xeo} excludes the factorisation scale from this discussion. We will come back to this point in section \ref{sec:multi-scale}.} to include the information on the perturbative scales into a Bayesian model is not the only one. 
Broadly speaking, we identify two different prescriptions of how one can incorporate information on perturbative scales into the construction of a probability distribution:
\begin{description}
\item{\bf Prescription 1: Scale-marginalisation (sm).} Despite the perturbative scales not being physical, one assigns a Bayesian degree of belief to them. In this way the scales become (hidden) parameters of a Bayesian model. It is then possible to construct $\cP(\Sigma| \bfSigma_n)$ through Bayesian inference and marginalisation over the perturbative scales. This is the approach advocated in ref.~\cite{Bonvini:2020xeo}.
\item{\bf Prescription 2: Scale-averaging (sa).} Since the perturbative scales are devoid of any physical meaning, they cannot be treated as hidden parameters of a Bayesian model, i.e., they cannot be treated as random variables with a statistical interpretation. Instead, the result of a perturbative calculation through N$^n$LO should be interpreted as a family of independent predictions for the true value $\Sigma$ parametrised by the scales. The probability distribution $\cP(\Sigma| \bfSigma_n)$ is then defined by adding coherently the probabilities computed for each family member.
\end{description}

As we will show later, the two prescriptions of obtaining scale-independent probabilities differ not only in their philosophy, but in practice they can produce different results due to implicit biases towards some scales over others.
In the next section we present the mathematical framework to construct the probability distribution $\cP(\Sigma|\bfSigma_n)$. In order to keep the notations concise, we focus first on the simplified situation where the perturbative coefficients $\bfSigma_n$ depend on a single perturbative scale $\mu$ (which we may think of as the renormalisation scale for now).

\subsubsection{Prescription 1: scale-marginalisation}
\label{sec:scale-marginalization_pres}
We start by reviewing the \emph{scale-marginalisation (sm)} prescription of ref.~\cite{Bonvini:2020xeo}. 
Since $\mu$ is a parameter of the model, the probability distribution $\cP(\Sigma|\bfSigma_n)$ is obtained by marginalising over $\mu$;
\beq\bsp\label{eq:P_marginalised}
\cP(\Sigma|\bfSigma_n) \,&= \int d\mu\, \cP(\Sigma,\mu|\bfSigma_n)= \int d\mu\, \cP(\Sigma|\bfSigma_n,\mu)\,\cP(\mu|\bfSigma_n)\,.
\esp\eeq
Let us interpret the probabilities in the integrand. For the first factor, we apply Bayes' formula to write
\beq\bsp
\cP(\Sigma|\bfSigma_n,\mu) &\,= \frac{\cP(\Sigma,\bfSigma_n|\mu)}{\cP(\bfSigma_n|\mu)} = \frac{\cPFS(\Sigma,\bfSigma_n(\mu))}{\cPFS(\bfSigma_n(\mu))} = \cPFS(\Sigma|\bfSigma_n(\mu))
\,.
\esp\eeq
The interpretation of this equation is as follows:
Since $\mu$ is a parameter of the model, whenever we evaluate a conditional probability \emph{for a fixed value of $\mu$}, we can replace the perturbative coefficients $\bfSigma_n$ (which are functions) by their values $\bfSigma_n(\mu)$ for fixed $\mu$ (which are numbers). In the following we always indicate by $\cP$ a probability distribution which takes are arguments the perturbative coefficients $\bfSigma_n$ (which are functions), while $\cPFS$ denotes a probability distribution on perturbative coefficients computed at a fixed scale $\mu$. 
For now we assume that $\cPFS$ is given.
In section~\ref{sec:no_scale} we discuss a framework how a distribution $\cPFS$ can be constructed.

The second factor in the integrand of eq.~\eqref{eq:P_marginalised} is the posterior distribution for $\mu$, and we have
\beq\bsp
\cP(\mu|\bfSigma_n)&\, = \frac{\cP(\mu,\bfSigma_n)}{\cP(\bfSigma_n)} = \frac{\cP(\bfSigma_n|\mu)\,P_0(\mu)}{\int d\mu'\,\cP(\bfSigma_n|\mu')\,P_0(\mu')} = \frac{\cPFS(\bfSigma_n(\mu))\,P_0(\mu)}{\int d\mu'\,\cPFS(\bfSigma_n(\mu'))\,P_0(\mu')}\,,
\esp\eeq
where $P_0(\mu)$ is an (arbitrary) prior distribution on the perturbative scale $\mu$. In ref.~\cite{Bonvini:2020xeo} it was argued that a physically well-motivated choice is a flat distribution in the logarithm of the scale,
\beq\label{eq:scale_prior}
P_0(\mu) := \frac{1}{2\mu\,\log F}\,\Theta\left(\log F - \left|\log\frac{\mu}{\mu_0}\right|\right)\,,
\eeq
where $F>1$ is a constant and $\Theta$ denotes the Heaviside step function;
\beq
\Theta(x) = \left\{\begin{array}{ll}
1\,,\,\quad \text{ if } x > 0\,,\\
0\,,\,\quad \text{ otherwise }\,.
\end{array}\right.
\eeq
The central scale $\mu_0$ is in principle arbitrary, but assumed to be of the order of the hard scale of the process. Putting everything together, we see that $\cP(\Sigma|\bfSigma_n)$ can be expressed in terms of the probability distribution $\cPFS$ and the prior $P_0$ on the scale:
\begin{align}
 \cP&(\Sigma|\bfSigma_n)  \equiv \cP_{\text{sm}}(\Sigma|\bfSigma_n) := \frac{\int d\mu\,\cPFS(\Sigma|\bfSigma_{n}(\mu))\,\cPFS(\bfSigma_{n}(\mu))\,P_0(\mu)}{\int d\mu'\,\cPFS(\bfSigma_n(\mu'))\,P_0(\mu')}\,.
\label{eq:P_scale-marginalization}
\end{align}
Equation~\eqref{eq:P_scale-marginalization} allows one to relate $\cP_{\text{sm}}(\Sigma|\bfSigma_n)$ to two other probability distributions (in addition to the prior on $\mu$). The first one is the probability distribution $\cPFS(\Sigma|\bfSigma_{n}(\mu))$ for the (true) value of the observable $\Sigma$ given the $n+1$ first perturbative orders calculated at the scale $\mu$. The second one is the probability distribution $\cPFS(\bfSigma_n(\mu))$, which is related through Bayes' theorem to $\cPFS(\Sigma^{(n)}(\mu)|\bfSigma_{n-1}(\mu))$. The latter computes the probability distribution for the N$^n$LO corrections $\Sigma^{(n)}(\mu)$ evaluated at the scale $\mu$, given the $n$ first perturbative orders evaluated at that scale.
The two distributions $\cPFS(\Sigma|\bfSigma_n(\mu))$ and $\cPFS(\Sigma^{(n+1)}(\mu)|\bfSigma_{n}(\mu))$ are the answers to two different mathematical questions, and there is a priori no reason why they should be related. However, if perturbation theory is applicable, then the true value of $\Sigma$ should be well approximated by $\Sigma_{n+1}(\mu)$ (at least for reasonable choices of $\mu$). We therefore expect that also the distribution $\cPFS(\Sigma|\bfSigma_n(\mu))$ is well approximated by $P(\Sigma^{(n+1)}(\mu)|\bfSigma_n(\mu))$. Substituting $\Sigma^{(n+1)}(\mu)\approx\Sigma-\Sigma_{n}(\mu)$ in the latter, we expect that, if perturbation theory is applicable, then
\beq\label{eq:perturbative_P}
\cPFS(\Sigma|\bfSigma_n(\mu))\approx P(\Sigma-\Sigma_{n}(\mu)|\bfSigma_n(\mu))\,.
\eeq
For a more detailed derivation of this approximation in a  Bayesian framework we refer to ref.~\cite{Bonvini:2020xeo}. Using this approximation, we can cast eq.~\eqref{eq:P_scale-marginalization} into the form:
\begin{align}
 \cP_{\text{sm}}&(\Sigma|\bfSigma_n)  \approx \frac{\int d\mu\,\cPFS(\Sigma-\Sigma_{n}(\mu)|\bfSigma_{n}(\mu))\,\cPFS(\bfSigma_{n}(\mu))\,P_0(\mu)}{\int d\mu'\,\cPFS(\bfSigma_n(\mu'))\,P_0(\mu')}\,.
\label{eq:P_scale-marginalization_approx}
\end{align}
We stress that, unlike indicated by the notation in eq.~\eqref{eq:perturbative_P}, the probability distributions in the left and right-hand sides of eqs.~\eqref{eq:P_scale-marginalization} and~\eqref{eq:perturbative_P} are actually \emph{distinct}: $\cPFS(\Sigma|\bfSigma_n(\mu))$ is the conditional distribution for the true value $\Sigma$ of the observable, while $P(\Sigma-\Sigma_{n}(\mu)|\bfSigma_n(\mu))$ is the distribution for the value of the N$^{n+1}$LO correction $\Sigma^{(n+1)}(\mu)\approx\Sigma-\Sigma_{n}(\mu)$ computed at the scale $\mu$. We will always assume that available N$^n$LO results sufficiently well approximate $\Sigma$ and that we are in the perturbative regime where eq.~\eqref{eq:perturbative_P} holds, so we do not distinguish the notation for the two distributions.

\subsubsection{Prescription 2: scale-averaging}
\label{sec:weighted-sum-model}

In this section we present an alternative way to include scale information into the model which does not require the perturbative scale $\mu$ to be promoted to a model parameter with a statistical or physical meaning. We call this method the \emph{scale-averaging (sa)} prescription.
The starting point is to interpret $\mu$ as a parameter defining a family of perturbative predictions which approximate $\Sigma$ at N$^n$LO in perturbation theory. For each member of this family, i.e., for each fixed value $\mu$ of the scale, we can compute the probability $\cPFS(\Sigma|\bfSigma_{n}(\mu))$ (as before we assume again that $\cPFS$ is given). We then combine this family of predictions into a single distribution $\cP(\Sigma|\bfSigma_n)$ as a weighted average of probabilities
\beq\bsp\label{eq:P_mixture-distribution}
\cP(\Sigma|\bfSigma_n) \equiv\cP_{\text{sa}}(\Sigma|\bfSigma_n) &\,:= \int d\mu \,w(\mu)\,\cPFS(\Sigma|\bfSigma_n(\mu))\,.
\esp\eeq
The form of the weighting function $w(\mu)$ is arbitrary, but it must be chosen such that $\cP(\Sigma|\bfSigma_n)$ satisfies all the axioms of a normalised probability distribution. In particular, $w(\mu)$ must itself be normalised:
\beq
\int d\mu\,w(\mu) = 1\,.
\eeq
Since one expects that the perturbative scale $\mu$ should be chosen close to the characteristic  hard scale of the process in order to ensure good perturbative convergence, we postulate a window function around a fixed scale $\mu_{0}\sim Q$:
\beq\label{eq:weight}
w(\mu) := \frac{1}{2\mu\,\log F}\,\Theta\left(\log F - \left|\log\frac{\mu}{\mu_0}\right|\right)\,,
\eeq 
where $F>1$ is a constant.
We note that our choice for $w(\mu)$ is identical to the prior on the scale in the scale marginalisation in eq.~\eqref{eq:scale_prior}. The interpretation, however, is different: in the previous section the scale $\mu$ is a model parameter with a statistical interpretation that leads to eq.~\eqref{eq:P_scale-marginalization}. In eq.~\eqref{eq:P_mixture-distribution}, the scale has no statistical interpretation, and is simply a parameter defining a family of distributions, which are then combined into a single distribution with weighting function $w(\mu)$ according to eq.~\eqref{eq:P_mixture-distribution}. 
With the weight given in eq.~\eqref{eq:weight}, this corresponds to a uniform average in the logarithm of the scale over the interval $\mu_0/F<\mu< F\mu_0$.
Using the approximation in eq.~\eqref{eq:perturbative_P} we obtain the final expression
\begin{equation}
\cP_{\text{sa}}(\Sigma|\bfSigma_n) \approx \int d\mu \,w(\mu)\,\cPFS(\Sigma-\Sigma_n(\mu)|\bfSigma_n(\mu))\,.
\label{eq:P_scale-average_approx}
\end{equation}

\subsubsection{Extension to multiple perturbative scales}
\label{sec:multi-scale}

In the case where the perturbative coefficients $\bfSigma_n$ are independent of the unphysical perturbative scale $\mu$, i.e., $\bfSigma_n(\mu) = \bfSigma_n(\mu_0)$ for some $\mu_0$, the distributions $\cP$ reduce to $\cPFS$:
\beq
\cP(\Sigma|\bfSigma_n) = \cPFS(\Sigma|\bfSigma_n(\mu_0))\,, \quad \text{if} \quad \bfSigma_n(\mu) = \bfSigma_n(\mu_0)\,.
\eeq
To extend the discussion to more scales $\mu_i$, $1\le i\le s$ we introduce a multivariate prior $P_{0}(\mu_1,\ldots,\mu_s)$ or weighting function $w(\mu_1,\ldots,\mu_s)$, and we integrate over each scale $\mu_i$ separately. 

We need to comment on hadronic observables, which always depend on the factorisation scale $\mu_F$. In ref.~\cite{Bonvini:2020xeo} the factorisation scale was explicitly excluded from the marginalisation procedure described in section~\ref{sec:scale-marginalization_pres}. Indeed, the factorisation scale also appears in the PDFs. It is not clear what is the correlation and how the PDF uncertainty impacts the probability distribution $\cP(\Sigma|\bfSigma_n)$, which is supposed to represent the probability distribution for the true value of the observable and should supersede and incorporate all other uncertainties. Here we aim for something much simpler. We want to use the distribution $\cP(\Sigma|\bfSigma_n)$ to construct credibility intervals that serve as estimators for the MHO uncertainty. In this sense, the role of $\cP(\Sigma|\bfSigma_n)$ is similar to scale variation as a tool to estimate the size of the MHOs. Like in the case of scale variation, we will assume that the MHO uncertainty is independent of the PDF uncertainty, and the two can be added in quadrature.

\subsection{Connection to prescriptions for choosing the perturbative scales}
\label{sec:scale_discussion}
In the previous section we have presented two methods to construct a scale-independent probability distribution $\cP(\Sigma|\bfSigma_n)$. The two methods differ not only by a different philosophy of how the unphysical perturbative scales are included, but can also lead to different results for the inferred probability distributions. 
Here we show that under rather general conditions  the position of the peak of
$\cP(\Sigma|\bfSigma_n)$  as a function of $\Sigma$ will be biased towards  two different scales $\mu_\text{FAC}$ and $\mu_\text{PMS}$:
\begin{itemize}
    \item 
The point of \emph{fastest apparent convergence} (FAC) \cite{Grunberg:1980ja,Stevenson:1981vj} of an observable computed at N$^n$LO is the scale $\mu_{\text{FAC}}$ such that
\beq\label{eq:FAC_condition}
\Sigma_n(\mu_{\text{FAC}}) = \Sigma_{n-1}(\mu_{\text{FAC}})\,,
\eeq
where we used the notation $\Sigma_n(\mu):=\Sigma_n(\mu,\mu)$.
Note that the value of $\mu_{\text{FAC}}$ depends on the order at which it is computed, and there may be more than one solution to eq.~\eqref{eq:FAC_condition} at a given order in perturbation theory.

\item 
The \emph{principle of minimal sensitivity} (PMS) \cite{Stevenson:1981vj} aims at selecting a scale $\mu_{\text{PMS}}$ at N$^n$LO that maximises the stability of the observable under changes of the scale
\beq
\frac{\partial}{\partial\mu}\Sigma_n(\mu)\big|_{\mu=\mu_{\text{PMS}}} = 0\,.
\eeq
Just like the FAC point, the PMS point depends on the order on which it is computed and may not be unique at a given order in perturbation theory.
\end{itemize}
Consider  now a broad class of Bayesian inference models satisfying the following conditions:
\begin{enumerate}
\item We work with the prior in eq.~\eqref{eq:scale_prior} or the weighting function in eq.~\eqref{eq:weight}.
\item $P(\Sigma^{(n+1)}(\mu)|\bfSigma_n(\mu))$ is \emph{symmetric}, i.e., it is insensitive to the signs of the correction:
\beq\label{eq:sym_P}
P(\Sigma^{(n+1)}(\mu)|\bfSigma_n(\mu)) = P(|\Sigma^{(n+1)}(\mu)|\,\big|\,|\Sigma^{(0)}(\mu)|,\ldots,|\Sigma^{(n)}(\mu)|)\,.
\eeq 
\item The distributions $P(\Sigma^{(n+1)}(\mu)|\bfSigma_n(\mu))$ have a single peak in $\Sigma^{(n+1)}(\mu)$ for each fixed value of $\mu$, and the peak is reached for some $\mu\in  [\mu_0/F,\mu_0F]$.
\item The peak is more and more pronounced as $n$ increases, i.e., as more information on the progression of the perturbative series becomes known. 
\end{enumerate}
Note that these assumption imply
that the peak of $P(\Sigma^{(n+1)}(\mu)|\bfSigma_n(\mu))$ is located at $\Sigma^{(n+1)}(\mu) = 0$, for all $\mu$. 
The aforementioned assumptions are reasonable, and they are satisfied for some of the models considered in ref.~\cite{Bonvini:2020xeo}, in particular by the geometric model and the scale-variation model of ref.~\cite{Bonvini:2020xeo} (see also section~\ref{sec:geometric-model}). 
In appendix~\ref{app:peak} we show that any model satisfying the assumptions above will also satisfy the following property for the peak of the distribution $\cP(\Sigma|\bfSigma_n)$:
\begin{itemize}
\item If there is a single $\mu_{\text{FAC}}\in[\mu_0/F,\mu_0F]$, then  $\cP_{\text{sm}}(\Sigma|\bfSigma_n)$ peaks at $\Sigma = \Sigma_n(\mu_{\text{FAC}})$, i.e., at the N$^n$LO prediction for the observable evaluated at the point of fasted apparent convergence.
\item If there is a single $\mu_{\text{PMS}}\in[\mu_0/F,\mu_0F]$, then $\cP_{\text{sa}}(\Sigma|\bfSigma_n)$ peaks at $\Sigma = \Sigma_n(\mu_{\text{PMS}})$, i.e., at the N$^n$LO prediction for the observable evaluated at the point of minimal sensitivity.
\end{itemize}
As a consequence, the sensitivity of the probability distributions and CIs on the choice of the scale variation interval (given by $F$) depends on the presence (or absence) of FAC and PMS points in the interval for the sm- and sa-prescriptions respectively. We will illustrate this on concrete examples in section~\ref{sec:hadronic_examples}.

Let us conclude this section with an important comment about the sm-prescription. In the sm-prescription, the perturbative scales are treated as model parameters with a statistical interpretation. Bayesian inference will then select the `best' value for the perturbative scales according to the model, see also the discussion in ref.~\cite{Bonvini:2020xeo}. At high enough perturbative orders, the dependence on the priors is expected to be mild, so that the conclusions derived from the inference are robust and (to a certain extent) prior-independent. The results of this section indicate that such a Bayesian interpretation of the perturbative scales should be taken with a grain of salt: For symmetric models with the prior in eq.~\eqref{eq:scale_prior} (which applies in particular to some of the models of ref.~\cite{Bonvini:2020xeo}), as soon as the range $[\mu_0/F,\mu_0F]$ contains an FAC point, the scale selected by the inference is biased: it will \emph{automatically} select the value of the observable at the FAC point as the most likely value. In other words, such a model will \emph{always} select the FAC point as its `best' estimate for the scales, independently of the actual progression of the perturbative series. Note that the previous discussion does not apply to the sa-prescription, because the perturbative scales are not assigned a probabilistic interpretation (in particular, the weight $w(\mu)$ in eq.~\eqref{eq:weight} is not a probability distribution). 
In fact we will see examples in section~\ref{sec:hadronic_examples} where the CIs for the sa-prescription increase monotonically with $F$.

\subsection{Connection to MHO estimates from scale variation} 
\label{sec:MHO-connection-7pt}
One may wonder about the practicality of the prescriptions in sections~\ref{sec:scale-marginalization_pres} and \ref{sec:weighted-sum-model}, because evaluating the integrals of the scale could be rather slow, as it may require to evaluate the perturbative approximation of $\Sigma$ for many values of the scales $\mu_i$. In many examples of higher-order computations for the LHC, the dependence on the factorisation and renormalisation scales is rather smooth. We can then approximate the integrals over scale in  eqs.~\eqref{eq:P_scale-marginalization_approx}  and~\eqref{eq:P_scale-average_approx} by a low-order quadrature rule.

\begin{figure}
    \centering
    \includegraphics[width=0.45\linewidth]{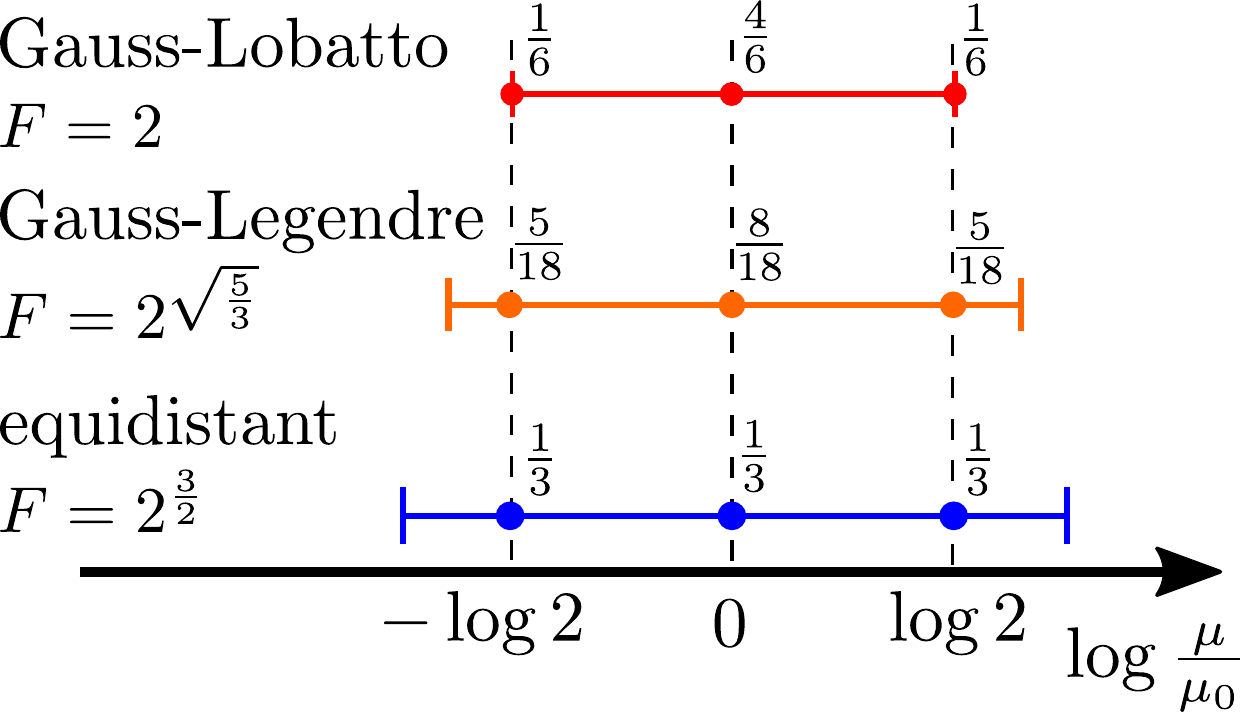}
    \includegraphics[width=0.45\linewidth]{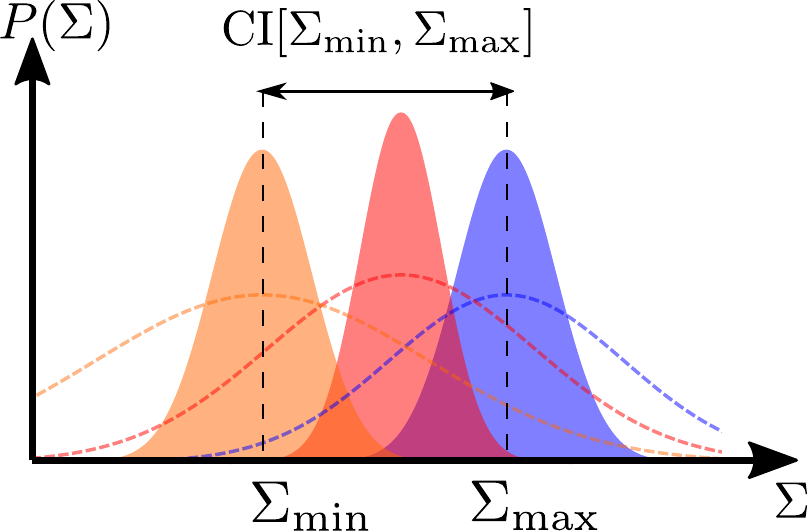}
    \caption{Left: The integration regions for different quadrature rules using scale variation points. Right: sketch of probability distributions at scale variation points when the width is smaller than scale variation interval (shaded) and when it is larger (dashed lines).}
    \label{fig:quadratures}
\end{figure}

We focus for now on observables depending on a single scale $\mu$, and we comment on multiple scales later. 
It is particularly convenient to choose the quadrature points to coincide with the ones used in the conventional scale variation, namely 
$\mu/\mu_0\in\{1/2,1,2\}$.%
\footnote{We note that the functional dependence on $\mu_R$ is determined by the running of the coupling constant and the 9-point $\mu_F, \mu_R$ grid can also be reconstructed from the 7-points given in eq.~\eqref{eq:7points} in some cases.
For a dynamical scale choice, this reconstruction is typically only feasible for distributions that allow for the determination of the numerical value of the scale that contributes to a specific bin within the uncertainties of the bin size.
}
For the three-point Gauss-Legendre quadrature rule this can be achieved by
 using the value of $F=2^{\sqrt{5/3}}\approx 2.45$ for $w(\mu)$ in eq.~\eqref{eq:weight}. Then the integral in  eq.~\eqref{eq:P_scale-average_approx} can be approximated as
 \beq\label{eq:Gauss-Legendre}
\int d\mu\,w(\mu)\,f(\mu) 
\approx w_{-1}\,f(\mu_0/2) + w_0\,f(\mu_0) + w_{+1}\,f(2\mu_0) \,,
\eeq
  where 
    the weights $w_i$  for the Gauss-Legendre quadrature are 
\beq\label{eq:weights}
w_0 = \frac{8}{18} \text{~~~and~~~} w_{\pm1} = \frac{5}{18}\,.
\eeq
This quadrature achieves high accuracy for smooth functions in the interval $-\log F<\log \mu/\mu_0< \log F$ (it is in fact exact if $f(\mu)$ is polynomial in $\log \mu$ up to degree 5; note that hadronic observables are never of this form, because some logarithms have been resummed to all orders into the PDFs and $\alpha_s(\mu)$). Similar formulas can of course be derived for quadrature rules of higher order and multi-dimensional $\mu$ integrals.
The advantage of the three-point rule in eq.~\eqref{eq:Gauss-Legendre} is that it allows one to easily convert the available data from the scale-variation approach into an approximation for the Bayesian probability distribution, which is scale-marginalised or scale-averaged over the interval $\mu_0/F < \mu <\mu_0 F$ with $F\approx2.45$. This approximation is valid as long as the three-point quadrature rule gives a good approximation of the integral. In applications, the dependence on the scale is usually smooth, so that low-order quadrature rules are expected to give reasonable approximations. We will illustrate this on examples in section~\ref{sec:hadronic_examples}.
One could also restrict the integration region to the interval $[\mu_0/2,2\mu_0]$, i.e. $F=2$, by using a quadrature rule where the nodes coincide with the end-points of the integration, e.g., the Gauss-Lobatto quadrature rule (accurate at cubic order). We compare the $\log \mu/\mu_0$ ranges of different quadrature rules using the same scale variation points in the left panel of figure~\ref{fig:quadratures}.

The quadrature rule in eq.~\eqref{eq:Gauss-Legendre} 
allows us to obtain an approximate bound on the probability that the true value lies in the range obtained from conventional scale variation. We will first discuss the case of an observable depending on a single perturbative scale $\mu$, and generalise the result later. Define the lower and upper values of the scale variation around some central scale $\mu_0$ as
\beq\bsp
\Sigma_{\text{min}}&\,:= \Sigma_{n}(k_{\textrm{min}}\mu_0) = \min_{k\in\{1/2,1,2\}}\Sigma_n(k\mu_0)\,,\\\Sigma_{\text{max}}&\,:=\Sigma_{n}(k_{\textrm{max}}\mu_0) = \max_{k\in\{1/2,1,2\}}\Sigma_n(k\mu_0)\,.
\esp\eeq
Consider the sa-prescription and a model such that
\begin{enumerate}
\item the model is symmetric,
\item $P(\Sigma^{(n+1)}(\mu)|\bfSigma(\mu))$ peaks only at $\Sigma^{(n+1)}(\mu)=0$, for all values of $\mu$,
\item the three-point quadrature in eq.~\eqref{eq:Gauss-Legendre} gives a good approximation of the integral.
\end{enumerate}
If these assumptions are satisfied, then we can show that the probability that $\Sigma\in [\Sigma_{\text{min}},\Sigma_{\text{max}}]$ is bounded from above:
\beq\label{eq:bound}
\cP_{\text{sa}}(\Sigma\in [\Sigma_{\text{min}},\Sigma_{\text{max}}]|\bfSigma_n) \lesssim 1-\frac{1}{2}(w_{\textrm{min}} + w_{\textrm{max}})\,,
\eeq
where $w_{\textrm{min}}$ is the quadrature weight associated to the scale $\mu=k_{\textrm{min}}\mu_0$ through eq.~\eqref{eq:Gauss-Legendre}, and similarly for $w_{\textrm{max}}$. The $\lesssim$ sign reminds us that the bound is only as accurate as the assumptions above, in particular the validity of eq.~\eqref{eq:Gauss-Legendre}. To understand the origin of this approximate bound, we first note that
\beq\label{eq:cumulative_version}
\cP_{\text{sa}}(\Sigma\in [\Sigma_{\text{min}},\Sigma_{\text{max}}]|\bfSigma_n) = \Phi(\Sigma_{\text{max}}) - \Phi
(\Sigma_{\text{min}})\,,
\eeq
where $\Phi(\Sigma)$ is the cumulative distribution $\Phi(\Sigma) := \int_{-\infty}^{\Sigma}d\Sigma'\,\cP_{\text{sa}}(\Sigma'|\bfSigma_n)$.
Using eqs.~\eqref{eq:P_scale-average_approx} and~\eqref{eq:Gauss-Legendre}, we express $\Phi(\Sigma)$ as a sum of cumulative distributions $\varphi(\Sigma,\mu) := \int_{-\infty}^{\Sigma}d\Sigma'\,P(\Sigma'-\Sigma_n(\mu)|\bfSigma_n(\mu))$ for a fixed choice of the scale:  
\beq\bsp  \Phi(\Sigma_{\text{min}})&\, \approx w_{-1}\,\varphi(\Sigma_{\text{min}},\mu_0/2)+w_{0}\,\varphi(\Sigma_{\text{min}},\mu_0)+w_{+1}\,\varphi(\Sigma_{\text{min}},2\mu_0)\\
 &\,\gtrsim w_{\textrm{min}}\,\varphi(\Sigma_{\text{min}},k_{\textrm{min}}\mu_0)\,,
 \esp\label{eq:Phimin}
\eeq
 where the last step follows from the fact that all three terms are positive. An illustration of the three functions $P(\Sigma-\Sigma_n(k\mu_0)|\bfSigma_n(k\mu_0))$, $k\in\{1/2,1,2\}$, is shown in figure~\ref{fig:quadratures} for two scenarios: when the distributions are narrow compared to the scale variation interval $[\Sigma_\text{min},\Sigma_\text{max}]$ (shaded areas) and when the distributions are wider than $[\Sigma_\text{min},\Sigma_\text{max}]$ (dashed lines).
  Note that we do not know which value of $k$ corresponds to which distribution in figure~\ref{fig:quadratures}. However, from the assumptions above it follows that only one of them peaks at $\Sigma_{\text{min}}$, namely the one corresponding to the value of $k=k_{\textrm{min}}$.  Since the distribution is by assumption symmetric around $\Sigma=\Sigma_{\text{min}}$, we must have $\varphi(\Sigma_{\text{min}},k_{\textrm{min}}\mu_0) = \frac{1}{2}$.
 Hence, we obtain
 \beq\label{eq:cumulative_low_bound}
 \Phi(\Sigma_{\text{min}}) \gtrsim \frac{1}{2}\,w_{\textrm{min}}\,.
 \eeq
 Using a similar argument, we can conclude
 \beq\label{eq:cumulative_upper_bound}
 \Phi(\Sigma_{\text{max}}) \lesssim 1-\frac{1}{2}\,w_{\textrm{max}}\,.
 \eeq
 Equation~\eqref{eq:bound} then follows upon inserting the bounds for $ \Phi(\Sigma_{\text{min}})$ and  $\Phi(\Sigma_{\text{max}})$ into eq.~\eqref{eq:cumulative_version}.

We can ask when the approximate bound in eq.~\eqref{eq:bound} is saturated. It is easy to see that the bound for the cumulative distribution in eq.~\eqref{eq:cumulative_low_bound} is saturated if the three distributions $P(\Sigma-\Sigma_n(k\mu_0)|\bfSigma_n(k\mu_0))$, $k\in\{1/2,1,2\}$ are narrow compared to their separation, because in that scenario we have $\Phi(\Sigma_{\textrm{min}}) \approx \frac{1}{2}\,w_{\textrm{min}}$. (see figure~\ref{fig:quadratures}). A similar argument holds for the bound in eq.~\eqref{eq:cumulative_upper_bound}, $\Phi(\Sigma_{\textrm{max}}) \approx 1-\frac{1}{2}\,w_{\textrm{max}}$. Hence, we conclude that the approximate bounds on the probability for $[\Sigma_{\text{min}},\Sigma_{\text{max}}]$ are saturated for observables and models for which the distributions $P(\Sigma-\Sigma_n(k\mu_0)|\bfSigma_n(k\mu_0))$, $k\in\{1/2,1,2\}$ are well separated.

 \begin{table}[!t]
   \begin{center}
 \begin{tabular}{c|c|c}
 \hline\hline
 Quadrature rule & $\beta_1$ &$\beta_2$ \\
 \hline
 Gauss-Legendre &$ 13/18\approx 72\%$ &  $ \approx 92\%$\\
 Gauss-Lobatto & $\phantom{11}5/6\approx 83\%$ & $ \approx 97\%$\\
 Equidistant & $\phantom{11}2/3\approx 66\%$ & $ \approx 89\%$\\
 \hline\hline
 \end{tabular}
 \caption{\label{tab:bounds}Approximate bounds on the probability for interval $[\Sigma_{\text{min}},\Sigma_{\text{max}}]$.}
 \end{center}
 \end{table}

 We can repeat exactly the same argument for observables depending on multiple scales $(\mu_1,\ldots,\mu_s)$ by using the same quadrature rule $s$ times for scale-averaging for each scale $\mu_i$ in eq.~\eqref{eq:Phimin}.
Each cumulative distribution $\varphi\left( \Sigma, \mu_1, \ldots,\mu_s \right)$ in the sum is multiplied by a product of $s$ quadrature weights. It is then straightforward to show that the bound becomes
 \beq\label{eq:bound_2}
\cP_{\text{sa}}(\Sigma\in [\Sigma_{\text{min}},\Sigma_{\text{max}}]|\bfSigma_n) \lesssim \beta_s\,,\textrm{  with  }\beta_s := 1-\frac{1}{2^{s}}\,(w_{\textrm{min}} + w_{\textrm{max}})^s\,.
\eeq

 Let us discuss some implications of the approximate bound in eq.~\eqref{eq:bound_2}. Under the assumptions spelled out above, eq.~\eqref{eq:bound_2} establishes a connection between the interval obtained by varying the scale up and down by a factor of $2$ and the Bayesian probability distribution constructed from the sa-prescription. The value of the approximate bound depends on the quadrature weights used to approximate the integral. For the three quadrature rules illustrated in figure~\ref{fig:quadratures}, the values of the bound are shown in table~\ref{tab:bounds} for $s=1,2$. Note that different quadratures correspond to different intervals over which the scale-averaging is done. In addition, the accuracy of the approximation in eq.~\eqref{eq:Gauss-Legendre} also depends on the order of the quadrature rule. 
Therefore the approximate bounds in table~\ref{tab:bounds} 
depend on how well the assumptions stated at the beginning of this section are satisfied. In particular, it \emph{does not} allow one to attach a definite probability that the true value lies in the range given by the scale variation. However, we can show that the $(100\beta_s)\%$ credibility interval  $\text{CI}_{100\beta_s} =[\Sigma_{100\beta_s}^\text{low},\Sigma_{100\beta_s}^\text{up}]$ will contain $[\Sigma_\text{min},\Sigma_\text{max}]$
and the two intervals (approximately) agree whenever the approximate bound is saturated. 
Indeed, from the definition of the credibility interval in eq.~\eqref{eq:CIdef} it follows that
\beq
\Phi(\Sigma_{100\beta_s}^\text{low}) = \frac{1}{2}(1- \beta_s) \textrm{  and  }\Phi(\Sigma_{100\beta_s}^\text{up}) = \frac{1}{2}(1+ \beta_s)\,.
\eeq
Equations~\eqref{eq:cumulative_low_bound} and~\eqref{eq:cumulative_upper_bound} and their generalisations for $s>1$ then imply:
\beq
 \Phi(\Sigma_{100\beta_s}^\text{low})\lesssim \Phi(\Sigma_{\text{min}})\le \Phi(\Sigma_{\text{max}})\lesssim \Phi(\Sigma_{100\beta_s}^\text{up})\,,
 \eeq
 and so the monotonicity of the cumulative distribution implies 
 \beq\label{eq:bound_sv}
 \Sigma_{100\beta_s}^\text{low} \lesssim \Sigma_{\text{min}}\le \Sigma_{\text{max}}\lesssim \Sigma_{100\beta_s}^\text{up}\,,
 \eeq
  and the claim follows.
  
 The previous discussion has an important consequence: Assume that we are in a situation with $s=2$ scales. Then we conclude from table~\ref{tab:bounds} that the 9-point scale variation interval satisfies
 \beq
 [\Sigma_{\text{min}},\Sigma_{\text{max}}] \subset \textrm{CI}_{100\beta_2} \approx \textrm{CI}_{89-97}\,.
 \eeq
 Clearly, the 7-point interval is also contained in the same interval.
 In the case where the approximate bound is saturated, we expect that for two scales  $(\mu_F,\mu_R)$ the intervals obtained from scale variation are comparable to $\approx 95\%$ credibility intervals. In practise, we often find that  observables are much less sensitive to one of the scales, namely, $\mu_F$, and we are effectively dealing with $s=1$ case, where $\beta_1\approx 66$--$83\%$. We will illustrate in section~\ref{sec:hadronic_examples} that the conventional scale variation intervals are often similar in size to CI$_{68}$ using Bayesian methods. The arguments presented in this section show that this is not a coincidence. Nevertheless we reiterate that the derivation of these bounds relies on specific assumptions. In particular, the sm-prescription and models with asymmetric probability distributions do not need comply with these requirements and the bounds can be broken. However, as we will see in the examples in section~\ref{sec:hadronic_examples}, in many cases the size of the CIs is similar for symmetric and asymmetric models.

\section{Modelling probability distributions at fixed scales}
\label{sec:no_scale}

In the previous section we have discussed two ways to define a probability distribution $\cP(\Sigma|\bfSigma_n)$ in the context of perturbation theory. In both approaches the computation of the distribution $\cP(\Sigma|\bfSigma_n)$ is reduced to an integral over the perturbative scale $\mu$. The integrand involves, on the one hand, the prior $P_0(\mu)$  in eq.~\eqref{eq:scale_prior} or the weighting function $w(\mu)$ in eq.~\eqref{eq:weight}, and, on the other hand, the probability $\cPFS(\Sigma|\bfSigma_n(\mu))$ evaluated at a fixed value $\mu$ of the perturbative scale. The latter is related through eq.~\eqref{eq:perturbative_P} to the probability distribution $\cPFS(\Sigma^{(n+1)}(\mu)|\bfSigma_n(\mu))$ for the size of the next perturbative order at the scale $\mu$, given the $n+1$ first orders computed at the same scale. Various models for $\cPFS(\Sigma^{(n+1)}(\mu)|\bfSigma_n(\mu))$ have been presented in the literature, in particular the CH-method of ref.~\cite{Cacciari:2011ze} and its generalisations of ref.~\cite{Bonvini:2020xeo}. All theses examples define symmetric distributions in the sense of eq.~\eqref{eq:sym_P}. In section~\ref{sec:scale_discussion} we have argued that symmetric distributions always lead to very special distributions $\cP(\Sigma|\bfSigma_n)$ after integration over $\mu$, with peaks that are correlated to the FAC and PMS points. This may introduce unwanted bias into the model. 

The goal of this section is to introduce a model that is not symmetric, which is a necessary condition for distributions not necessarily peaked at the FAC or PMS points. We start by presenting a framework to define different models for $\cPFS(\Sigma^{(n+1)}(\mu)|\bfSigma_n(\mu))$. This framework contains, in particular, the CH-method of ref.~\cite{Cacciari:2011ze} and its generalisations of ref.~\cite{Bonvini:2020xeo} as special cases. We then use this framework to define a new model that takes into account the signs of the corrections and leads to asymmetric distributions.

\subsection{The mathematical setup}
\label{sec:math_setup}

Consider the set consisting of elements $(\bfdelta,\bfp)$, where $\bfp\in\mathbb{R}^p$ is a vector and $\bfdelta = (\delta_0,\delta_1,\ldots)$ is an infinite sequence of real numbers representing the infinite series $\sum_{k=0}^\infty\delta_k$.
We prefer to think of the sequence $\bfdelta$ rather than the infinite series, because the order of the $\delta_k$'s matters. 

We make the following assumptions:
\begin{enumerate}
\item The parameters $\bfp$ are hidden, and we do not have access to them. Instead, they are distributed according to a known prior distribution $P_0(\bfp)$.
\item The sequence $\bfdelta$ and the hidden parameters $\bfp$ are not independent. For a given value of $\bfp$, we know the probability that the first $n+1$ elements of $\bfdelta$ take a certain value, i.e., for each $n\ge 0$, if $\bfdelta_n := (\delta_0,\delta_1,\ldots,\delta_n)$, we know the conditional probability $P^{(n)}(\bfdelta_n|\bfp)$.
\item The conditional probabilities $P^{(n)}(\bfdelta_n|\bfp)$ for different values of $n$ are related by marginalisation:
\beq\label{eq:marginalisation_constraint}
P^{(n-1)}(\bfdelta_{n-1}|\bfp) = \int d\delta_n\,P^{(n)}(\bfdelta_n|\bfp)\,,\text{ for all } n > 0\,.
\eeq
\end{enumerate}
We refer to a choice of $(P_0,P^{(0)},P^{(1)},P^{(2)},\ldots)$ as a \emph{model} for sequences $(\bfdelta,\bfp)$. We always assume that all probabilities are normalised, and we will often keep the superscript on $P^{(n)}$ implicit and use the simplified notation $P(\bfdelta_n|\bfp) := P^{(n)}(\bfdelta_n|\bfp)$.

Assume that we are given the first $n+1$ terms $\bfdelta_n$ of $\bfdelta$. The main idea is  that, {within a given model}, we can use Bayesian inference to obtain a conditional probability distribution $P(\delta_{n+1}|\bfdelta_n)$ for the next term $\delta_{n+1}$ in the sequence $\bfdelta$, given the first $n+1$ terms $\bfdelta_n$. The probability distribution $P(\delta_{n+1}|\bfdelta_n)$ is given by Bayes' formula:
\beq\label{eq:P_delta|delta}
P(\delta_{n+1}|\bfdelta_n) = \frac{P(\bfdelta_{n+1})}{P(\bfdelta_{n})}\,.
\eeq
The probability distributions for $P(\bfdelta_{n})$ and $P(\bfdelta_{n+1})$ are themselves obtained from Bayes' formula:
\beq\label{eq:P_delta}
P(\bfdelta_{n}) = \int d^p\bfp\,P(\bfdelta_{n}|\bfp)\,P_0(\bfp)\,.
\eeq
We refer to a model as \emph{symmetric} if it satisfies (cf.~eq.~\eqref{eq:sym_P})
\beq
P(\delta_0,\ldots,\delta_n) = P(|\delta_0|,\ldots,|\delta_n|)\,.
\eeq
A symmetric model always predicts a vanishing median and expectation value for $\delta_{n+1}$ given $\bfdelta_n$,
\beq
\langle \delta_{n+1}\rangle_{\bfdelta_n} = \delta_{n+1}^{\text{median}} = 0\,,\quad \text{ if the model is symmetric}\,,
\eeq
and all CIs computed from a symmetric model are symmetric around $\delta_{n+1}=0$.

\subsection{The geometric model} 
\label{sec:geometric-model}

The CH model assumes a universal bound $c$ for all coefficients $\delta_k(\mu)=\Sigma^{(k)}(\mu)/\alpha_s^k(\mu)$ for a fixed value of $\mu$.
In ref.~\cite{Bonvini:2020xeo} a variation of the CH model, called the \emph{geometric model}, was introduced to take into account the fact that the expansion parameter does not need to be equal to $\alpha_s$.%
\footnote{In ref.~\cite{Bagnaschi:2014wea} another modification of the CH model was introduced to take this effect into account. We do not explore the method of ref.~\cite{Bagnaschi:2014wea} in this paper.} The geometric model considers a sequence of coefficients $\delta_k(\mu)=\Sigma^{(k)}(\mu)/\Sigma^{(0)}(\mu)$ (so that $\delta_0(\mu)=1$),
whose magnitude is bounded from above by a geometric series. Consequently, the model has two hidden parameters $\bfp = (a,c)$, and the coefficients of the sequence $\bfdelta(\mu)$ must satisfy the constraint (for simplicity we suppress the dependence on the scale): 
\beq\label{eq:geometric_constraint}
\frac{|\delta_k|}{a^k} \le c\,,\quad\text{ for all }k\ge 0\,.
\eeq
The constraints for $k=0,1$ imply $c>1$ and $a>0$. Since we do not make any other assumption about the coefficients, we choose a flat distribution that takes into account eq.~\eqref{eq:geometric_constraint}~\cite{Bonvini:2020xeo}:
\beq\bsp\label{eq:P_cond_geom}
P_{\text{geom}}(\bfdelta_n|a,c) &\,= \prod_{k=0}^m \frac{1}{2a^kc}\,\Theta\left(c-\frac{|\delta_k|}{a^k}\right)\\
&\,=\frac{1}{(2c)^{n+1}a^{n(n+1)/2}}\,\Theta\left(c-\max\left(\frac{|\delta_0|}{a^0},\ldots, \frac{|\delta_n|}{a^n}\right)\right)\,.
\esp\eeq
The model is clearly symmetric.
For the prior on the hidden parameters, $a$ and $c$ are assumed independent, $P_0(a,c) = P_0(a)P_0(c)$. The prior on $c$ is given by 
\beq\label{eq:c_prior_CH}
P_0(c) = \frac{\varepsilon}{c^{1+\varepsilon}}\,\Theta(c-1)\,.
\eeq
Here $\varepsilon>0$ is a constant introduced to obtain a normalisable prior. The prior on $a$ is~\cite{Bonvini:2020xeo}:
\beq\label{eq:a_prior}
P_0(a) = (1+\omega)\,(1-a)^\omega\,\Theta(a)\,\Theta(1-a)\,,
\eeq
where $\omega>0$ suppresses the prior at $a=1$. Note that $\varepsilon$ and $\omega$ are constants, and they are not considered hidden parameters of the model. For comparisons with the asymmetric model defined in the next section, we will use $\epsilon=0.1$ and $\omega=1$.
The form of eq.~\eqref{eq:P_cond_geom} and the priors are simple enough that the integral in eq.~\eqref{eq:P_delta} can be performed in closed form, see ref.~\cite{Bonvini:2020xeo}.

Let us conclude this section with some comments. First, we note that the geometric model contains the original CH model as a special case, where $a=\alpha_s(\mu)$, i.e., the prior for $a$ is 
\beq
P_{0,\textrm{CH}}(a) = \delta(a-\alpha_s(\mu))\,,
\eeq
so that
\beq
P_{\textrm{CH}}(\bfdelta_n|c) = \int da\,P_{\text{geom}}(\bfdelta_n|a,c)\,P_{0,\textrm{CH}}(a)  = P_{\text{geom}}(\bfdelta_n|\alpha_s(\mu),c)\,.
\eeq
Second, so far we have chosen the coefficients $\delta_k(\mu)$ to be real numbers. We can assume that this sequence was obtained by evaluating some sequence of functions $\Sigma^{(k)}$ at some fixed value $\mu$, i.e., $\delta_k(\mu) = \Sigma^{(k)}(\mu)/\Sigma^{(0)}(\mu)$. 
Reference~\cite{Bonvini:2020xeo} proposes a variant where the probability distribution takes input from other values of the functions $\Sigma^{(k)}$. 
The \emph{scale variation model} of ref.~\cite{Bonvini:2020xeo} considers sequences $\bfdelta(\mu)$ whose coefficients satisfy a bound of the form:
\beq
\frac{|\delta_k(\mu)|}{r_k(\mu)} \le c\,, \quad\text{ for all }k\ge 0\,,%\, := \frac{c}{|\Sigma_k(\mu)|}\,\max_{\mu/f\le \nu\le f\mu}\left|\frac{\Sigma_k(\nu)-\Sigma_k(\mu)}{\log(\nu/\mu)}\right|
\eeq
where $r_k(\mu)$ encodes information on the derivative of $\Sigma_k(\mu)$ with respect to $\mu$. 
The probability distribution for \emph{fixed} $\mu$ is related to $P_{\textrm{CH}}$ via the relation (with $\alpha_s$ replaced by unity):
\beq
P_{\text{sc.-var.}}(\bfdelta_n(\mu)|c) = P_{\text{CH}}\left(\frac{\delta_0(\mu)}{r_0(\mu)},\ldots, \frac{\delta_n(\mu)}{r_n(\mu)}\Big|c\right)\,\prod_{k=0}^n \frac{1}{r_k(\mu)}\,.
\eeq
In this case the prior for $c$ is chosen in ref.~\cite{Bonvini:2020xeo} to be a Poisson distribution.

\subsection{An asymmetric geometric model: the $abc$-model}
\label{sec:abc_model}

The symmetry in the geometric model can be traced back to the fact that the constraint in eq.~\eqref{eq:geometric_constraint} defining the model only depends on the magnitude of the coefficients, and it is insensitive to their signs.
Alternative models that lead to asymmetric distributions and  include information about coefficient's signs are discussed in appendix B  of ref. \cite{Bonvini:2020xeo} (see in particular B.2 and B.5). 
Here, we consider a new model where the upper and lower bounds are different in magnitude.
More precisely, we have three hidden parameters $\bfp=(a,b,c)$, and we consider a sequence of coefficients $\delta_k=\Sigma^{(k)}/\Sigma^{(0)}$ such that (again, we suppress the dependence on the scale $\mu$)
\beq\label{eq:asym_geometric_constraint}
b-c\,\le \frac{\delta_k}{a^k}\le b+c\,,\quad\text{ for all }k\ge 0\,.
\eeq
We refer to this model as the \emph{asymmetric geometric model}, or the \emph{$abc$-model}.
The conditional probability distribution takes a form very similar to the one in the geometric model in eq.~\eqref{eq:P_cond_geom}:
\begin{align}\label{eq:P_cond_asym}
P_{abc}(\bfdelta_n|a,b,c) &\,= \prod_{k=0}^m \frac{1}{2|a|^kc}\,\Theta\left(c-\left|\frac{\delta_k}{a^k}-b\right|\right)\\
\nonumber&\,=\frac{1}{(2c)^{n+1}|a|^{n(n+1)/2}}\,\Theta\left(c-\max\left(\left|\frac{\delta_0}{a^0}-b\right|,\ldots, \left|\frac{\delta_n}{a^n}-b\right|\right)\right)\,.
\end{align}
Like in the geometric model, our choice of priors assumes that $a$ is independent of the parameters $b$ and $c$. However, since $b$ and $c$ jointly define the upper and lower bounds, we do not see any reason to consider them independent. Instead, we use a joint probability for $(b,c)$:
\beq
P_0(a,b,c) = P_0(a)\,P_0(b,c)\,,
\eeq
where the factors in the right-hand side are given by:
\beq\bsp\label{eq:abc_priors}
P_0(a) &\,= \frac{1}{2}(1+\omega)\,(1-|a|)^{\omega}\,\Theta(1-|a|)\,,\\
P_0(b,c) &\,= \frac{\varepsilon\,\eta^{\epsilon}}{2\xi c^{2+\varepsilon}}\,\Theta(c-\eta)\,\Theta(\xi c-|b|)\,.
\esp\eeq
Unlike in the geometric model, we do not constrain $a$ to be positive and only require that $|a|<1$. The prior $P_0(b,c)$ is characterised by three constants $\epsilon,\eta$ and $\xi$. As in eq.~\eqref{eq:c_prior_CH}, $\epsilon>0$ is used to make the prior normalisable. The second parameter $\eta>0$ determines
the minimal difference between the upper and lower bounds of the series, see eq.~\eqref{eq:asym_geometric_constraint}. Finally, $\xi>0$ restricts the allowed range of $b$, with $\xi = 0$ forcing $b$ to vanish and we recover a symmetric geometric model.
\begin{figure}
    \centering
    \includegraphics[width=0.9\linewidth]{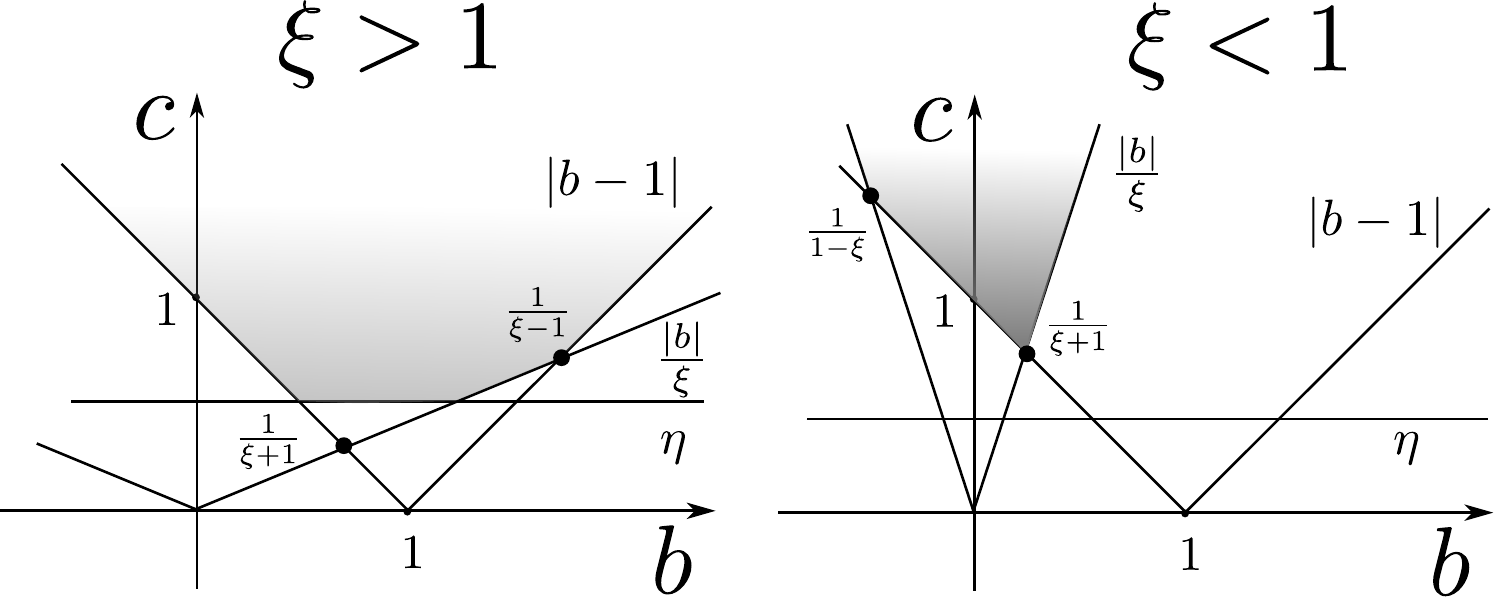}
    \caption{The shaded area indicates the ranges for $b$ and $c$ allowed by the prior $P_0(b,c)$, in eq.~\eqref{eq:abc_priors} and the inference from the universal input $\delta_0=1$ in eq~\eqref{eq:P_cond_asym} for $\xi>1$ (left) and $\xi<1$ (right). The position of $c=\eta>0$ line is arbitrary.}
    \label{fig:bc_ranges}
\end{figure}
We note that the constraints in eq.~\eqref{eq:asym_geometric_constraint} for $k=0$ imply $c>|b-1|$. Combining this with the prior in eq.~\eqref{eq:abc_priors}, we obtain $c > \max \{\eta, \frac{1}{1+\xi}\}$.
In figure~\ref{fig:bc_ranges} we show the ranges for $b$ and $c$ allowed by the prior in eq.~\eqref{eq:abc_priors} and $c>|b-1|$ for the two cases $\xi>1$ and $\xi<1$. We can clearly see from figure~\ref{fig:bc_ranges} that if $\eta< \frac{1}{1+\xi}$, then $\eta$ no longer affects the model inference. Similarly if $\xi > \frac{\eta+1}{\eta}>1$, i.e., $\eta> \frac{1}{\xi-1}$, then the inference becomes independent of $\xi$. Every subsequent term $\
\delta_k$ in the series adds the  constraint $c>\left|\frac{\delta_k}{a^k}-b\right|$, whose position depends on the values of $\delta_k$ and $a$.

We conclude this section by stressing that the quantities $\omega$, $\xi$, $\epsilon$ and $\eta$ are not hidden parameters of the model. Rather, they are (positive) constants defining a family of $abc$-models. In the next section we will study the model sensitivity to the choice of these constants. However, one has to keep in mind that this only explores the robustness for the specific model defined by eqs.~\eqref{eq:asym_geometric_constraint}, \eqref{eq:P_cond_asym} and \eqref{eq:abc_priors}. There is an infinite-dimensional functional space of possible models and priors that one might consider. We believe that the $abc$-model offers a good compromise between the flexibility of the model (it gives asymmetric probability distribution and can be equally well applied to monotonic and simple alternating series) and the transparency and simplicity of the model.

\subsubsection{Sensitivity to the choice of priors}
\label{sec:sensitivity_to_priors}

\begin{figure}
 \centering
    \includegraphics[width=\linewidth]{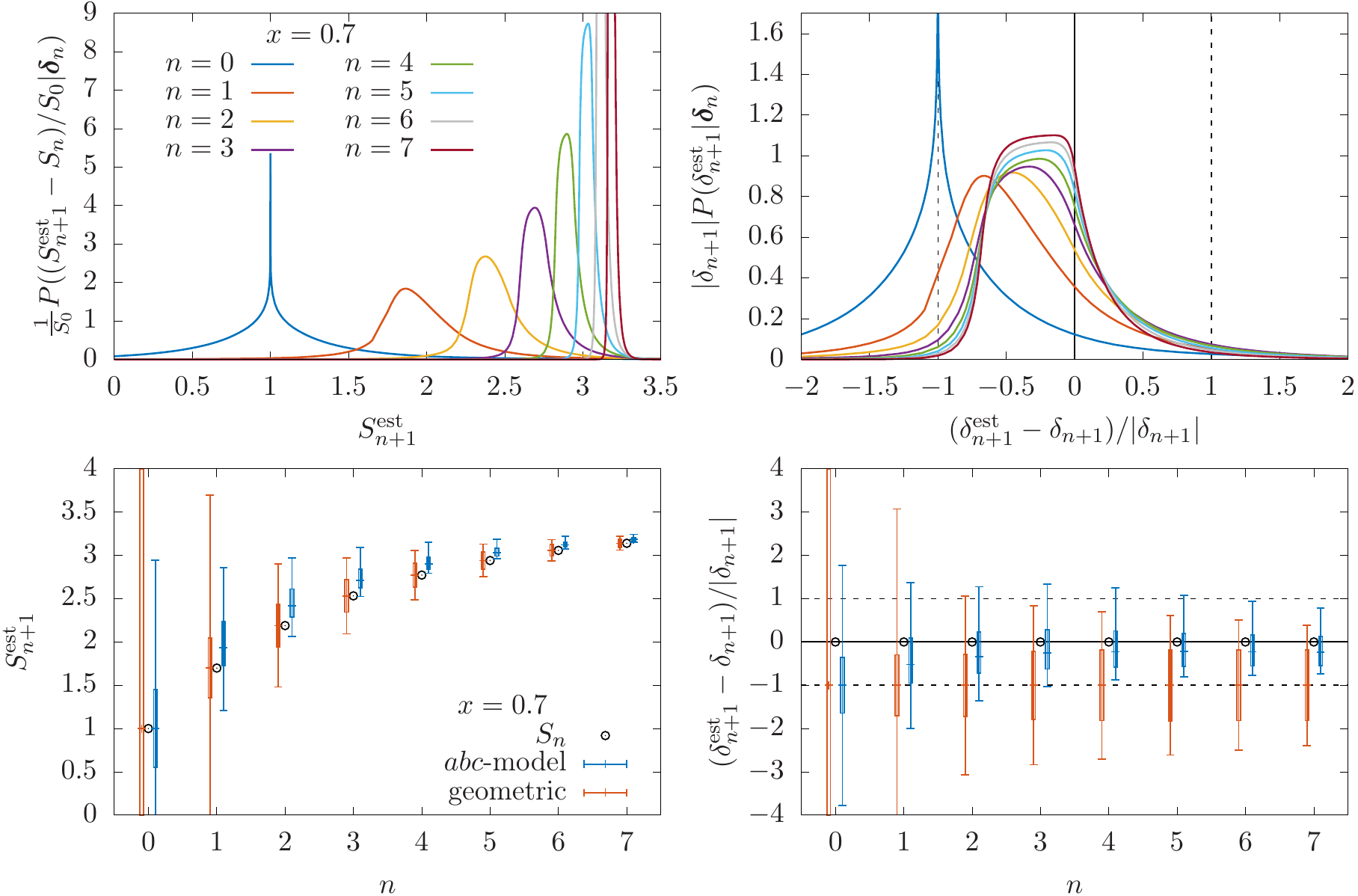}
    \caption{\label{fig:geom_progression}
     Top left panel: The probability distribution from the $abc$-model for $S_{n+1}^\text{est}$  for different values of $n$ for the geometric series with $x=0.7$. Top right panel: The same as the left panel, but we show the probability for the scaled deviation from the known correction $(S_{n+1}^\text{est}-S_{n+1})/|S_{n+1}-S_{n}|$.
    Bottom left panel: The median (plus), 68\% CI (errorbox) and 95\% CI (errorbar)  for the posterior of $S_{n+1}^\text{est}$ , computed from the $abc$ (blue) and geometric (red) models using information on the previous orders. The exact values of $S_n$ are shown as black circles. Bottom right panel: The same as the left panel, but the exact $S_n$ value is subtracted from $S_{n+1}^\text{est}$ and the difference is normalised by $|S_{n+1}-S_{n}|$.
      }
    \end{figure}

Let us illustrate the $abc$-model and its sensitivity to the choice of priors on the example of the geometric series $\delta_k = x^k$, for $|x|<1$. Then there exists $a=x$ for which the region of values of $(b,c)$ excluded by eq.~\eqref{eq:P_cond_asym} collapses to $c>|1-b|$. We see from figure \ref{fig:bc_ranges} that $\eta>0$ prevents $c$ from reaching zero. Setting $\eta\to0$ and $\xi\to\infty$ turns off any constraints by the prior $P_0(b,c)$, and the $abc$-model can describe such a geometric series arbitrary well with  $a\to x, b\to 1,c\to 0$  in eq.~\eqref{eq:P_cond_asym}.

We compute the conditional probability for the estimated next term in the series $\delta_{n+1}^\text{est}$:
\begin{equation}
    P_{abc}(\delta_{n+1}^\text{est}|\bfdelta_n)=\frac{P_{abc}(\bfdelta_{n+1})}{P_{abc}(\bfdelta_n)}\,,
\end{equation}
where
\begin{align}
P_{abc}(\bfdelta_n)&=\int da\, db\,dc\, P_{abc}(\bfdelta_n|a,b,c)P_0(a)P_0(b,c)\\
\nonumber&=\int da\frac{P_0(a)}{|a|^{n(n+1)/2}}\int db\,\frac{\epsilon\,\eta^{\epsilon}}{2^{n+2}\xi(n+2+\epsilon) } \max\left(\eta, \max_{0\leq k\leq n}\left|\frac{\delta_k}{a^k}-b\right|, \frac{|b|}{\xi}\right)^{-(n+2+\epsilon)}
\end{align}
For fixed $a$, the integral over $b$ can be done analytically and expressed in terms of maximum and minimum functions of $\frac{\delta_k}{a^k}$, $\eta$ and $\xi$.
In appendix~\ref{app:abc_model} we show how to evaluate the distribution $P_{abc}(\bfdelta_n)$ analytically in terms of Gauss' hypergeometric function, allowing for a fast and efficient numerical implementation of the model.

In the top left panel of figure~\ref{fig:geom_progression} we show the probability distributions for  the partial sums $S_n := \sum_{k=0}^n\delta_k = \sum_{k=0}^n x^k$ for $x=0.7$ and $n\leq7$ using the $abc$-model with parameter values $(\epsilon,\omega,\xi,\eta) = (0.1,1,2,0.1)$ (see the discussion below for the choice of these values). 
 For $n=0$, the probability distribution is symmetric and centred around $S_0=1$. For $n>0$ the distributions are clearly not symmetric and become more and more peaked as $n$ increases.  
In the top right panel we show the probability distributions for the scaled deviation from the known correction $(S_{n+1}^\text{est}-S_{n+1})/|S_{n+1}-S_{n}| =(\delta_{n+1}^\text{est}-\delta_{n+1})/|\delta_{n+1}| $.
This allows us to compare different orders without the suppression of the expansion parameter. In this plot the $S_{n+1}$ value corresponds to zero on the $x$-axis, while $S_n$  corresponds to $\pm 1$ (depending on the sign of $\delta_{n+1}$). Again, for $n=0$ the distribution is centred around the initial value $S_0$, but for each subsequent order, the distribution shifts towards the true value. We note that the shape of the distribution does not change significantly beyond $n=3$, so that the narrowness of the distributions in the left panel is solely due to the higher power of the expansion parameter.
In the simple case of a geometric series, the scaled distributions are controlled by the prior $P_0(b,c)$. For $\eta=0.1$ and $\xi=2$, the minimum of $c$ is at $(b,c)=(\frac{\xi}{1+\xi},\frac{1}{1+\xi})\approx (0.6,0.3)$ (see figure~\ref{fig:bc_ranges}), which gives a good approximation for the centre position and the (half) width of distributions seen in figure~\ref{fig:geom_progression}. As explained earlier, increasing $\xi$ and reducing $\eta$
can narrow down the posterior distribution around the exact geometric series value $x^{n+1}$.

In the bottom panels of figure~\ref{fig:geom_progression} we show the 68\% and 95\% CIs for the $abc$ (blue) and geometric models (red) (we use the same values $(\omega,\epsilon)=(1,0.1)$ for both models). On the left we show the CIs for $S_{n+1}^\text{est}$, while on the right we again scale to the known size of the next term in the series.
We see that for $n>0$ the 68\% CIs computed with the $abc$-model encompass the next value of $S_n$. This is not the case for the geometric model, even though the 68\% CIs are larger in magnitude. This is due to the fact that the geometric model is symmetric around the last known order, thereby not accommodating the fact that all the coefficients $\delta_k$ are positive. In contrast, the CIs of the $abc$-model are systematically shifted towards the exact values for $n>3$, and the 95\% CIs even exclude the previous value $S_n$.
However, one must remember that these features of the $abc$-model may not always be desired in practical applications. and can be controlled by the priors (see below).

For completeness, in appendix~\ref{app:plots} we have also included a similar plot showing the alternating series with $x=-0.7$. Here we just note that $abc$-model correctly predicts the negative sign of the expansion parameter and remains consistently shifted towards the next value in the geometric series $\delta_{n+1}$. In all other respects the distributions and the CIs are identical to the ones shown in figure~\ref{fig:geom_progression}.

\begin{figure}
 \centering
    \includegraphics[width=\linewidth]{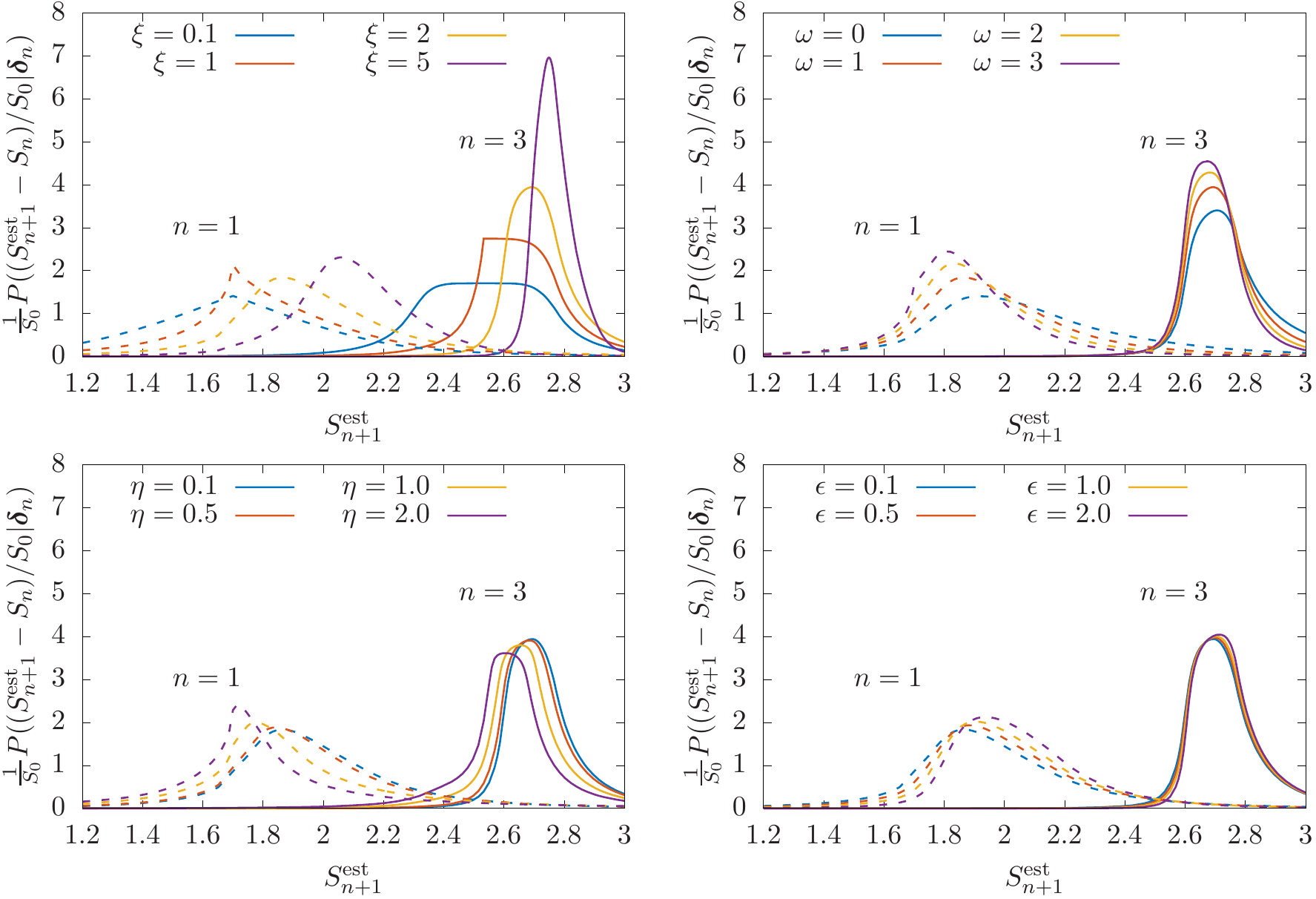} 
    \caption{\label{fig:geom_prior}The dependence of the probability distribution $1/S_0 P_{abc}( (S_{n+1}^\text{est}-S_n)/S_0|\bfdelta_{n})$ for geometric series for $n=1$ (dashed) and $n=3$ (solid) on the parameters $(\epsilon,\omega,\eta,\xi)$ for 4 selected values, with the others held fixed at $(\epsilon,\omega,\eta,\xi)=(0.1,1, 0.1,2)$.
    }
    \end{figure}

In figure~\ref{fig:geom_prior} we study the dependence of $P_{abc}(\delta_{n+1}|\bfdelta_n)$ on the choice of the values for $(\epsilon,\omega,\eta,\xi)$ that define the priors on $(a,b,c)$ in eq.~\eqref{eq:abc_priors}. We vary each parameter independently, with the other held fixed at their default values $(\epsilon,\omega,\xi,\eta) = (0.1,1,2,0.1)$. We see that, as expected, the dependence  on $\epsilon$, $\omega$, and $\eta$ reduces as $n$ increases. The $\xi$-dependence, however, does not disappear completely with increasing $n$. This is due to the fact that $\xi$ controls the amount of asymmetry of $P_{abc}(\delta_{n+1}|\bfdelta_n)$. For small values of $\xi$, the distribution approaches a symmetric distribution around the last known value $S_n$, while for large $\xi$ it becomes much more peaked around the next value $S_{n+1}$. Empirically, we find that our choice $\xi=2$ allows distributions to be shifted towards the next term, but keeping the width of the distribution of the same size as the shift. Next we look at $\omega$, which controls the suppression of the prior for $a$ around $|a|=1$  ($\omega=0$ corresponds to a flat prior). Because here we consider a rather large value of the expansion parameter $x=0.7$, increasing $\omega$ slightly reduces the expectation value of $a$ and shifts the distribution to smaller values of $S_{n+1}^\text{est}$.
Our default choice of $\omega=1$ biases the hidden model parameter inference towards small $|a|$ values.
In the bottom left panel of figure~\ref{fig:geom_prior} we see that larger values of $\eta$ shift the posterior distribution towards the previous order. $\eta$ plays a similar role to $\xi$ in controlling the allowed minimum value of $c$. We use a small value of $\eta=0.1$, which is in fact superseded by the constraint from $\xi$ (see figure~\ref{fig:bc_ranges}).
Finally, in contrast to $\eta$, for larger values of $\epsilon$ the model becomes slightly more peaked and shifted towards the next term in the series. We choose a small value $\epsilon=0.1$ to maintain fairly flat distribution in the prior of $\log c$.

\section{Quantities without explicit scale dependence}
\label{sec:qft_examples}
The goal of this section and the next is to illustrate the concepts and models from the previous sections on various examples of observables in QFT. We start by investigating several quantities whose perturbative expansion does not have an explicit dependence on the perturbative scales, or for which scale dependence is not relevant. This allows us to illustrate the application of the $abc$-model to genuine perturbative expansions from QFT.
The sequence $\bfdelta_n$ is the sequence of the $n+1$ first perturbative coefficients, normalised such that $\delta_0=1$
\beq
\label{eq:deltannorm}
\delta_k := \Sigma^{(k)}/\Sigma_0\,.
\eeq
The rescaling introduces a Jacobian into eq.~\eqref{eq:perturbative_P}, which now takes the form
\beq\label{eq:perturbative_P_delta}
P(\Sigma|\bfSigma_n)\approx \frac{1}{\Sigma_0}P_{abc}\left(\frac{\Sigma}{\Sigma_0}-\sum_{k=0}^n\delta_k\Big|\bfdelta_n\right).
\eeq

We then use this probability distribution to compute  CIs for $\Sigma$ within the $abc$-model using perturbative input through N$^n$LO. These intervals estimate the size of the missing N$^{n+1}$LO terms, and so they serve as measures of the MHO uncertainty at N$^n$LO. To assess the validity of this procedure, we show in each case how the size of the CIs computed at different perturbative orders compares to the actual size of the next order whenever it is available. By abuse of notation, we will use both
$P_{abc}(\Sigma_{n+1}^\text{est}|\bfdelta_n)$ and $P_{abc}(\delta_{n+1}^\text{est}|\bfdelta_n)$,  where the relation to eq.~\eqref{eq:perturbative_P_delta}  has to be inferred from the argument.

Before discussing explicit examples of perturbative expansions in QFT, let us make some general comments. First, we stress that the approximation in eq.~\eqref{eq:perturbative_P_delta} is only valid if we are in the perturbative regime. It is in particular not valid once we enter the regime where the perturbative series starts to diverge. In that situation the right-hand side of eq.~\eqref{eq:perturbative_P_delta} may still represent a valid probability distribution for the next perturbative coefficient, but this distribution is no longer a good approximation of the left-hand side, which is the probability distribution for the true value of $\Sigma$. We note that for most applications only very few perturbative orders are available, and this will not be a problem.
Second, it is possible to be in a situation where the model happens to describe the input data extremely well. The $abc$-model strongly favours a perfect geometric series. If the available perturbative data closely follow a geometric progression up to a certain order by some numerical coincidence, then the resulting distributions can be strongly peaked. In such a scenario, the 68\% or 95\% CIs may even exclude the highest available order $\Sigma_n$. This is a \emph{design feature} of the $abc$-model that can be controlled by the priors (see section~\ref{sec:abc_model}).

We stress that regardless of how well the
truncated perturbative series is described by a chosen model,
there is no guarantee that the actual next order in perturbative QFT or the measured experimental value of $\Sigma$ will fall into a specific choice of CI. Therefore, the choices of the model, the priors and the CI used to estimate the MHO uncertainty of an N${}^n$LO computation should be carefully scrutinised for each new observable separately.

\subsection{The cusp anomalous dimension}

Our first example is the light-like cusp anomalous dimension. It is known through four loops in QCD and $\mathcal{N}=4$ Super Yang-Mills (SYM)~\cite{Brandt:1981kf,Korchemsky:1987wg,Moch:2004pa,Grozin:2014hna,Grozin:2015kna,Henn:2016men,vonManteuffel:2016xki,Davies:2016jie,Lee:2016ixa,Boels:2017skl,Boels:2017ftb,Moch:2017uml,Moch:2018wjh,Grozin:2018vdn,Lee:2019zop,Henn:2019rmi,Bruser:2019auj,vonManteuffel:2019wbj,Henn:2019swt,vonManteuffel:2020vjv}. In the planar limit of the $\mathcal{N}=4$ SYM theory, it is even known to all orders in perturbation theory from integrability~\cite{Beisert:2006ez}. The cusp anomalous dimension is therefore an ideal candidate to study the applicability of Bayesian techniques to derive MHO uncertainties for perturbative field theory quantities.

\paragraph{Planar $\mathcal{N}=4$ SYM.}
The cusp anomalous dimension in planar $\mathcal{N}=4$ SYM admits a perturbative expansion:
\beq\label{eq:planar_cusp}
\gamma_K(\lambda) = \sum_{k=1}^\infty \gamma_K^{(k)}\,\lambda^k\,.
\eeq
Here the expansion parameter is the 't Hooft coupling $\lambda = \frac{g_s^2\,N_c}{8\pi^2}$, where $g_s$ is the Yang-Mills coupling and $N_c$ is the number of fundamental SU$(N_c)$ colours. Note that $\mathcal{N}=4$ SYM is a conformal field theory (even away from the planar limit), and so the coupling is not UV renormalised and does not depend on any renormalisation scale. The first few expansion coefficients are shown in table~\ref{tab:cusp_N=4_planar}. It was observed that the series in eq.~\eqref{eq:planar_cusp} is convergent if  $|\lambda|\lesssim \frac{1}{8}$.

\begin{table}[!th]
\begin{center}
\begin{tabular}{c|ccccccccccccccc}
\hline\hline
$k$ & 1 & 2 & 3 & 4 & 5 & 6 & 7 
\\
\hline
$\gamma_{K}^{(k)}$ & $4$ & $-6.57$ & $23.81$ & $-117.17$ & $662.62$ & $-4028.78$ & $25618.7$ 
\\
$\gamma^{(k)}_{\rm QCD}$ & $1$ & $1.73$ & $2.80$ & $2.89$ & $-$ & $-$ & $-$
\\
\hline\hline
\end{tabular}
\caption{\label{tab:cusp_N=4_planar}The first few terms in the perturbative series of the cusp anomalous dimension for planar $\mathcal{N}=4$ SYM,  eq.~\eqref{eq:planar_cusp}, and QCD, eq.~\eqref{eq:QCD_cusp}. }
\end{center}
\end{table}

\begin{figure}
    \centering
    \includegraphics[width=\linewidth]{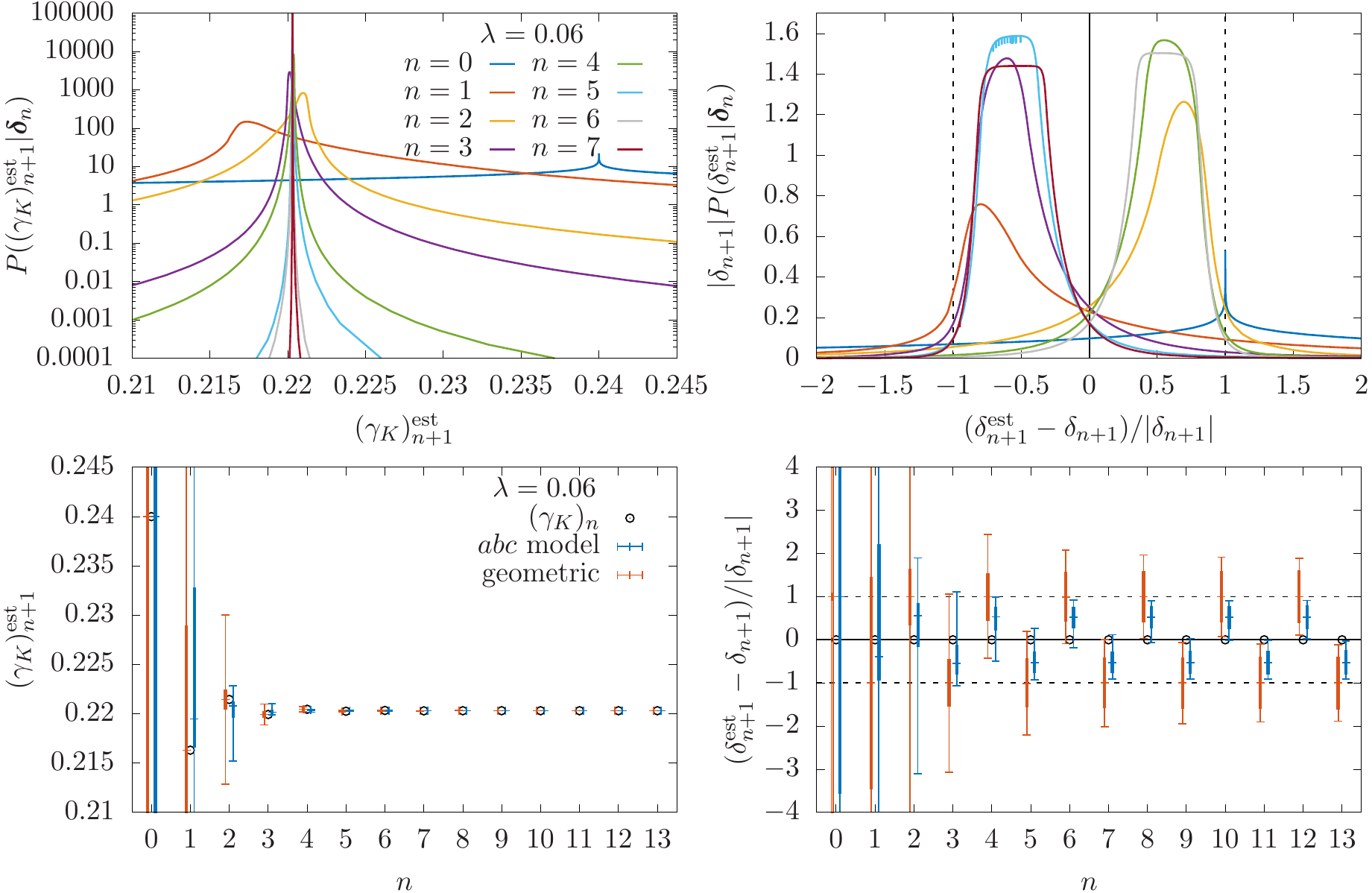}
    \caption{Top left panel: The probability distribution from the $abc$-model for the cusp anomalous dimension $(\gamma_K)_{n+1}^\text{est}$ in planar $\mathcal{N}=4$ SYM  for $\lambda=0.06$ and for different values of $n$. Top right panel: The same distributions
    normalised to the exact N${}^{n+1}$LO  correction.
    Bottom left panel: the median (plus), 68\% CI (errorbox) and 95\% CI (errorbar) for the posterior of $(\gamma_K)_{n+1}^\text{est}$ , computed from the $abc$ (blue) and geometric (red) models using information on the previous orders. The exact value of $(\gamma_K)_n$ is shown as black circles. Bottom right panel: CIs scaled to the exact N${}^{n+1}$LO correction.}
    \label{fig:SYM_CUSP_AD2}
\end{figure}
\begin{figure}
    \centering
    \includegraphics[width=\linewidth]{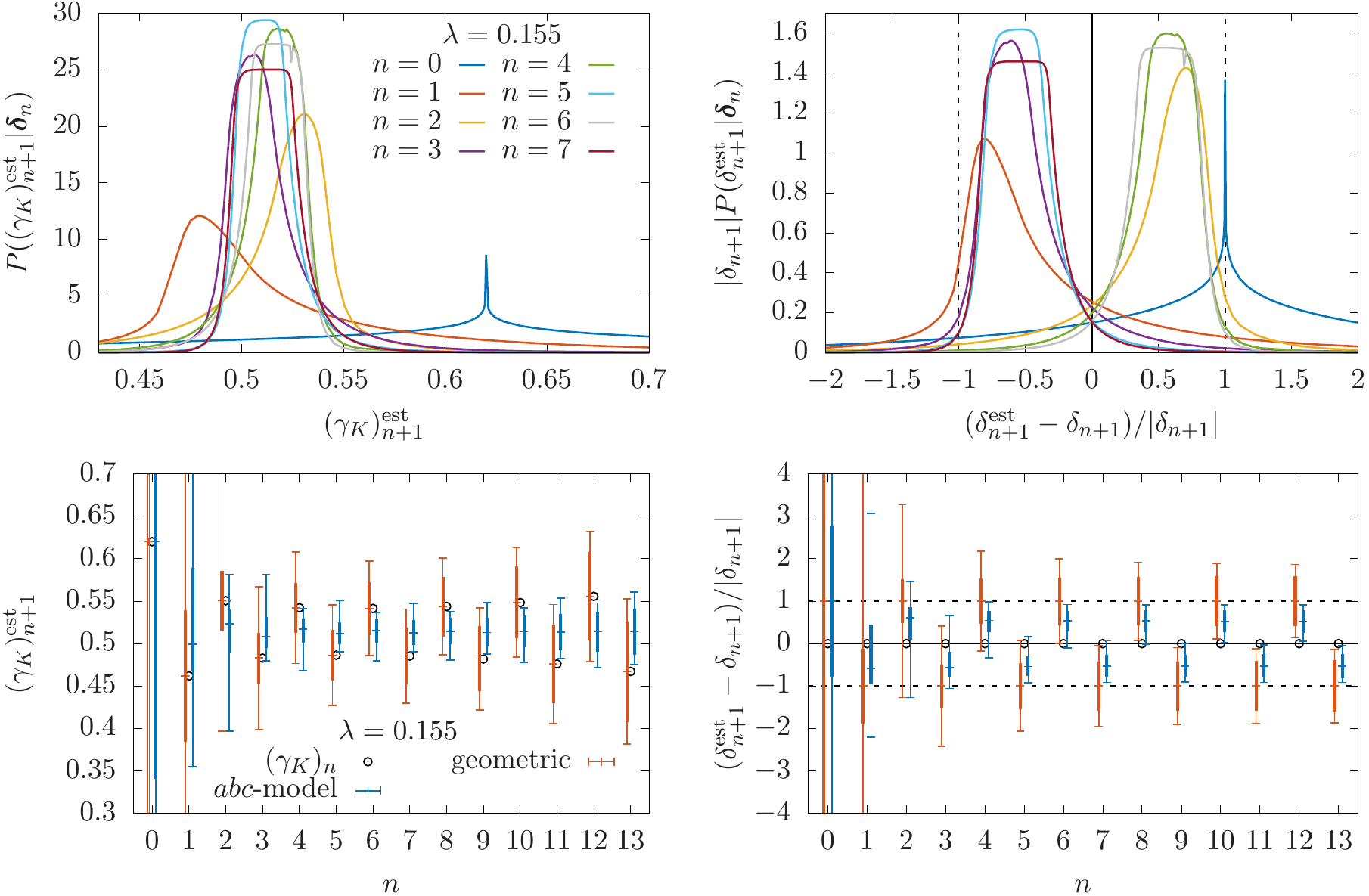}
    \caption{
     The probability distributions and CIs for the cusp anomalous dimension $(\gamma_K)_{n+1}^\text{est}$ in planar $\mathcal{N}=4$ SYM  for $\lambda=0.155$ and for different values of $n$. See the caption of figure~\ref{fig:SYM_CUSP_AD1} for details.}
    \label{fig:SYM_CUSP_AD1}
\end{figure}

In figures~\ref{fig:SYM_CUSP_AD2} and~\ref{fig:SYM_CUSP_AD1} we show the probability distributions obtained from the $abc$-model for the partial sums
\begin{equation}
    (\gamma_K)_n:= \sum_{k=1}^{n+1} \gamma_K^{(k)}\,\lambda^k\,,
\end{equation}
with $\lambda = 0.06$ and $\lambda=0.155$. The series $\bfdelta_n$ is normalised according to eq.~\eqref{eq:deltannorm}.

For $\lambda=0.06<\frac{1}{8}$, the perturbative series (which is alternating) is expected to converge. In figure~\ref{fig:SYM_CUSP_AD2} we clearly see a convergent pattern in the left top and bottom panels. The distributions quickly become very peaked around the limiting value with very small $68\%$ and $95\%$ CIs. In the right panels, we show the distributions and CIs normalised to the exact N${}^{n+1}$LO correction. This effectively cancels the suppression due to increasing power of the expansion parameter $\lambda$.  We see that after such a rescaling, the distributions do not change significantly for $n>2$ (up to a trivial reflection around the vertical axis, because of the alternating nature of the expansion). We note that
in contrast to the case of an exact geometric series (see figure \ref{fig:geom_progression}), here the posterior distribution for $\delta_{n+1}^\text{est}$ is centred around the half of the exact value $|\delta_{n+1}|$. As the series progresses, the scaled  distribution reaches a peak for $n=5$. After that it becomes steeper and broader.
In the bottom right panel of figure~\ref{fig:SYM_CUSP_AD2}, we see that the 95\% CIs shrink more than the 68\% CIs. For comparison we also show the CIs for the geometric model, which are significantly wider, but are no better in capturing the next order than the narrower CIs from the $abc$-model. In both cases the 95\% CIs appear to give a reasonable estimate for the size of $\delta_{n+1}$, even though the series is clearly not a perfect geometric series.

For $\lambda=0.155$ the series is asymptotic and starts to diverge after the seventh perturbative order, i.e. $n>6$. This behaviour is reflected in the probability distributions in the left panel of fig.~\ref{fig:SYM_CUSP_AD1}: they become peaked around the limiting value up to the inclusion of the sixth order (the sixth and seventh order corrections are numerically almost identical). After that a plateau develops which becomes broader as we increase the perturbative order.
Consequently, the width of the CIs increases steadily after the seventh order (bottom left panel). However, the rescaled distributions (right panels) look essentially identical to the convergent $\lambda=0.06$ case. This is not surprising, because the two series differ only by the value of the expansion parameter, which is scaled away in the right panels.
Therefore, like before the $95\%$ CIs give a good estimate of the size of the next order corrections, even in the case of an asymptotic series. We note, however, that for $n>6$ we enter the regime where the next order is not expected to give a good approximation of the true value of $\gamma_{K}$. Hence, eq.~\eqref{eq:perturbative_P_delta} is not valid, and so the credibility intervals are not expected to give a good estimate of the uncertainty attached to the perturbative result.

\paragraph{The cusp anomalous dimension in QCD.}
In QCD, the cusp anomalous dimension is known up to four loops \cite{Brandt:1981kf,Korchemsky:1987wg,Moch:2004pa,Grozin:2014hna,Grozin:2015kna,Henn:2019swt}. It depends on the representation of the Wilson loop and the matter content of the theory. Here we focus on the fundamental representation, and we work with 5 massless flavours.  In this case we have
\beq\label{eq:QCD_cusp}
\Gamma^{\rm QCD}_{\rm cusp }(\mu_R)= C_F  \sum_{k=1}^\infty \left( \frac{\alpha_s(\mu_R)}{\pi}\right)^k \gamma^{(k)}_{\rm QCD}\,,
\eeq
with $C_F = 4/3$ and the numerical values of coefficients $\gamma^{(k)}_{\rm QCD}$ are summarised in table~\ref{tab:cusp_N=4_planar}.
We note that the scale dependence enters solely through the coupling constant $\alpha_s$. Without any context, the scale is completely arbitrary and there exists no physically motivated choice for the central scale without specifying the physical process.

\begin{figure}
    \centering
    \includegraphics[width=\linewidth]{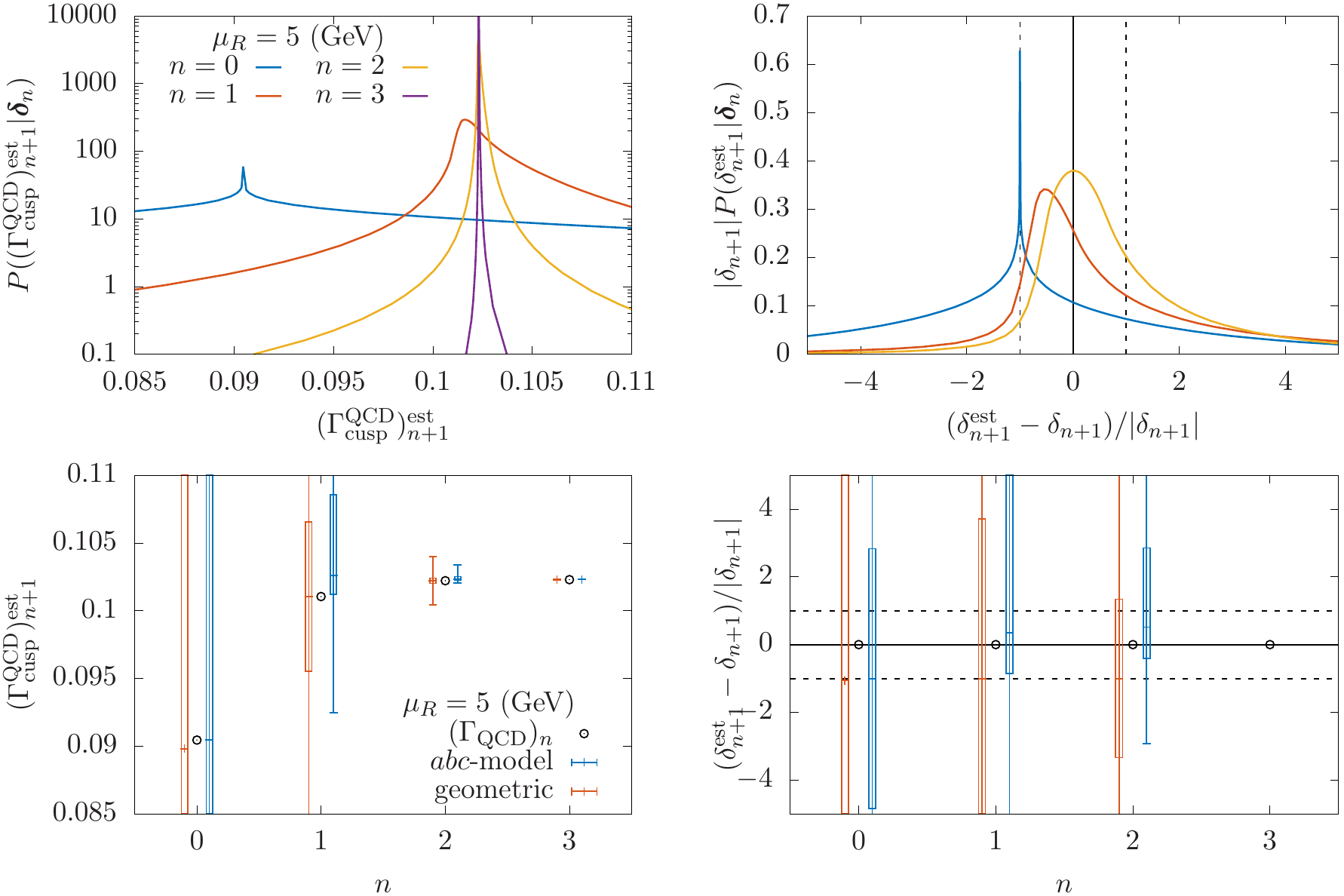}
    \caption{
    Top left panel: The probability distribution from the $abc$-model for the QCD cusp anomalous dimension $(\Gamma^\text{QCD}_\text{cusp})_{n+1}^\text{est}$ evaluated with $\alpha_s(\mu_R=5\,\text{GeV})\approx 0.213$ and for different values of $n$. Top right panel: The same distributions
    normalized to the exact N${}^{n+1}$LO  correction.
    Bottom left panel: the median (plus), 68\% CI (errorbox) and 95\% CI (errorbar) for the posterior of $(\Gamma_\text{cusp}^\text{QCD})_{n+1}^\text{est}$ , computed from the $abc$ (blue) and geometric (red) models using information on the previous orders. The exact values of $(\Gamma_\text{cusp}^\text{QCD})_n$ are shown as black circles. Bottom right panel: CIs scaled to the exact N${}^{n+1}$LO correction.
    }
    \label{fig:QCD_cusp_summary}
\end{figure}

In figure~\ref{fig:QCD_cusp_summary} we show (with $\alpha_s(\mu_R=5\,\text{GeV})\approx 0.213$) the probability distributions and CIs for the partial sums $(\Gamma^{\rm QCD}_{\rm cusp })_n= C_F  \sum_{k=1}^{n+1} \left( \frac{\alpha_s}{\pi}\right)^k \gamma^{(k)}_{\rm QCD}$ obtained from the $abc$-model.
In the top left panel we see that the series is rapidly convergent, and the probability distributions become progressively more peaked. From the rescaled distribution in the top right panel we see that the probability distribution for $n=2$ peaks at the exact value of the correction. We do not show scaled results for $n=3$, because the exact five-loop result is not known yet.
In the bottom left panel we show the 68\% and 95\% CIs for the $abc$-model and the geometric model. We see that, unlike in the case of planar $\mathcal{N}=4$ SYM, the $68\%$ CIs always include the next correction, thus giving a reliable estimate of the MHOs (at least for the first few orders considered here).
We emphasise that this is particularly non-trivial at four loop order, where Casimir scaling is violated for the first time, and a new quartic Casimir enters.
Estimates of the size of the four-loop corrections based on Pad\'e approximants from the first three orders
were within $5\%$ of the four-loop contributions that admit Casimir scaling for $n_f=5$, but were 57\% off from the complete result~\cite{Moch:2005ba}.
The $68\%$ CI obtained from the first three loop orders in the $abc$-model correctly captures the size of the four-loop result, including the contribution from the quartic Casimir.

From the bottom right plot we see that the median of the posterior distribution for the $abc$ model (marked by a plus) is closer to the exact value than for the geometric model,
although the $abc$-model appears to expect larger higher-order correction, and the distributions are skewed towards larger corrections.
For $n=4$ the CIs are not visible in the bottom left panel due to the suppression of the expansion parameter.
Using the $abc$-model we estimate that CIs for MHO corrections given the 4-loop order result for $\mu_R=5\,\text{GeV}$, i.e., $\Sigma=\Gamma^\text{QCD}_\text{cusp}-(\Gamma^\text{QCD}_\text{cusp})_4$, is
\begin{equation}
\text{CI}_{68}=C_F \left( \frac{\alpha_s(\mu_R)}{\pi}\right)^5 [2.1, 9.5],\quad
\text{CI}_{95}=C_F \left( \frac{\alpha_s(\mu_R)}{\pi}\right)^5 [-0.38, 21],
\end{equation}
where we scaled out $C_F \left( \frac{\alpha_s(\mu_R)}{\pi}\right)^5$ for visibility. We see that the $abc$-model at 68\% credibility level would expect a positive higher-order correction. In contrast, the geometric model estimates a symmetric interval  CI$_{68}=[-5.4,5.4]$ with the same normalisation .

\subsection{On-shell and $\overline{\text{MS}}$-scheme quark masses}
\label{sec:quark_masses}
Our next example is the relation between the heavy quark mass $m$ in the on-shell scheme  and the mass $\mMS$ in  the $\overline{\text{MS}}$-scheme.
The bare mass $m_0$ is related to the on-shell and $\overline{\text{MS}}$  quark masses via the renormalisation factors $Z_m$
\beq
m_0 = Z_m^{\overline{\rm MS}}(\mu_R)\, \mMS(\mu_R) =Z_m^{\rm OS}\, m\, .
\eeq
For heavy quarks it is possible to perturbatively compute these renormalisation factors. They can be used to define the scheme-conversion factor between the two schemes,
\beq
z_m(\mu_R) = \frac{Z_m^{\rm OS}}{Z_m^{\overline{\rm MS}}(\mu_R)}= \frac{\mMS(\mu_R)}{m}=\sum_{k=0}^\infty \left( \frac{\alpha_s(\mu_R)}{\pi} \right)^k z_m^{(k)}(\mu_R)\;.
\eeq
For heavy quarks $z_m(\mu_R)$ is currently known up to four loops \cite{Melnikov:2000qh,Marquard:2015qpa,Marquard:2016dcn}. We can express the on-shell mass as a series calculated in perturbative QFT
\beq
m = z^{-1}_m(\mu_R) \mMS(\mu_R)=  \sum_{k=0}^\infty m^{(k)},
\eeq
In practice, one does not consider arbitrary $\mu_R$ but rather work with a self-consistent definition of the $\mMS$ mass $\mMS(\mMS)$. Thus for the quark masses we consider the scale as fixed number rather than as a free parameter.

\begin{figure}
 \centering
    \includegraphics[width=\linewidth]{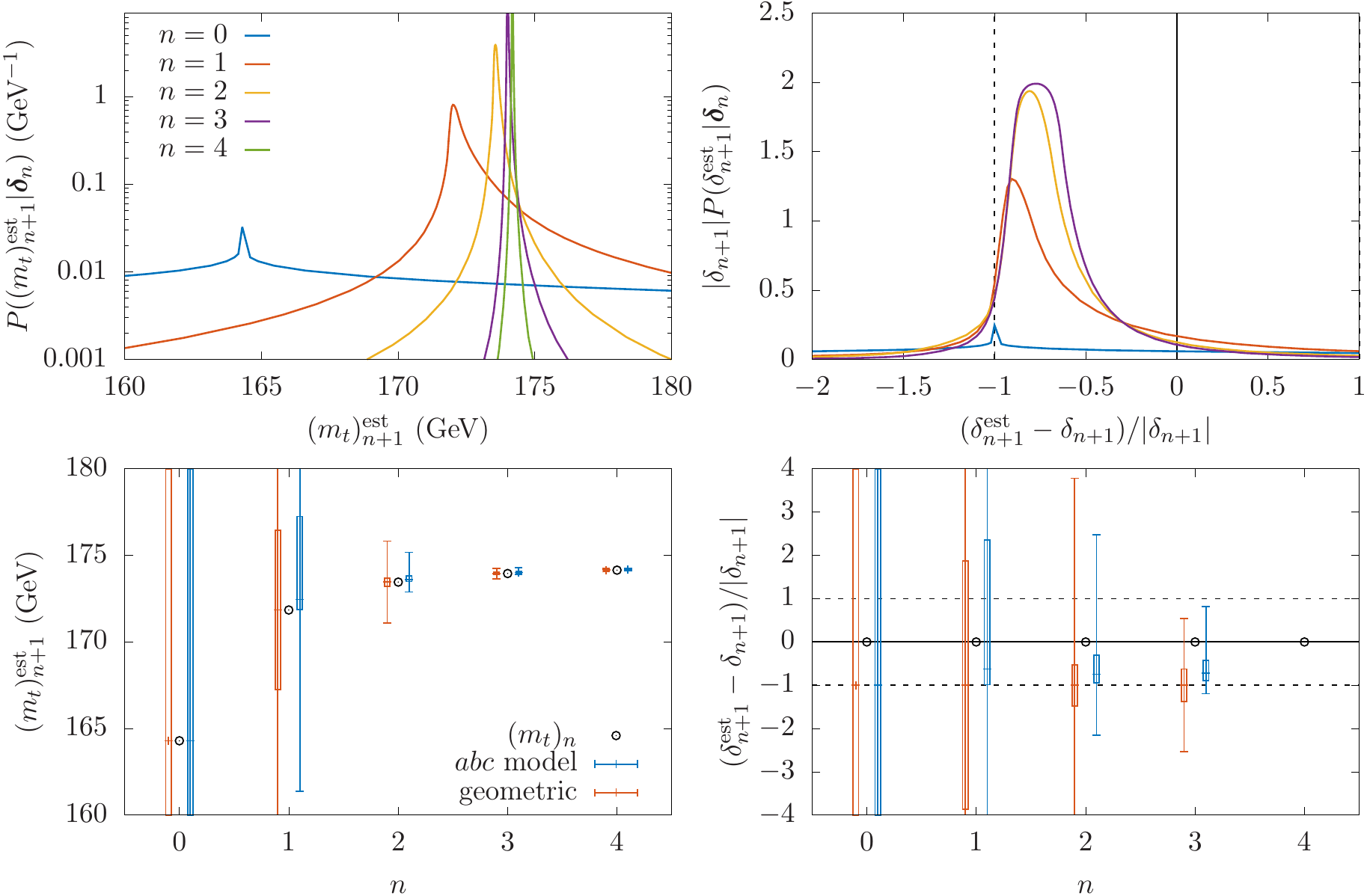}
    \caption{\label{fig:mt_progression}
Top left panel: The probability distribution from the $abc$-model for the on-shell top quark mass $(m_t)^\text{est}_{n+1}$ evaluated at $\mu_R=\mMS_t$ and for different values of $n$. Top right panel: The same distributions
    normalised to the exact N${}^{n+1}$LO  correction.
    Bottom left panel: the median (plus), 68\% CI (errorbox) and 95\% CI (errorbar) for the posterior of $(m_t)_{n+1}^\text{est}$ , computed from the $abc$ (blue) and geometric (red) models using information on the previous orders. The exact values of $(m_t)_n$ are shown as black circles. Bottom right panel: CIs scaled to the exact N${}^{n+1}$LO correction.
    }
    \end{figure}

In figure~\ref{fig:mt_progression}  we show the probability distributions and CIs of the on-shell top quark mass $(m_t)_{n+1}^\text{est}= \sum_{k=0}^{n+1} m^{(k)}$  given the first $n+1$ perturbative orders of
the scheme-conversion factor $z^{k}_{m_t}(\mu_R)$ evaluated at $\mu_R=\mMS_t$.
In the top left panel we see that the distributions become more and more peaked as perturbative information becomes available. On the top right we see that the $abc$-model correctly estimates the positive sign of the corrections, but the estimated corrections are in general smaller than the exact value.
In the bottom left panel we see that the CIs shrinks rapidly as $n$ increases.  In the bottom right we see that, unlike in the case of the QCD anomalous dimension, the $68\%$ CIs of both the $abc$ and geometric models
do not include the next perturbative order, although the 95\% CIs are sufficiently wide to include it.
We note that the $abc$-model distributions are asymmetric, with a tail towards positive perturbative corrections showing that, although the $abc$-model does not peak at the exact value of N${}^n$LO contribution, it (correctly) anticipates that larger positive corrections are more likely. This is a result of the fact that the perturbative coefficients $z_m^{(k)}$ are all negative (at least through four loops).
Using the $abc$-model we estimate the CIs for the MHOs given the 4-loop result, i.e., $\Sigma=m_t-(m_t)_4$,  to be (in GeV): 
\begin{equation}
\text{CI}_{68}= [0.008, 0.046]\,,\quad \text{CI}_{95}= [-0.027, 0.112]\,.
\end{equation}
It is interesting to compare the size of the CIs obtained from the $abc$-model to the range five-loop estimates (in GeV) obtained in ref.~\cite{Kataev:2018gle}: 

\begin{align}
(m_t)_0\delta_{5}^\text{ECH} &=  0.073\,,\quad&
(m_t)_0\delta_{5}^\text{ECH direct} &=   0.083\,,\quad&
(m_t)_0\delta_{5}^\text{FL} &=  0.126\,,\quad &\nonumber\\
(m_t)_0\delta_{5}^{\text{FL}, M\rightarrow \bar{m}} &=  0.086\,,\quad,&
(m_t)_0\delta_{5}^{r-n} &=  0.112\,. & &
\end{align}
We see that the $68\%$ CI from the $abc$-model are lower than the estimates of ref.~\cite{Kataev:2018gle}, but the $95\%$ CI includes all values of ref.~\cite{Kataev:2018gle}, except one.

For completeness, in appendix~\ref{app:plots}, we include plots for the bottom and charm quarks.
The scaled probability distributions and CIs look similar to the ones in the right panels of figure~\ref{fig:mt_progression}.
We note, however, that the MHO estimate using Bayesian inference is less reliable for lighter quarks, as in this case corrections become large and the perturbative expansion itself fails. This can be seen especially in the case of the charm quark mass.

\begin{figure}
 \centering
    \includegraphics[width=\linewidth]{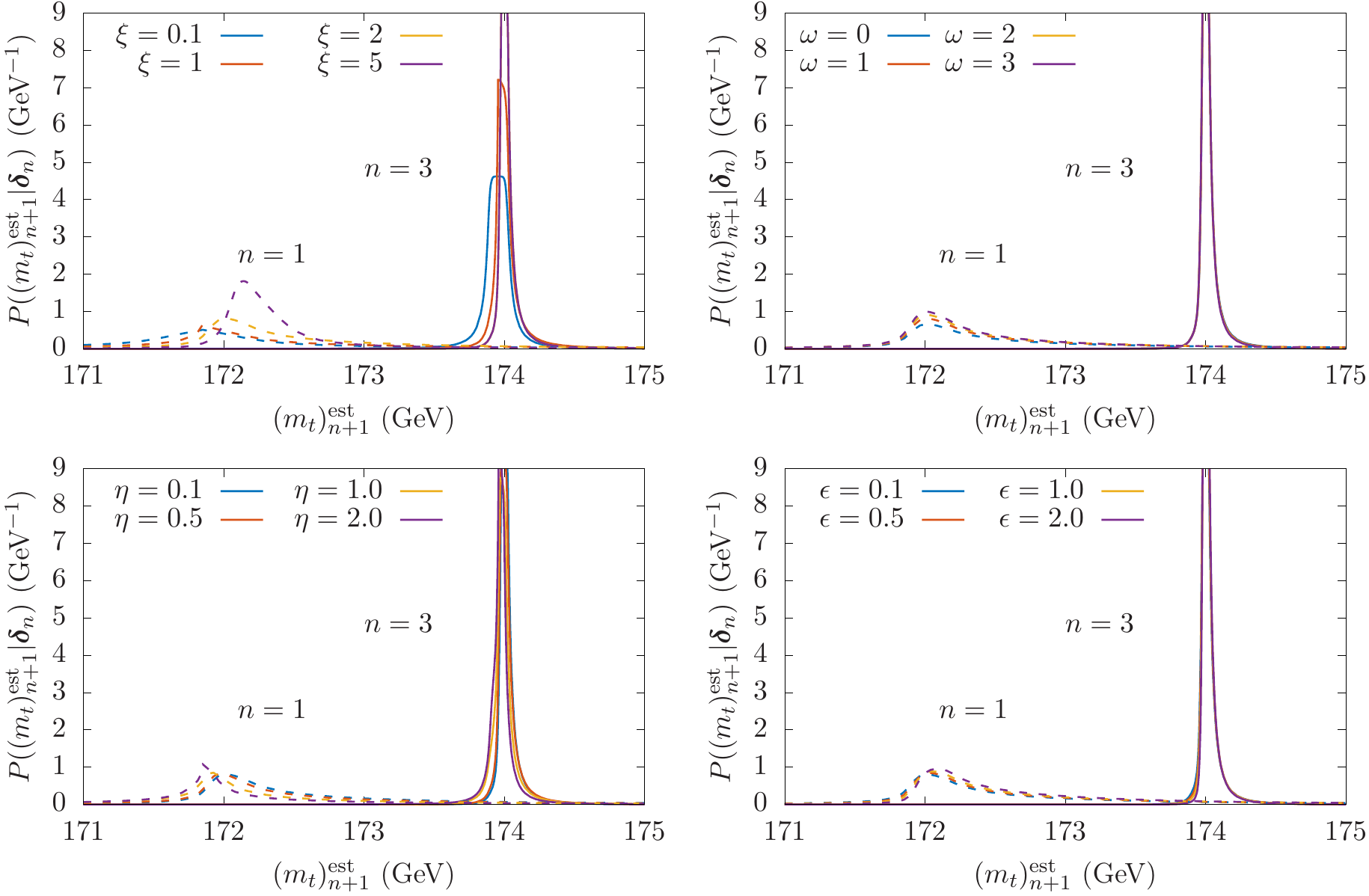}
    \caption{\label{fig:mt_prior}
    The sensitivity of the probability distribution $1/(m_t)_0 P_{abc}( ((m_t)_{n+1}^\text{est}-(m_t)_n)/(m_t)_0|\bfdelta_{n})$ for on-shell top quark mass for $n=1$ (dashed) and $n=3$ (solid) on the parameters $(\epsilon,\omega,\eta,\xi)$ for 4 selected values with the others held fixed at $(\epsilon, \omega,\eta,\xi)=(0.1, 1, 0.1,2)$.
    }
    \end{figure}

In figure~\ref{fig:mt_prior} we show the dependence of the probability distributions for the $(m_t)_{n+1}^\text{est}$ for $n=1,3$ from the $abc$-model on the  parameters $(\eta,\xi,\epsilon,\omega)$. As already discussed in section~\ref{sec:abc_model}, we see that the dependence on $(\eta,\epsilon,\omega)$ reduces as more perturbative information becomes available, while $\xi$ controls the amount of asymmetry of the distribution.
However, after the fourth perturbative order, the distributions become very narrow and peaked, and the shape is largely independent of the values of $(\eta,\xi,\epsilon,\omega)$ if $\xi\geq 2$.

\subsection{Electron anomalous magnetic moment and positronium hyperfine-splitting}

In this section we illustrate how Bayesian techniques work in the context of two classical results in QED which are known to relatively high order in the coupling, namely the electron anomalous magnetic moment and the positronium hyperfine splitting. QED corrections are typically believed to be rapidly convergent due to the smallness of the QED coupling $\alpha = 1/137.035 \,998\, 995(85) $ \cite{Mohr:2018hvt}. Further, these results are evaluated in the on-shell renormalisation scheme, thus they do not depend on any unphysical scale.

Let us start by discussing the electron anomalous magnetic moment.
The QED corrections take the form
\beq
\frac{g-2}{2} \equiv a_e = \sum_{k=1}^\infty \left( \frac{\alpha}{\pi}\right)^k C_e^{(2k)}\,.
\eeq
Here we focus only on the mass-independent part of the corrections, which are known up to 5 loops~\cite{Schwinger:1948iu,Petermann:1957hs,Sommerfield:1957zz,Laporta:1996mq,Aoyama:2012wj,Aoyama:2014sxa,Laporta:2017okg,Aoyama:2017uqe,Volkov:2018jhy,Aoyama:2019ryr}. The numerical values for the coefficients $C_e^{2k}$ are collected in table~\ref{tab:QED}.

\begin{table}[h]
\begin{center}
\begin{tabular}{c|ccccccccccccccc}
\hline\hline
$k$ & 1 & 2 & 3 & 4 & 5 &
\\
\hline
$C_e^{(2k)}$ & $0.5$ & $-0.328$ & $1.181$ & $-1.912$ & $6.675(192)$ &
\\
\hline\hline
$ij$ &$00$& $10$& $21$ &$20$ &$32$&$31$\\
\hline
$C_{\rm HFS}^{(ij)}$& $0.583$ & $-1.24$ & $-2.06 $& $-3.88$ & $-8.64 $& $14.97$\\
\hline\hline
\end{tabular}
\caption{\label{tab:QED} Known coefficients of mass-independent corrections to electron anomalous magnetic moment and to positronium hyperfine splitting.}
\end{center}
\end{table}

\begin{figure}
    \centering
    \includegraphics[width=\linewidth]{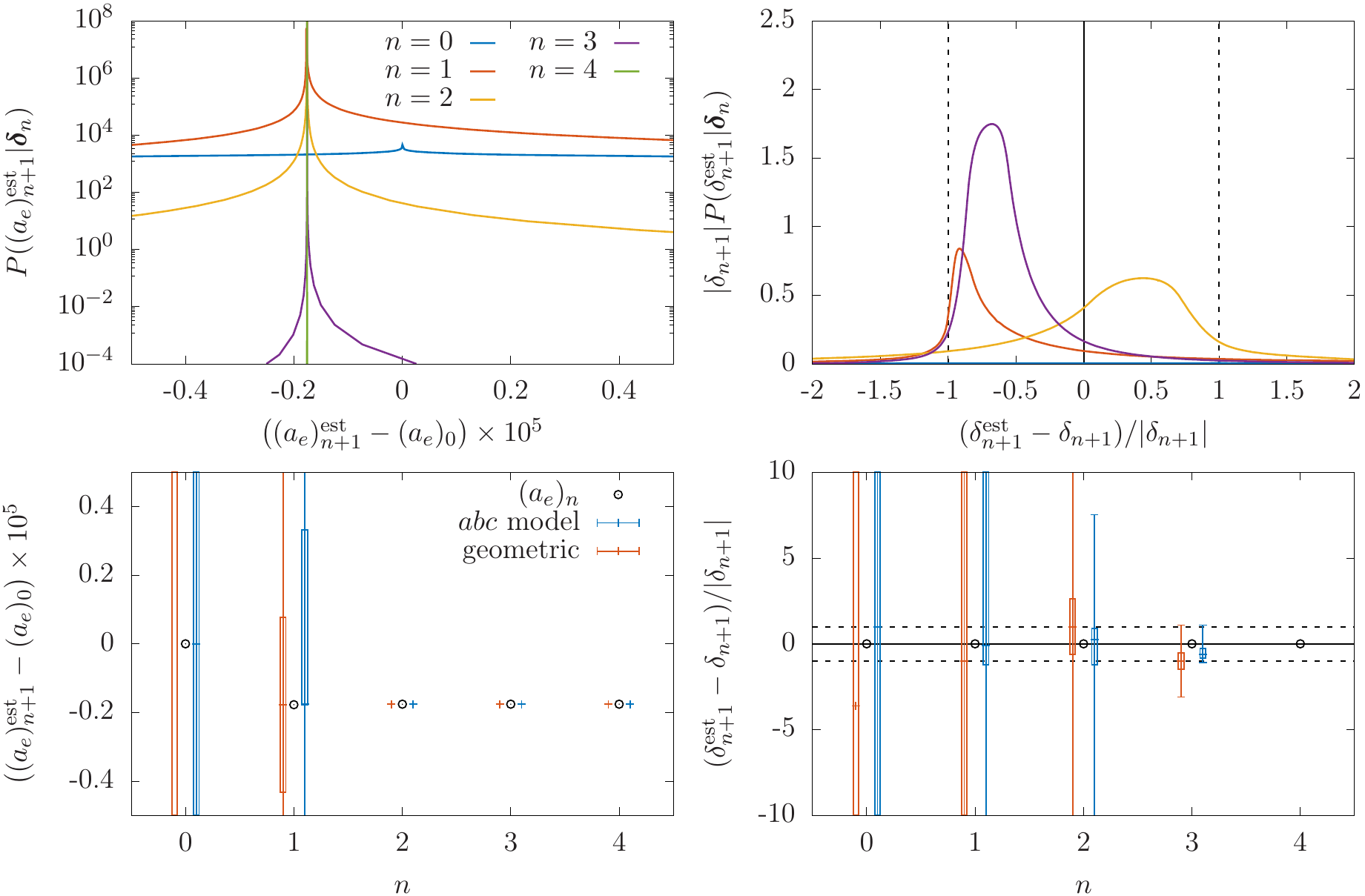}
    \caption{Top left panel: The probability distribution from the $abc$-model for $(a_e)_{n+1}^\text{est}$ and for different values of $n$. Top right panel: The same distributions
    normalised to the exact N${}^{n+1}$LO  correction.
    Bottom left panel: the median (plus), 68\% CI (errorbox) and 95\% CI (errorbar) for the posterior of $(a_e)_{n+1}^\text{est}$ , computed from the $abc$ (blue) and geometric (red) models using information on the previous orders. The exact values of $(a_e)_n$ are shown as black circles. Bottom right panel: CIs scaled to the exact N${}^{n+1}$LO correction.}
    \label{fig:QED_g-2}
\end{figure}

Figure \ref{fig:QED_g-2} illustrates the $abc$-model estimates for the probability distributions and the CIs for the partial sums
$(a_e)_{n+1}^\text{est} =\sum_{k=1}^{n+1} \left( \frac{\alpha}{\pi}\right)^k C_e^{(2k)} $.
Due to the very small values of the expansion parameter, the absolute distributions and CIs are hardly visible for $n>1$. The distribution for $n=1$ is rather broad, resulting in CIs  that overestimate the actual size of the higher-order corrections. At $n=2$, the 68\% CI captures the size of the next correction, while for $n=3$ only the $95$\% CI includes the next order.

\begin{figure}
    \centering
    \includegraphics[width=\linewidth]{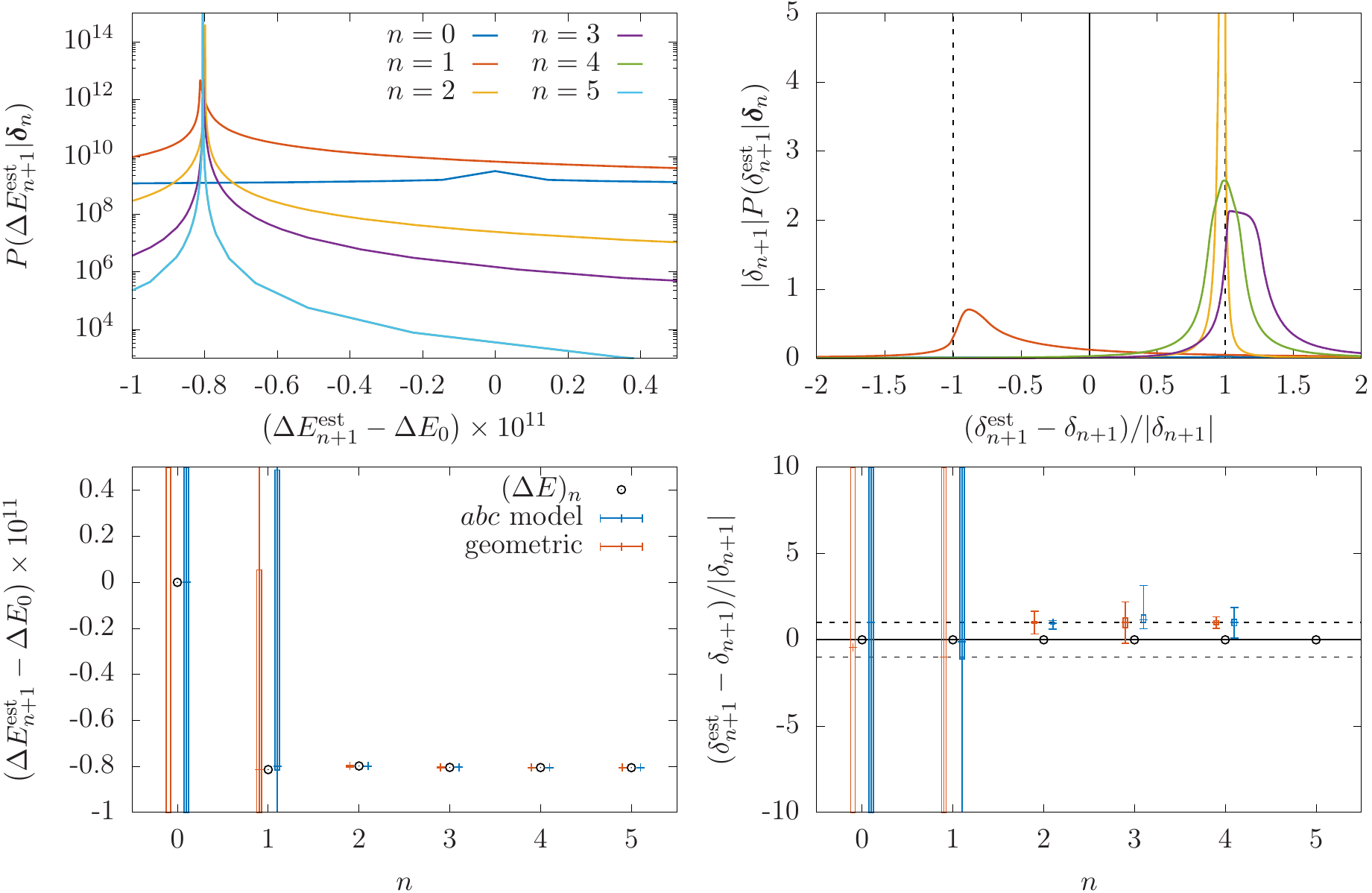}
    \caption{The probability distribution and the CIs for the hyperfine-splitting $\Delta E$ in units of the electron mass $m_e$. See the caption of figure~\ref{fig:QED_g-2}.    }
    \label{fig:QED_HFS}
\end{figure}

Our second QED example is the positronium hyperfine-splitting. This is a non-perturbative quantity and admits a double expansion in $\alpha$ and $\ln \alpha$:
\beq
\Delta E = m_e \alpha^4 \sum_{i,j=0}^{\infty} C^{(ij)}_{\rm HFS} \left(\frac{\alpha}{\pi}\right)^i \ln^j \alpha\,.
\eeq
Only terms with $j<i$ are non-zero with the exception of the coefficient $C_{00}$. This introduces a natural ordering among the coefficients, which are ordered first by $i$ then by $j$. For the known terms this ordering is also equivalent to ordering the contributions by their size, since $\ln \alpha \sim -5 $.
The currently known values  \cite{Karplus:1952wp,Lepage:1977gd,karshenboim1993new,Czarnecki:1998zv,Kniehl:2000cx} are shown in table~\ref{tab:QED}.

The presence of the logarithmic corrections in the series clearly violates the geometric progression of the series assumed in both the geometric and $abc$-models. Our Bayesian inference for such
series is therefore less reliable as an estimate of the MHOs in comparison to a purely perturbative quantity such as $a_e$. This can be also seen in  figure~\ref{fig:QED_HFS}: At the lowest two perturbative orders the CIs overestimate the size of the perturbative corrections. As the logarithmically enhanced corrections appear, both geometric and $abc$ models fail to provide reliable estimates of the next-order correction.
This is expected, since the models consider only a single expansion parameter $a$ and expect the next-order correction to be suppressed by an additional power of $a$.

In principle, we can consider even more complicated QED bound-state observables, for example, the bound electron $g$-factor \cite{Pachucki:2005px,Pachucki:2017xfd,Czarnecki:2017kva,Czarnecki:2020kzi}. In this case, one has a triple expansion in powers of $(Z\alpha)$, $\alpha$ and $\ln(Z\alpha)$, where $Z$ is atomic charge. We expect that such series require more attuned models that take into account non-perturbative terms and double or triple expansions. We conclude this section by observing that  QED corrections require dedicated studies within Bayesian framework that we leave for future works.

\section{Hadronic observables with explicit scale dependence}
\label{sec:hadronic_examples}

In this section we discuss the application of Bayesian techniques to estimate MHOs for a selection of hadronic observables known to NNLO, or even N$^3$LO, in the strong coupling constant. 
In all cases, we compute 
scale-independent probability distributions and CIs using the scale-marginalisation (sm) and scale-averaging (sa) prescriptions introduced in sections~\ref{sec:scale-marginalization_pres} and \ref{sec:weighted-sum-model}. Throughout this section we work with the geometric and $abc$-models of section~\ref{sec:no_scale}. We start by discussing inclusive cross-sections and discuss differential distributions towards the end of this section. In contrast to the examples studied in ref.~\cite{Bonvini:2020xeo}, we take into account the variation of the factorisation scale.

\subsection{Higgs production in gluon fusion}

We start by discussing the inclusive Higgs production cross-section in gluon fusion:
\begin{equation}
 (\sigma_{ggH})_n(\mu_F,\mu_R)= \sum_{k=0}^n\sigma^{(k)}_{ggH}(\mu_F,\mu_R)\,.
\end{equation}
The inclusive cross-section is known through NLO in QCD including all finite top-mass effects~\cite{Georgi:1977gs,Dawson:1990zj,Djouadi:1991tka,Graudenz:1992pv,Spira:1995rr,Harlander:2005rq,Aglietti:2006tp,Aglietti:2004nj,Anastasiou:2006hc,Anastasiou:2009kn}, and through N$^3$LO in an effective theory expansion where the top quark is infinitely heavy~\cite{Harlander:2002wh,Anastasiou:2002yz,Ravindran:2003um,Anastasiou:2015ema,Anastasiou:2016cez,Mistlberger:2018etf}. All results in this section were obtained with \texttt{iHixs~2}~\cite{Dulat:2018rbf}, neglecting finite top-mass effects, with the \verb+PDF4LHC15_nnlo_100+ PDF set and the on-shell top quark mass $m_t=172.5 \,{\rm GeV}$. 
We choose as value for the central scale $\mu_0 = m_H/2=62.5\,\text{GeV}$ (both for scale-variation results and in the priors in eqs.~\eqref{eq:scale_prior} and~\eqref{eq:weight}).

\begin{figure}
 \centering
    \includegraphics[width=\linewidth]{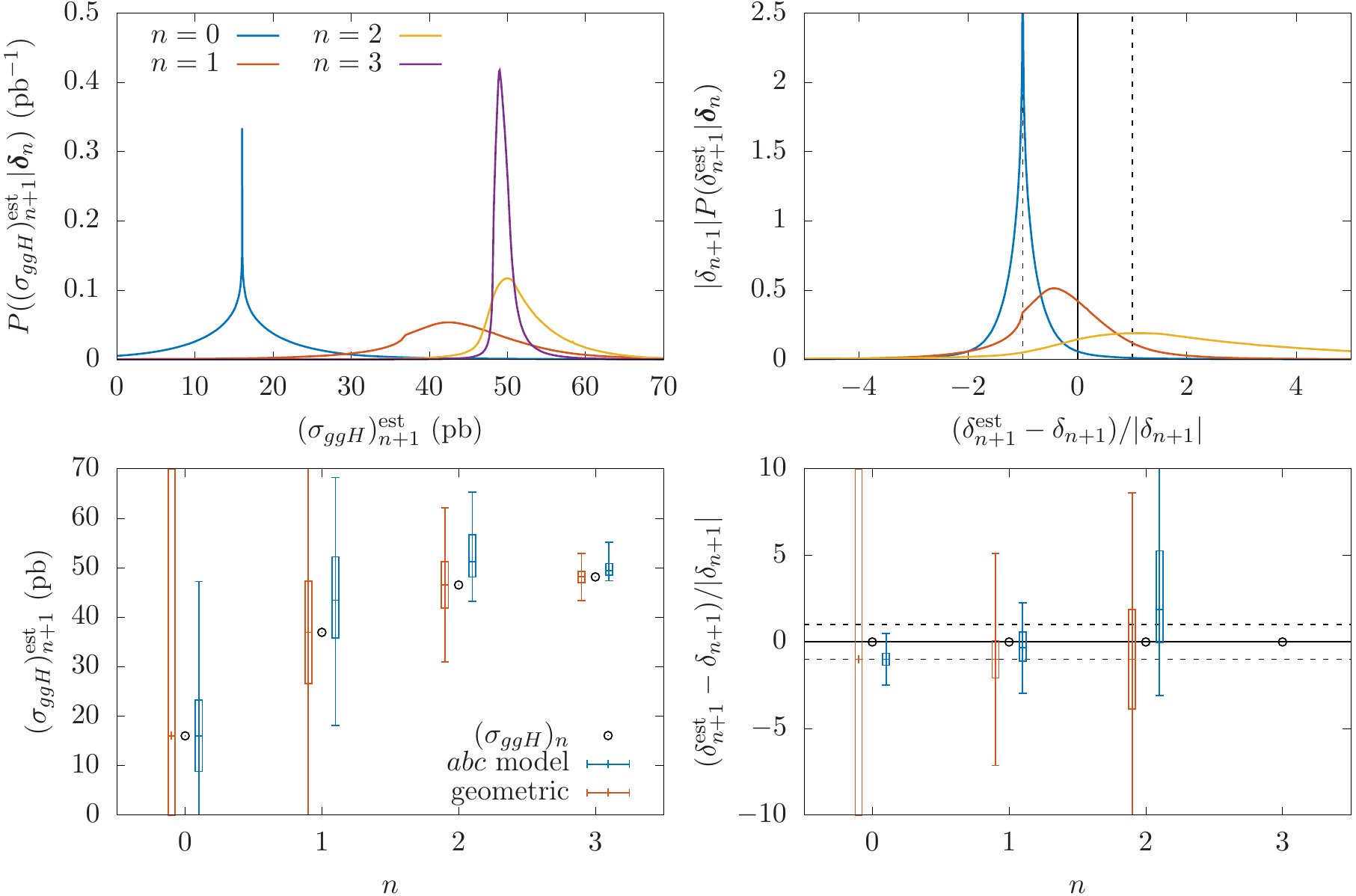}
    \caption{\label{fig:higgs_progression}
Top left panel: The probability distributions from the $abc$-model for the inclusive gluon-fusion Higgs production cross-section $(\sigma_{ggH})^\text{est}_{n+1}$ evaluated at $\mu_R=\mu_F=m_H/2$ and for different values of $n$. Top right panel: The same distributions
    normalised to the exact N${}^{n+1}$LO  correction.
    Bottom left panel: the median (plus), 68\% CI (errorbox) and 95\% CI (errorbar) for the posterior of $(\sigma_{ggH})_{n+1}^\text{est}$, computed from the $abc$ (blue) and geometric (red) models using information on the previous orders. The exact values of $(\sigma_{ggH})_{n}$ are shown as black circles. Bottom right panel: CIs scaled to the exact N${}^{n+1}$LO correction.
    }
    \end{figure}

First, in figure~\ref{fig:higgs_progression} we show the probability distributions and CIs for $ (\sigma_{ggH})_n^\text{est}$ for fixed $\mu_R=\mu_F=\mu_0$.
We see that for $n=1,2$, the CIs for both the $abc$ and geometric models include the next order at 68\% credibility level. We note, however, that for $\mu_0=m_H/2$ the N${}^3$LO corrections are particularly small, and the CIs of the $abc$-model overshoot the computed value. We will see below that for other scale choices, the CIs of the $abc$-model are centred around the N${}^3$LO values.

\begin{figure}
    \centering
    \includegraphics[width=0.7\linewidth]{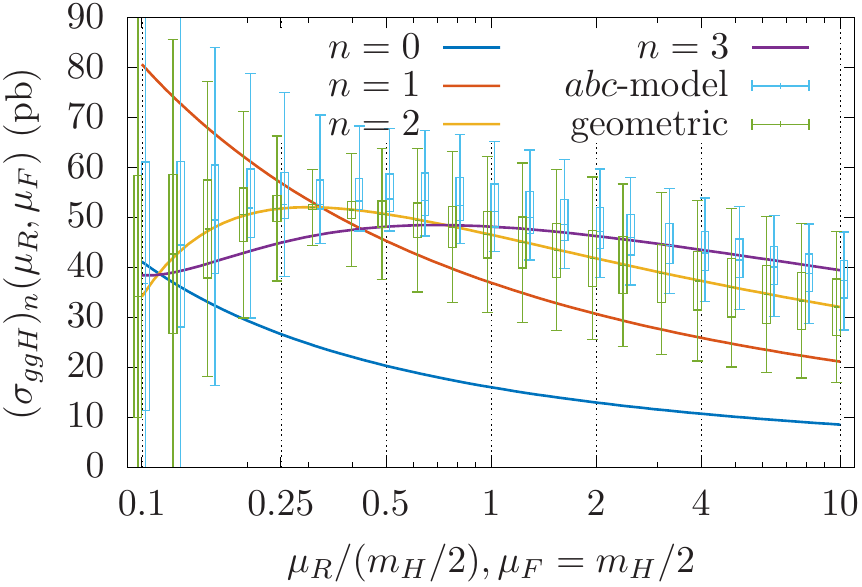}
    \caption{The inclusive Higgs production cross-section as a function of the renormalisation scale (with $\mu_F=m_H/2$ held fixed) at different orders $n=0,1,2,3$. The errorbars and erroboxes correspond to the NNLO 95\% and 68\% CIs for the geometric model (light green) and the $abc$-model (light blue), given the $n<3$ results.}
    \label{fig:higgs_scale_dependence}
\end{figure}

Next, in figure~\ref{fig:higgs_scale_dependence} we plot the renormalisation scale dependence of $(\sigma_{ggH})_{n}$ at different orders (for this plot only we keep $\mu_F=\mu_0$ fixed).  We see that in the conventional scale variation window, $F=2$, the cross-section depends quite strongly on the scale, except for the highest order. The N${}^3$LO corrections vanish for $\mu_R\approx 0.8 \mu_0$, and the lines representing the NNLO and N$^3$LO results as a function of the scale cross, i.e., we reach a FAC point. The FAC point at NNLO is at a smaller value, $\mu_R\approx 0.3 \mu_0$. Also, the crossing points visible in figure~\ref{fig:higgs_scale_dependence} approximately coincide with the local maxima for $n=2,3$, i.e., the FAC and PMS points are surprisingly close to each other.

In figure~\ref{fig:higgs_scale_dependence} we further show the CIs  in the geometric and  $abc$-models with LO, NLO and NNLO cross-sections (at fixed scale) used as an input. We observe that the CIs for the geometric model are centred around the $n=2$ line due to the symmetry of the geometric model, but the $abc$-model assigns a higher probability to a positive correction. Indeed, for $\mu_R>2\mu_0$ the CIs are almost centred around the N${}^3$LO line. For $\mu_R<0.8\mu_0$, the N${}^3$LO corrections become negative, while the $abc$-model still anticipates a positive correction. We note that the CIs for the geometric model shrink dramatically as one approaches the NNLO FAC point at $\mu_R=0.3\mu_0$, i.e., large corrections are deemed unlikely by the geometric model close to that scale. Once the NNLO corrections become negative, the $abc$-model also becomes centred around the NNLO result, i.e., the $abc$-model no longer expects a positive correction. The CIs in this region do not shrink as dramatically as for the geometric model, although generally the $abc$-model leads to narrower CIs than the geometric model.

\subsubsection{Sensitivity to the scale interval and scale prescription}

We begin by analysing the dependence of our sm- and sa-prescriptions on the range $F$ that is chosen in eqs.~\eqref{eq:P_scale-marginalization_approx} and \eqref{eq:P_scale-average_approx}, i.e., $\mu_0/F \leq \mu \leq F \mu_0$.

\begin{figure}
    \centering
        \includegraphics[width=\linewidth]{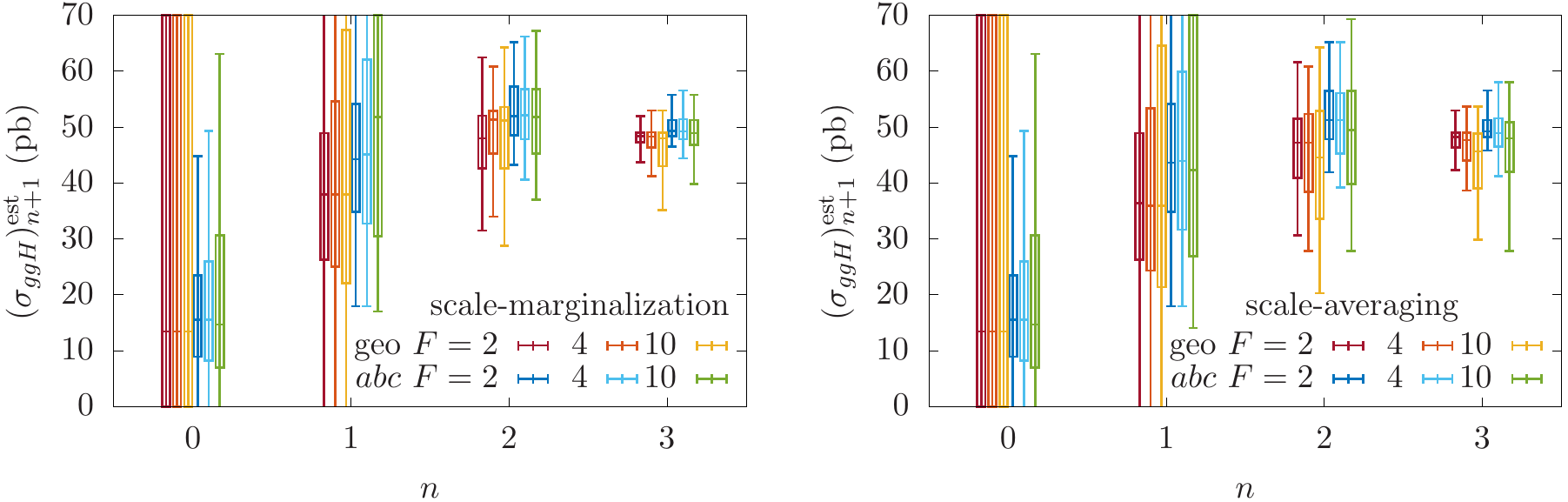}
        \caption{Left: The dependence of the CIs on the size $F$ of the scale interval ($\mu_0/F \leq \mu \leq F \mu_0$) for gluon-fusion Higgs cross-section for the geometric and $abc$-models using the sm-prescription Right: The same plot using the sa-prescription.}
    \label{fig:higgsF}
\end{figure}

Figure~\ref{fig:higgsF} shows the CIs  for the values $F=2,\,4,\,10$ using the geometric and $abc$-models.
The left and right panels summarise the results from the sm- and sa-prescriptions respectively. For $n=2$ the sm-prescription exhibits a somewhat smaller dependence on the choice of $F$, while the CIs for the sa-prescription grow with $F$. This can be understood from the discussion in section~\ref{sec:scale_discussion}: for symmetric models like the geometric model, the sm-prescription tries to adapt to the point where the higher-order corrections are minimised, i.e., the FAC point.
Once this point is covered by the range in the marginalisation, a further increase does not have a substantial impact on the uncertainties. 
From  figure~\ref{fig:higgs_scale_dependence} we see that this is precisely what happens for the geometric model for $n=2$, when $F$ is increased from 2 to 4, and even for $F=10$ the CIs change only very little. The discussion of section~\ref{sec:scale_discussion} does not apply to the $abc$-model, which is asymmetric. Recall from figure~\ref{fig:higgs_scale_dependence} that the $abc$-model does not have such dramatic reduction in the size of the CIs when reaching an FAC point. Therefore it is also less biased to the inclusion of FAC point into the integration range for $\mu$ in the sm-prescription. For $n=3$ the second FAC point at very small scales becomes relevant only for the largest $F$ values.

The sa-prescription, on the other hand, is biased to the regions where the scale dependence is flat, i.e., it is biased towards the PMS point. However, for this particular process we observe the peculiar situation that FAC  and PMS points are very close to each other (at least for the orders we considered). Due to the coherent addition of probabilities, for the sa-prescription increasing the range of the $\mu$-integration inevitably will also increase the CIs.

Let us conclude this discussion with an important point. From a Bayesian perspective, one expects that the predictions of the model become independent of the priors once enough data have become available. In particular, in the sm-prescription, the perturbative scales are treated as model parameters, and so one expects that at high enough orders the model predictions should only mildly depend on the prior in eq.~\eqref{eq:scale_prior}, and the choice of $F$, in agreement with our findings in figure~\ref{fig:higgsF}. However, it would be premature to conclude that the probability distributions are independent of the prior and $F$: we have shown in section~\ref{sec:scale_discussion} that for the geometric model in the sm-prescription, the preferred scale is the FAC point. The fact that the CIs are insensitive to the choice of $F$ is likely to be related to the fact that the distributions will be highly peaked at the values of the cross-section at the FAC point, which is not necessarily related to the prior independence. Note that the argument of section~\ref{sec:scale_discussion} does not apply to the $abc$-model in the sm-prescription (because the $abc$-model is not symmetric).

\subsubsection{Accuracy of Gauss-Legendre quadrature rule}

As already discussed in section~\ref{sec:MHO-connection-7pt}, performing the numerical integral over the scale can become computationally expensive.
Quadrature rules, on the other hand, allow one to approximate the full integral using only a small number of input points. In particular, a set of points for a 9-point scale variation may be readily turned into an integral over $(\mu_R,\mu_F)$.

\begin{figure}
    \centering
       \includegraphics[width=\linewidth]{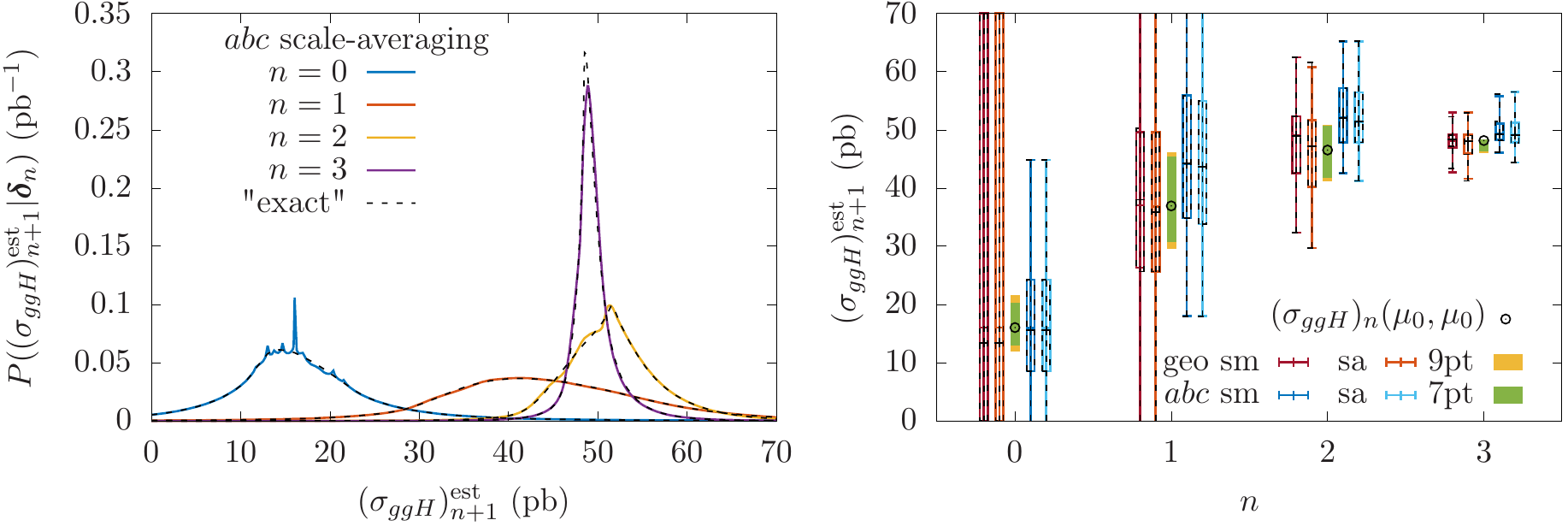}
       \caption{Left: The coloured lines show the probability distributions for the  $abc$-model at different orders using the sa-prescription and the 3-point Gauss-Legendre quadrature on $\mu_F/\mu_0,\mu_R/\mu_0 \in \{1/2,1,2\}$. The black dashed line indicates the ``exact'' numerical integration in the interval with $F=2.45$. Right: The CIs for the geometric and $abc$ models at different orders using the sm- and sa-prescriptions using the Gauss-Legendre quadrature. The result of the ``exact'' numerical integral are overlaid using black dashed lines. For comparison we also include the 9-point and 7-point variation intervals.}
    \label{fig:GL_higgs}
\end{figure}

The left panel of figure~\ref{fig:GL_higgs} illustrates the accuracy of this approximation for the probability distributions in the $abc$-model using the sa-prescription. The distributions for different values of $n$ (in colour) are obtained using the 3-point Gauss-Legendre quadrature for both $\mu_R$ and $\mu_F$. The results obtained by a finely-spaced equidistant numerical integration over the same scale interval with $F\approx 2.45$ is overlaid on top (black dashed lines). We see that, except for some features around the peak, the quadrature rules reproduce the ``exact'' numerical integration very well. In particular, this approximation has very little effect on the CIs, which are shown in the right panel of figure~\ref{fig:GL_higgs}, where we show the 68\% and 95\% CIs for the geometric and $abc$ models using both the sm- and sa-prescriptions. The dashed black lines indicate the results of the finely-spaced numerical integration. We note that after scale integration the probability distributions are no longer symmetric, even if the distributions
at fixed scale are. Overall, we observe that the Gauss--Legendre quadrature rule is able to reproduce both the 68\% and 95\% CIs from the ``exact'' numerical integration to a very good degree.
Therefore, in the subsequent sections we will only show the results obtained using the Gauss-Legendre quadrature.

For Higgs production in gluon fusion both the sm- and sa-prescriptions produce very similar CIs shown in the right panel of figure~\ref{fig:GL_higgs}. The CIs are also very similar in size for both the geometric and the $abc$-models. The results for the latter are, however, systematically shifted upwards due to the sequence of positive corrections at lower orders. For comparison we also display  the scale variation intervals using 9- and 7-point rules, which have no probabilistic interpretation.
We observe that the scale variation intervals are slightly smaller, but rather close, to the 68\% CIs obtained from the Bayesian methods. This illustrates the arguments of section~\ref{sec:MHO-connection-7pt} in a practical example. In particular, thanks to the slow $\mu_F$ variation, the scale variation intervals are close to saturating the $s=1$ bound in eq.~\eqref{eq:bound_sv}.

Finally, we conclude with the summary of 68\% and 95\% CIs and scale-variation intervals for the highest known order $n=3$ of $(\sigma_{ggH})_n$:
\begin{center}
\begin{tabular}{cccc|cc}
model & prescription & $\text{CI}_{68}$ (pb) & $\text{CI}_{95}$ (pb) & 7-point (pb) & 9-point (pb)\\
\hline
$abc$ 	& sa & [47.8, 51.2] & [44.4, 56.5] &\multirow{4}{*}{[46.3, 48.3]} & \multirow{4}{*}{[46.0, 48.3]}\\
$abc$	& sm & [48.3, 51.1] & [46.1, 55.8] & & \\
geo	& sa & [46.0, 49.2] & [41.6, 53.0] & & \\
geo	& sm & [47.0, 49.2] & [42.7, 53.0] & &
\end{tabular}
\end{center}
We note that as the N${}^3$LO cross-section reaches a local maximum near the central scale $\mu_0$, traditional scale variation produces a one-sided interval. The geometric model has approximately symmetric intervals around the central value, while the $abc$-model anticipates positive MHO terms.
Although all 68\% CIs and the scale-variation intervals are numerically close in size, we stress that they have completely different statistical meanings. Scale-variation intervals have no probabilistic interpretation and cannot be associated with a particular level of credibility. In contrast, the Bayesian CIs express a mathematically rigorous degree of belief for the MHOs from the posterior distribution for a given model and scale prescription.

\subsection{Vector-Boson Fusion Higgs and di-Higgs production}
Our next two examples are Higgs production in vector-boson fusion (VBF), and a similar process for double Higgs production. The inclusive VBF cross-section is known up to N${}^3$LO.  We denote VBF cross-sections for single Higgs~\cite{Bolzoni:2010xr,Bolzoni:2011cu,Dreyer:2016oyx,Dreyer:2020urf} and double-Higgs production~\cite{ Liu-Sheng:2014gxa,Dreyer:2018qbw,Dreyer:2020urf} by ($0 \leq n \leq 3$)
\begin{equation}\begin{split}
 (\sigma_{\rm VBF-H})_n(\mu_F,\mu_R)&\,= \sum_{k=0}^n\sigma^{(k)}_{\rm VBF-H}(\mu_F,\mu_R)\,,\\
 (\sigma_{\rm VBF-HH})_n(\mu_F,\mu_R)&\,= \sum_{k=0}^n\sigma^{(k)}_{\rm VBF-HH}(\mu_F,\mu_R)\,.
\end{split}\end{equation}
As before the centre-of-mass energy is $\sqrt{s}=13\,\rm TeV$. The central scale is given by the vector boson momentum~\cite{Han:1992hr} and we take into account the dependence on both factorisation  and renormalisation scales. Computations were performed with the \texttt{proVBFH} code \cite{proVBFH}.  

\begin{figure}
    \centering
    \includegraphics[width=0.49\linewidth]{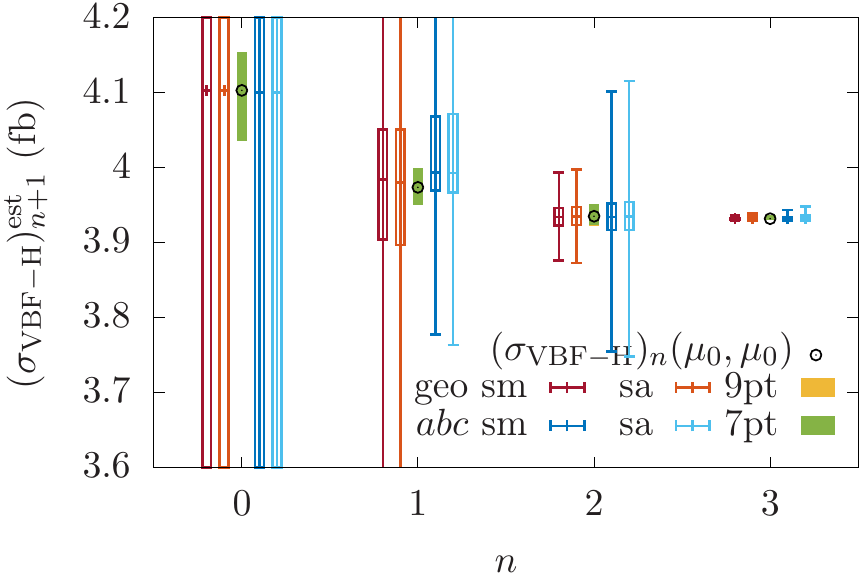}
    \includegraphics[width=0.49\linewidth]{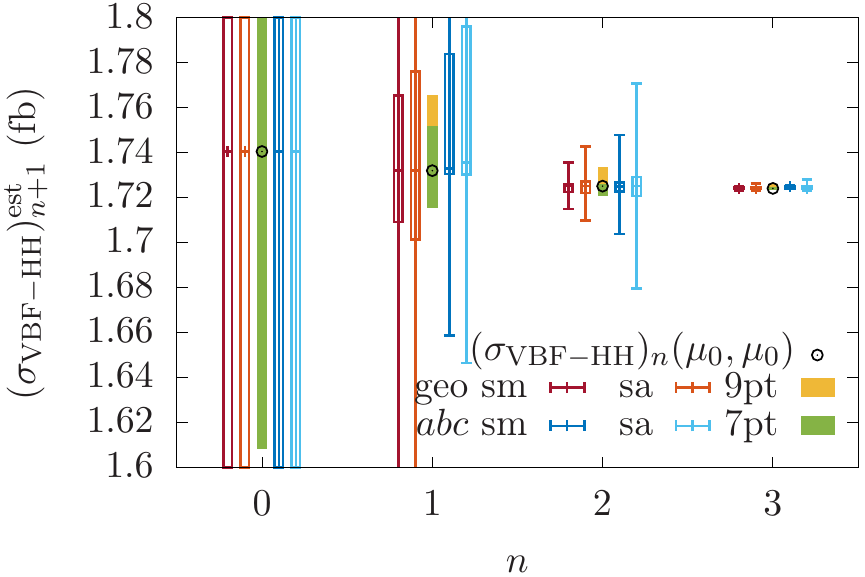}
    \caption{The 68\% and 95\% CIs for the VBF cross-sections for Higgs and di-Higgs production for the geometric and $abc$-models using the sa- and sm-prescriptions. The scale variation intervals using 7 and 9 points are shown for comparison.}    \label{fig:VBF}
\end{figure}

In the left panel of figure~\ref{fig:VBF} we display the CIs for different models and prescriptions for single Higgs VBF production. For $n < 2$ the Bayesian approach gives a larger uncertainty (68\% CIs) than the traditional scale variation. 
Because the NLO correction is negative, the $abc$-model anticipates an alternating series, and consequently the CIs for $n=1$ for the $abc$-model are positively shifted compared to the NLO result.
However, the NNLO corrections are again negative, and for $n=2$ all studied models and prescriptions give very similar 68\% CIs, although the $abc$-model has much larger 95\% CIs than the geometric model.
For $n=3$ the 68\% CIs shrink even further and become somewhat smaller than the scale variation intervals. For the single Higgs VBF cross-section $(\sigma_\text{VBF-H})_n$ at $n=3$ these CIs are:
{
\setlength{\tabcolsep}{3pt}
\begin{center}
\begin{tabular}{cccc|cc}
model & prescription& $\text{CI}_{68}$ (fb) & $\text{CI}_{95}$ (fb) & 7 point (fb) & 9 point (fb)\\
\hline
$abc$ 	& sa & [3.9306, 3.9357] & [3.9287, 3.9478] &\multirow{4}{*}{[3.9304, 3.9367]} & \multirow{4}{*}{[3.9304, 3.9367]}\\
$abc$	& sm & [3.9304, 3.9337] & [3.9290, 3.9430] & & \\
geo	& sa & [3.9305, 3.9343] & [3.9287, 3.9385] & & \\
geo	& sm & [3.9304, 3.9324] & [3.9293, 3.9355] & &
\end{tabular}
\end{center}
}
\noindent
We note that the sm-prescription gives much smaller CIs than the sa-prescription. In fact, the 95\% CIs of the scale-marginalised geometric model is smaller and does not contain the scale-variation interval, demonstrating that the bounds discussed in section~\ref{sec:MHO-connection-7pt} do not necessarily apply to the sm-prescription. In contrast the 95\% CIs for the geometric model in sa-prescription  contain the scale variation intervals, as expected.

In the right panel of figure~\ref{fig:VBF} we display the CIs for different models and prescriptions for di-Higgs VBF production. We observe very good convergence of the cross-section, and correspondingly the CIs from Bayesian inference shrink rapidly. We observe that the sa-prescription gives larger CIs than the sm-prescription, which is due to the presence of an FAC point in the scale integration interval. We also see a significant difference between the 7- and 9-point scale variation intervals. The 9-point interval for $n=2$ is larger than the 68\% CI for the geometric model with the sa-prescription, but still within the 95\% CI. This is a sign that both renormalisation and factorisation  scale dependencies are comparable and the one dimensional ($s=1$) bound estimated in section~\ref{sec:MHO-connection-7pt} is not applicable in this case. For $n=2$ and $n=3$ the 95\% CI for scale-marginalised geometric model is even smaller than 9-point scale variation interval, while sa-prescription 95\% still encompasses it, as expected. The CIs for di-Higgs VBF cross-section $(\sigma_\text{VBF-HH})_n$ at $n=3$ are:
{
\setlength{\tabcolsep}{3pt}
\begin{center}
\begin{tabular}{cccc|cc}
model & prescription & $\text{CI}_{68}$ (fb) & $\text{CI}_{95}$ (fb) & 7 point (fb) & 9 point (fb)\\
\hline
$abc$ 	& sa & [1.7237, 1.7248] & [1.7229, 1.7280] &\multirow{4}{*}{[1.7232, 1.7247]} & \multirow{4}{*}{[1.7232, 1.7258]}\\
$abc$	& sm & [1.7241, 1.7248] & [1.7238, 1.7252] & & \\
geo	& sa & [1.7237, 1.7247] & [1.7229, 1.7263] & & \\
geo	& sm & [1.7239, 1.7247] & [1.7232, 1.7249] & &
\end{tabular}
\end{center}
}
\noindent
We see that sa-prescription gives generally more conservative estimates for the MHOs than either the sm-prescription or scale variation.

\subsection{Drell-Yan processes}
Drell-Yan-type processes play a crucial role in hadron collider phenomenology. The inclusive cross-section for off-shell photon production with virtuality $Q^2$, $pp \to \gamma^{(*)}$, is known though N$^3$LO~ \cite{Altarelli:1978id,Hamberg:1990np,Duhr:2020seh}:
\begin{equation}
 \left( \sigma_{\rm DY-NC}\right)_n(Q^2,\mu_F,\mu_R)= \sum_{k=0}^n \sigma^{(k)}_{\rm DY-NC}(Q^2, \mu_F,\mu_R)\,.
\end{equation} 
N${}^3$LO corrections are also known for the charged-current Drell-Yan process
$pp \to W^{\pm}\to \ell^\pm \nu_{\ell}$ \cite{Altarelli:1979ub,Aurenche:1980tp,KubarAndre:1978uy,Hamberg:1990np,Anastasiou:2003ds,Melnikov:2006di, Duhr:2020sdp}. We focus here on the lepton charge asymmetry defined as
\begin{equation}
    A_W(Q^2) = \frac{\frac{d \sigma_{\rm DY-W^+}}{d Q^2}-\frac{d \sigma_{\rm DY-W^-}}{d Q^2}}
    {\frac{d \sigma_{\rm DY-W^+}}{d Q^2}+\frac{d \sigma_{\rm DY-W^-}}{d Q^2}}\,,
\end{equation}
and the $n$-th order approximation is
\begin{equation}
       \left(A_W\right)_n(Q^2,\mu_F,\mu_R)= \sum_{k=0}^{n}A_W^{(k)}(Q^2,\mu_F,\mu_R)\,.
\end{equation}
For both processes we only consider here $Q=m_W$ with the central scale set to $\mu_0=Q=m_W$.

\begin{figure}
    \centering
    \includegraphics[width=0.49\linewidth]{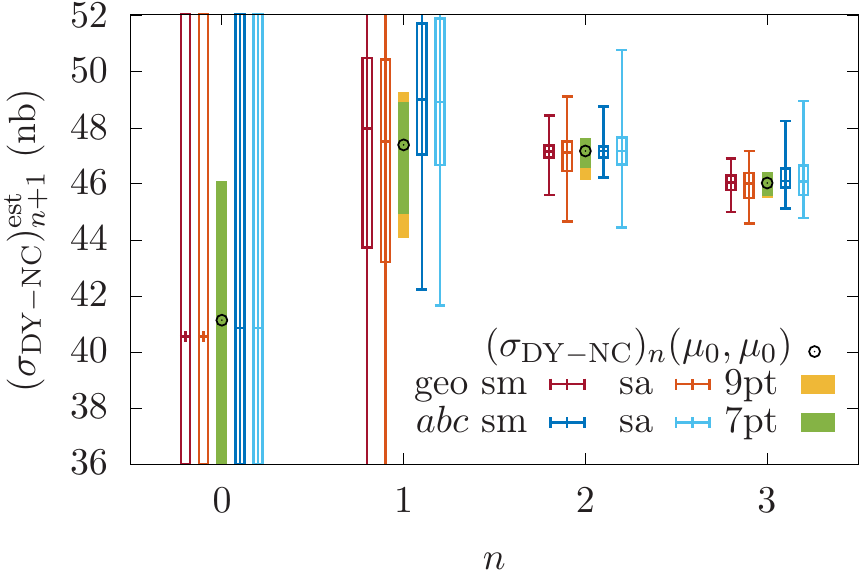}
    \includegraphics[width=0.49\linewidth]{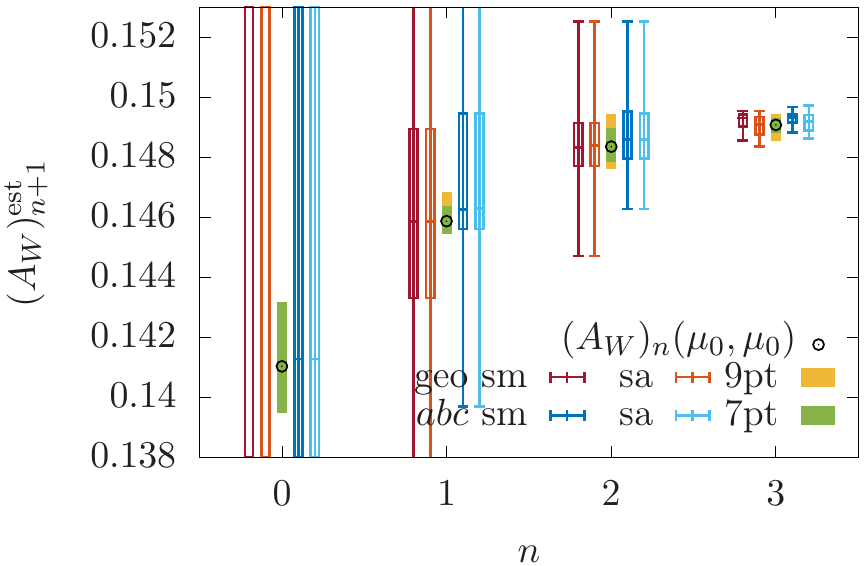}
    \caption{The 68\% and 95\% CIs for the neutral-current Drell-Yan cross-section and charged-current lepton-charge asymmetry for the geometric and $abc$-models using the sa- and sm-prescriptions. The scale variation intervals using 7 and 9 points are shown for comparison.}
    \label{fig:DY1}
\end{figure}

In the left panel of figure~\ref{fig:DY1} we show results for the neutral current Drell-Yan cross-section. We see that for $n=2,3$ the sa-prescription produces 68\% CIs which are similar in size to the regular scale variation intervals. The CIs from the sm-prescription for $n=2$ are much smaller, again due to the presence of an FAC point in the integration region for $\mu$.
The N${}^3$LO corrections are known to be sizeable, and they are not covered by the conventional 7-point scale variation at $n=2$~\cite{Duhr:2020seh}. 
The CIs for the neutral-current Drell Yan cross-section $(\sigma_\text{DY-NC})_n$ at $n=3$ are:

\begin{center}
\begin{tabular}{cccc|cc}
model & prescription & $\text{CI}_{68}$ (nb)  & $\text{CI}_{95}$ (nb)  & 7 point(nb)  & 9 point (nb) \\
\hline
$abc$ 	& sa & [45.6, 46.6] & [44.8, 49.0] &\multirow{4}{*}{[45.6, 46.4]} & \multirow{4}{*}{[45.5, 46.4]}\\
$abc$	& sm & [45.9, 46.5] & [45.1, 48.3] & & \\
geo	& sa & [45.5, 46.4] & [44.6, 47.2] & & \\
geo	& sm & [45.8, 46.3] & [45.0, 46.9] & &
\end{tabular}
\end{center}

We observe that the 68$\%$ CIs are similar in size among themselves, and to the scale-variation intervals, but the CIs from the $abc$-model are slightly shifted upwards in the anticipation of a positive MHO correction.

In the right panel of figure~\ref{fig:DY1} we show results for the lepton charge asymmetry for $\mu_0=Q=m_W$.
The perturbative expansion for $A_W(m_W^2)$ is quickly convergent with only a mild scale dependence, because some corrections cancel in the ratio. 
The perturbative coefficients feature a monotonic increase with the perturbative order, and the $abc$-model correctly anticipates positive contributions from MHOs. The CIs from the $abc$-model are slightly smaller than for the geometric model. We do not observe significant differences between the sm- and sa-prescriptions, except for $n=3$, where scale-marginalisation  gives more aggressive CIs. We note that the traditional 7-point scale variation intervals for $n=0,1$ fail to include the next correction, but for $n=2$ they are similar to the 68\% CIs obtained from Bayesian inference.
The results for the CIs for $A_W(m_W^2)$ at N$^3$LO are:
{
\setlength{\tabcolsep}{3pt}
\begin{center}
\begin{tabular}{cccc|cc}
model & prescription & $\text{CI}_{68}$  & $\text{CI}_{95}$  & 7 point  & 9 point \\
\hline
$abc$ 	& sa & [0.1489, 0.1494] & [0.1486, 0.1497] &\multirow{4}{*}{[0.1488, 0.1493]} & \multirow{4}{*}{[0.1485, 0.1494]}\\
$abc$	& sm & [0.1491, 0.1494] & [0.1488, 0.1497] & & \\
geo	& sa & [0.1487, 0.1493] & [0.1484, 0.1495] & & \\
geo	& sm & [0.1490, 0.1494] & [0.1485, 0.1495] & &
\end{tabular}
\end{center}
}
\noindent
We observe that the sm-prescription produces smaller CIs than the sa-prescription, indicating the presence of FAC point, which also pulls the CIs upwards with respect to the N${}^3$LO result. Scale-averaged CIs are more conservative and more centred around the N${}^3$LO results for both the geometric and $abc$-models.

\subsection{Deep inelastic scattering}

The Bayesian procedure of estimating MHOs can be readily  applied to differential spectra by considering individual points/bins of the associated observable.%
\footnote{Note that here no correlations are considered, i.e., cross-section entries in each bin of the distribution are assumed to be independent. The incorporation of correlations is left for future studies.}
In the following, we will investigate how the geometric and $abc$-models perform for various differential predictions. Results will be further contrasted between the sm- and sa-prescriptions, and compared to the 9-point scale variation intervals. 
Like for inclusive observables, we will perform scale integrals using the Gauss--Legendre method from section~\ref{sec:MHO-connection-7pt}. 
This setup allows us to accommodate already existing results based on 7- or 9-point scale variation, without the need for a full re-computation of these predictions. In this section we begin our discussion of differential spectra with the DIS process. We will present results for several benchmark processes at the LHC in the subsequent sections.

\begin{figure}
    \centering
    \includegraphics[width=\linewidth]{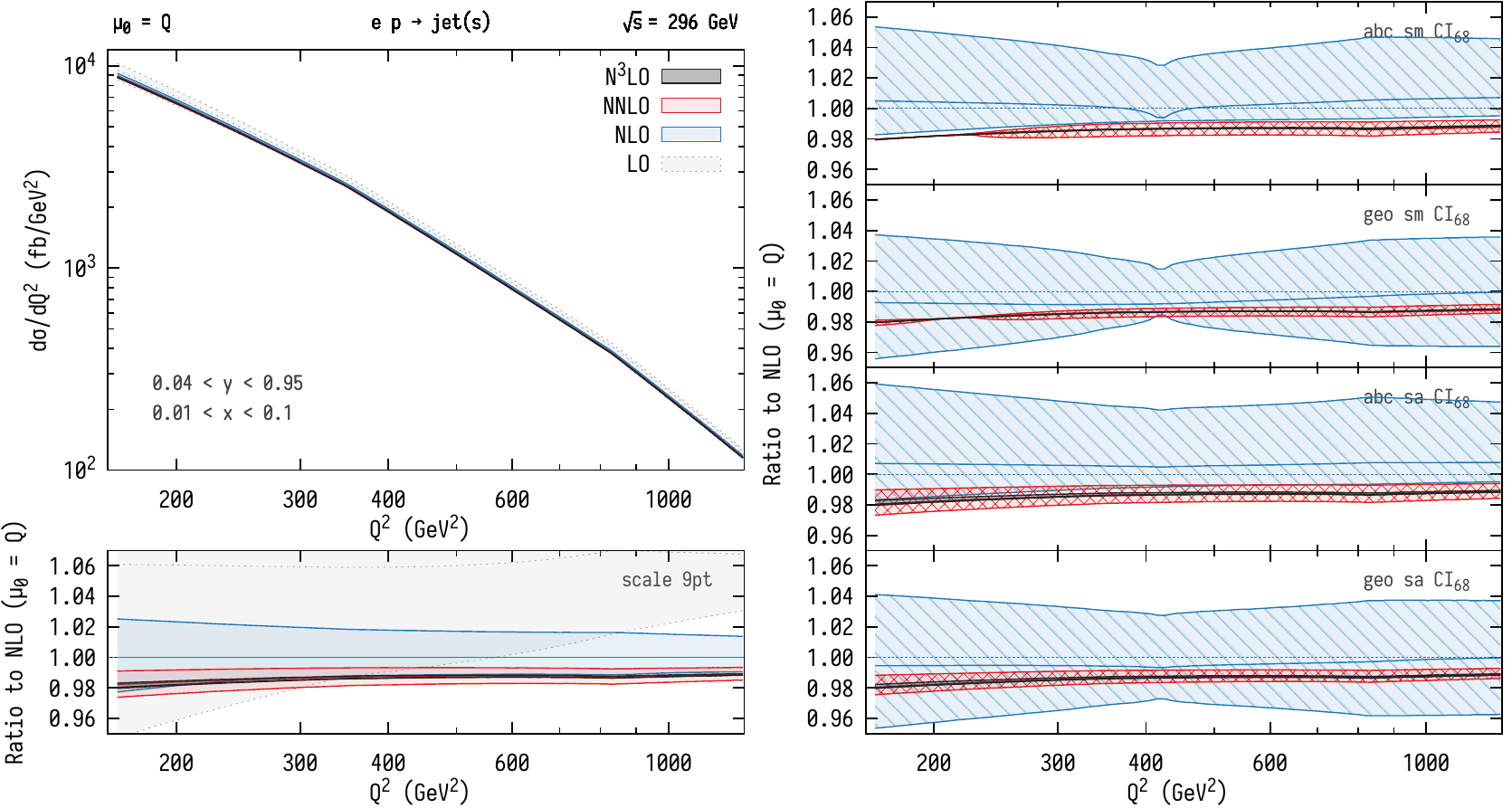}
    \caption{%
        Deep inelastic scattering distribution $d\sigma/dQ^2$ up to N${}^\text{3}$LO with $\mu_0=Q$ as the central scale choice. Left: absolute distributions and 9-point scale variation envelopes normalised to the NLO prediction at $\mu_0$. Right: 68\% CIs for the $abc$ and geometric models using the scale-marginalisation~(top) and scale-averaging~(bottom) prescriptions (also normalised to NLO).
    }
    \label{fig:dis_q2}
\end{figure}

Predictions up to N${}^\text{3}$LO for the DIS process were obtained from the calculation in ref.~\cite{Currie:2018fgr}, where we consider the distribution $d\sigma/dQ^2$ with respect to the virtuality $Q^2$ of the intermediate photon.
The results are presented in figure~\ref{fig:dis_q2}. 
The left panels show the absolute distributions (top) and the standard scale variation envelopes normalised by the NLO prediction evaluated at the central scale (bottom).
The four right panels contrast the MHO estimates (68\% CIs) obtained using the $abc$ and geometric models with the sa- (top two panels) and sm-prescriptions (bottom two panels).
All CIs are normalised to the NLO prediction evaluated at the central scale.%
\footnote{For clarity, we do not show 68\% CIs at LO, nor any of the 95\% CIs, which can be very large.}
Bands of different colours correspond to estimates at different perturbative orders: grey (LO), blue (NLO), red (NNLO) and dark grey (N${}^3$LO). 
Overall, we can observe that all models and prescriptions perform well in capturing the progression of the perturbative series with either overlapping or touching uncertainty bands. 
At NLO the Bayesian approaches give rise to 68\% CIs which are slightly larger than the 9-point scale variation interval, but are comparable in size at NNLO and N${}^\text{3}$LO.
In the case of the sm-prescription shown in the top two panels on the right in figure~\ref{fig:dis_q2}, the CI bands narrow dramatically at NLO for $Q^2\sim400~\mathrm{GeV}$, as well as at NNLO and N${}^\text{3}$LO at the lower end of the spectrum.
This can be traced back to the presence of a FAC points that are strongly emphasised by the sm-prescription.
In the case of the geometric model, we see that both scale prescriptions move the median of the NLO probability distribution towards the direction of the higher-order corrections, which slightly improves the overlap between the CIs at NNLO and N${}^3$LO.
For the $abc$-model, on the other hand, the central value at NLO moves in the opposite direction, with strongly asymmetric uncertainty bands that extend further towards the LO prediction. This is because after the negative NLO correction, the $abc$-model anticipates a positive NNLO correction, as would be the case for an alternating series. However, for this process the NNLO correction is again negative. The geometric model, which ignores the signs of the corrections, appears to perform better. We note that at higher orders the $abc$-model captures the perturbative progression well with only slightly larger CIs than the geometric model.

\subsection{Di-photon production}

We now consider predictions for di-photon production at the LHC at $\sqrt{s}=8~\mathrm{TeV}$ using the results of ref.~\cite{Gehrmann:2020oec}.
The production of a pair of photons, ${p}\,{p}\to\gamma\gamma$, provides a clean and well-measured final state that is an important background in Higgs measurements and can be used in the search for New Physics resonances.
One striking feature of this process is the apparent slow convergence of the perturbative QCD series, as partially induced by the staggered transverse-momentum cuts on the photons. 
As a consequence, lower-order predictions are found to be inadequate to describe the measurement, and only starting from NNLO a sensible comparison to the data can be performed. 

\begin{figure}
    \centering
        \includegraphics[width=\linewidth]{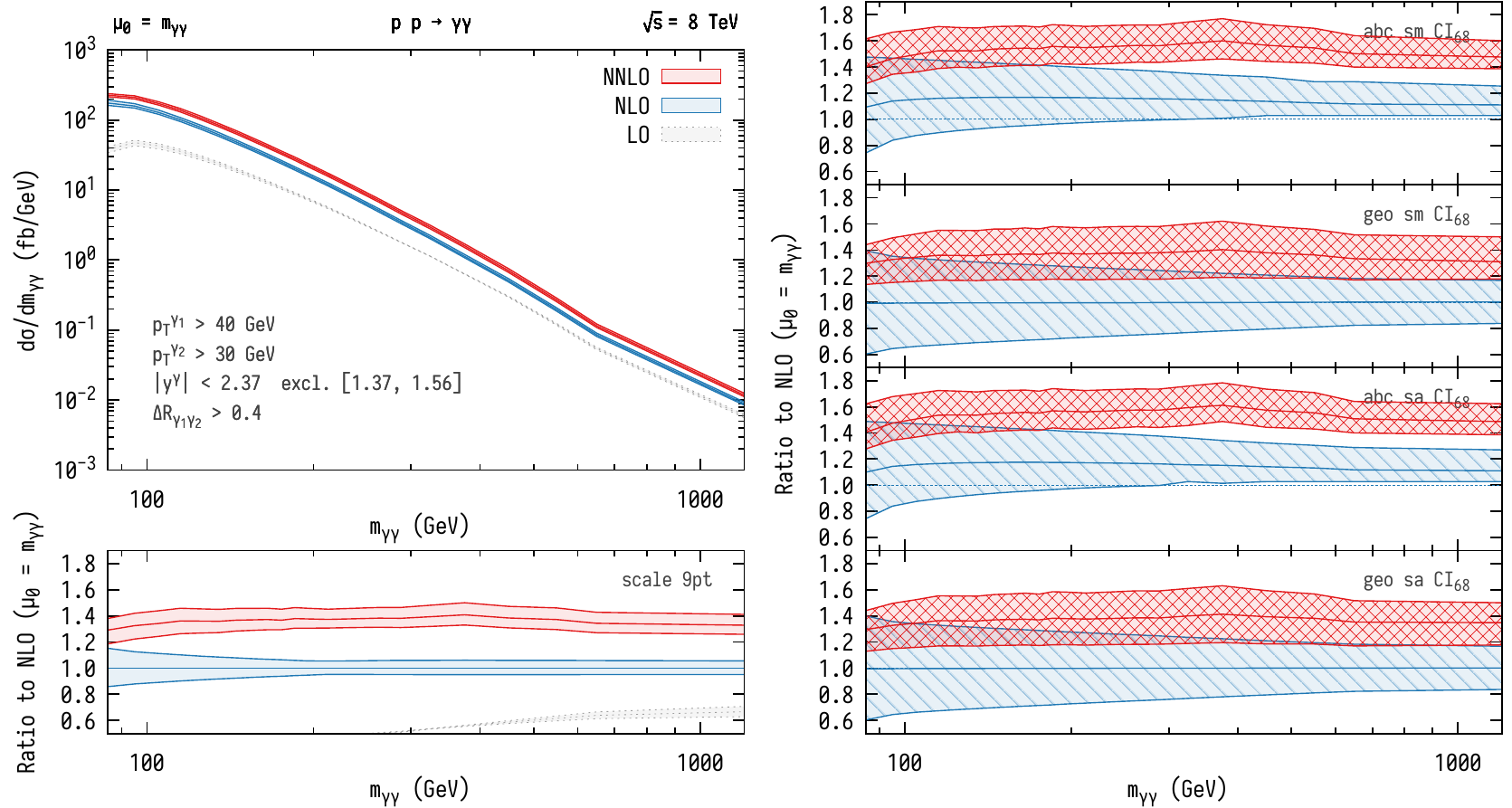}
    \caption{%
        Di-photon production up to NNLO with $\mu_0=m_{\gamma\gamma}$ as the central scale choice. Left: absolute $d\sigma/dm_{\gamma\gamma}$ distributions and 9-point scale variation envelopes normalised to the NLO prediction at $\mu_0$. Right: 68\% CIs for the $abc$ and geometric models using the scale-marginalisation~(top) and scale-averaging~(bottom) prescriptions (also normalised to NLO).
    }
    \label{fig:yy_myy}
\end{figure}

Results for the di-photon invariant-mass distribution $d\sigma/dm_{\gamma\gamma}$ are shown in figure~\ref{fig:yy_myy} following the same pattern as in figure~\ref{fig:dis_q2}.
We note that  the sizeable higher-order corrections put the respective predictions far outside from the traditional scale variation intervals of the previous order. 
Moreover, the scale variation envelope at NNLO is found to be of similar size, or even larger, than at NLO, with no convincing sign of convergence. 
The Bayesian models, on the other hand, produce 68\% CIs that are generally larger, and the CIs do not increase when going to higher orders.
With the large positive corrections prohibiting any FAC points, the two scale prescriptions produce CIs that are almost identical.
While the CI bands between NLO and NNLO touch in the case of the geometric model, for the $abc$-model the NNLO CIs are outside of the NLO band for $m_{\gamma\gamma} \gtrsim 200~\mathrm{GeV}$. This is likely due to large positive corrections encountered at each perturbative order. The $abc$-model anticipates this pattern to continue at higher orders and shifts the probability distribution towards larger cross-section values.
Finally, in view of the noticeable differences for the size and position of the bands at the highest order (NNLO) using different approaches, we can conclude that N${}^\text{3}$LO corrections will be highly relevant for this process to obtain more robust MHO estimates, which the traditional scale variation is likely substantially underestimating.

\subsection{Gauge boson production in association with a jet}

The production of electroweak gauge bosons that recoil against hard QCD emissions is among the most important standard candle processes at the LHC with a wide range of applications that span from detector calibration, precision QCD studies, to New Physics searches.
This process is now known up to NNLO for all gauge boson types. Here we focus on $W^+$+jet production using the predictions from ref.~\cite{Gehrmann-DeRidder:2017mvr}.

\begin{figure}
    \centering
       \includegraphics[width=\linewidth]{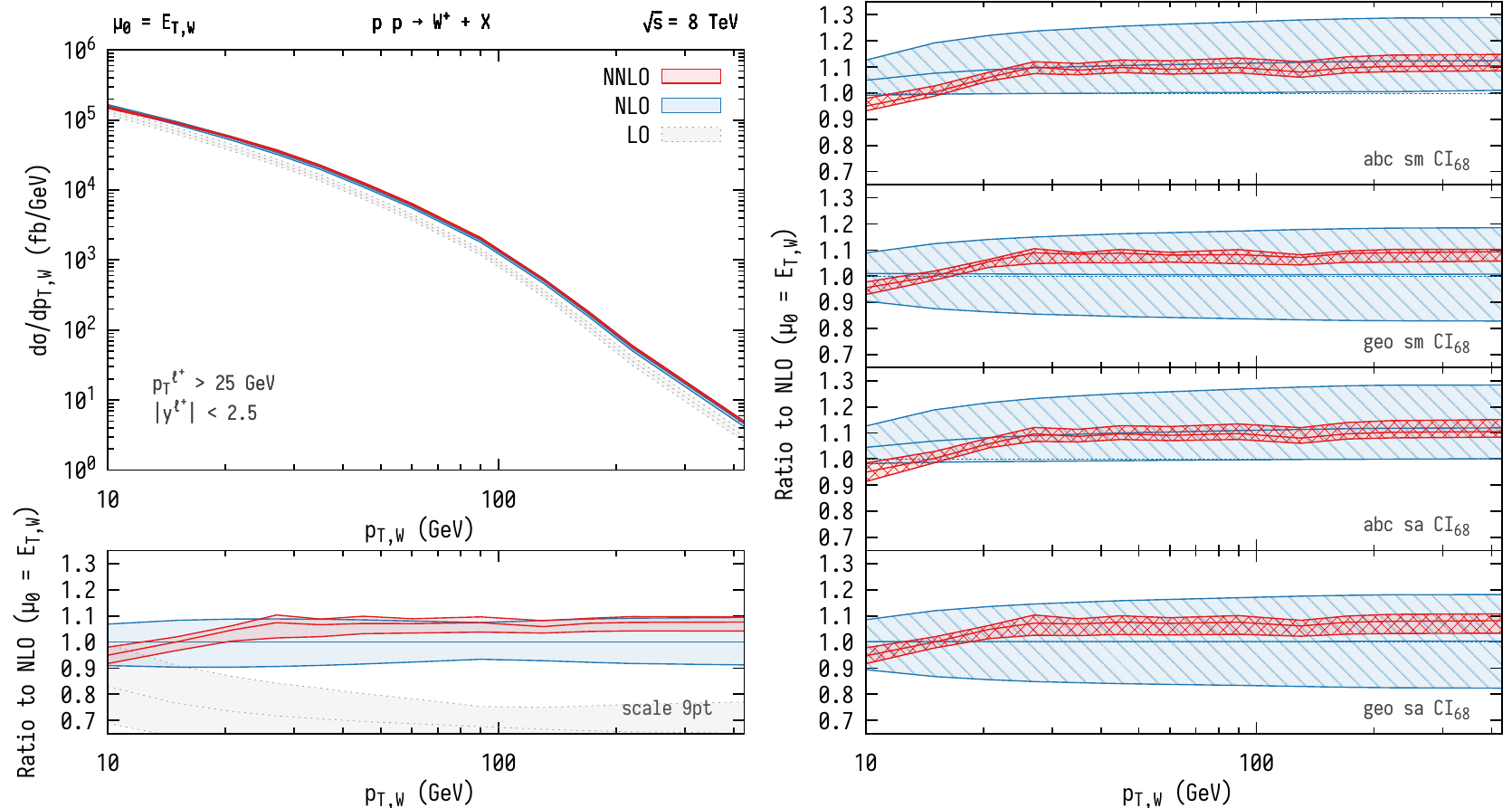}
    \caption{%
        $W^+$+jet production up to NNLO with $\mu_0=E_{T,W}$ as the central scale choice. Left: absolute $d\sigma/dp_{T,W}$ distributions and 9-point scale variation envelopes normalised to the NLO prediction at $\mu_0$. Right: 68\% CIs for the $abc$ and geometric models using the scale-marginalisation~(top) and scale-averaging~(bottom) prescriptions (also normalised to NLO).
    }
    \label{fig:wj_pt}
\end{figure}

Figure~\ref{fig:wj_pt} presents the results for the inclusive transverse momentum distribution $d\sigma/dp_{T,W}$ of the $W$ boson following the same layout as in figure~\ref{fig:dis_q2}.
We can observe a good perturbative behaviour for this process with NNLO corrections that are well captured by the traditional scale variation at one order lower. 
The Bayesian approaches give rise to 68\% CIs that are wider at NLO, while the size of the bands at NNLO are comparable between all approaches.
The sm- and sa-prescriptions give rise to almost identical CIs, similar to the observations made for the di-gamma process.
In the case of the $abc$-model, we can appreciate the flexibility of the asymmetric probability distribution. The positive offset at NLO centres the CIs at this order around the NNLO band. 
Finally, we note that once NNLO corrections are included, all the different prescriptions of estimating MHOs considered here give almost identical bands indicating that the uncertainties assigned at this order are very robust.

\subsection{Higgs boson production in association with a jet}

The study of the Higgs boson and its properties is among the highest priorities of the LHC programme and the transverse momentum distribution, in particular, plays an important role in this endeavour not only for boosted analyses but also to resolve the details of the production dynamics for this process.  
We focus here on Higgs boson production in association with a jet at $\sqrt{s}=13~\mathrm{TeV}$ in the four-lepton decay channel, $H\to4\ell$, and use the predictions for the ``ATLAS~II'' setup computed in ref.~\cite{Chen:2019wxf}.

\begin{figure}
    \centering
        \includegraphics[width=\linewidth]{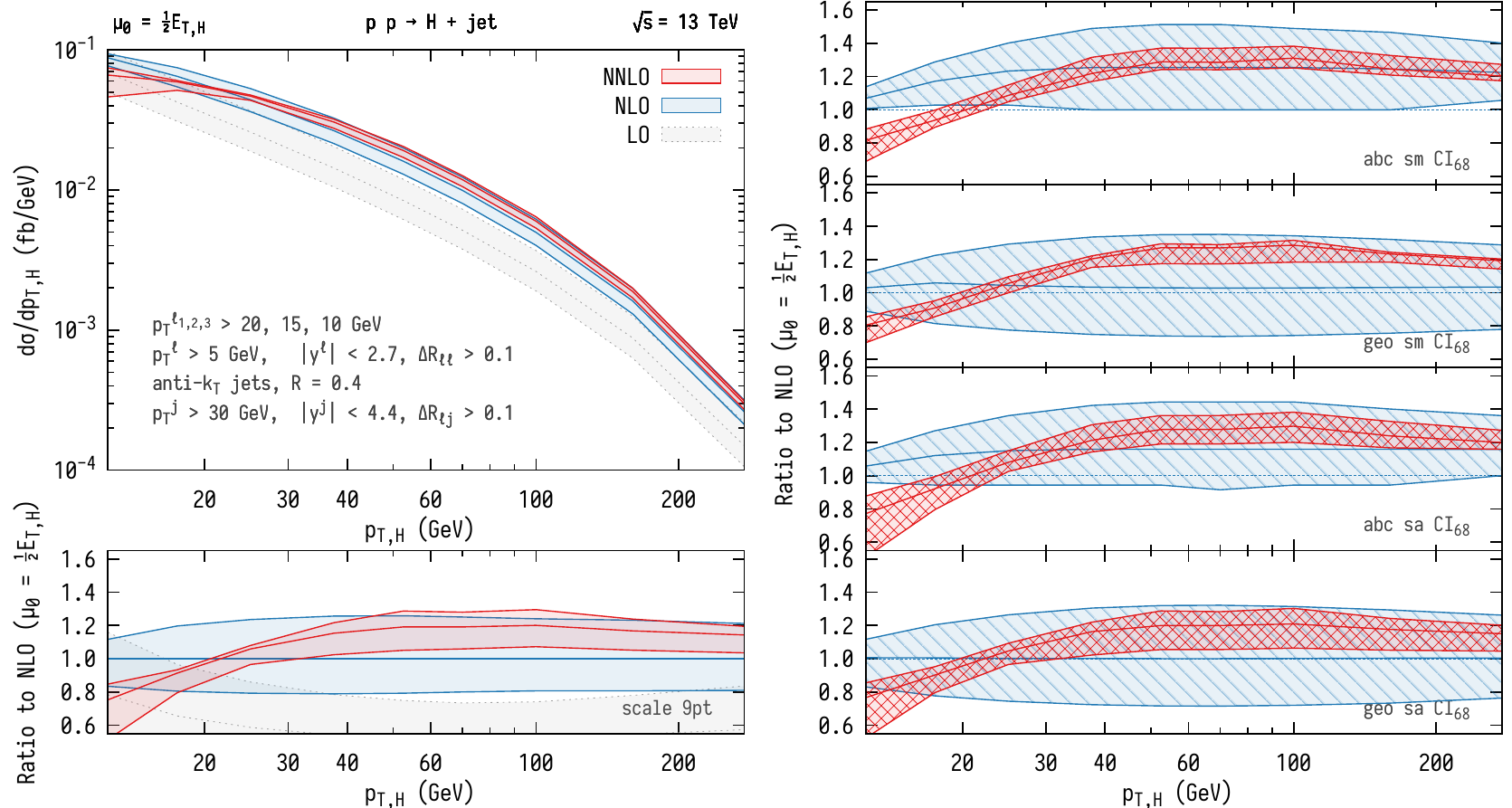}
    \caption{%
        Higgs+jet production up to NNLO with $\mu_0=\tfrac{1}{2}E_{T,H}$ as the central scale choice. Left: absolute $d\sigma/dp_{T,H}$ distributions and 9-point scale variation envelopes normalised to the NLO prediction at $\mu_0$. Right: 68\% CIs for the $abc$ and geometric models using the scale-marginalisation~(top) and scale-averaging~(bottom) prescriptions (also normalised to NLO).
    }
    \label{fig:hj_pt}
\end{figure}

The results up to NNLO for the transverse momentum distribution  $d\sigma/dp_{T,H}$ of the Higgs boson are shown in figure~\ref{fig:hj_pt}. We observe large positive corrections at each perturbative order with the exception of the lowest $p_T$ bins.
In contrast to the other differential distributions considered so far, we notice that the Bayesian models are not giving rise to substantially larger uncertainty bands (68\% CI) at NLO compared to the traditional scale variation envelope, but they are similar in size. 
Going to NNLO, the uncertainty estimates are found to be comparable between the different approaches, with the sm-prescription giving somewhat narrower bands. 
This compatibility between the different approaches again supports the robustness of the uncertainties assigned to these predictions.
With the monotonic pattern of positive corrections at each order, the $abc$-model is again able to adapt to this feature in the intermediate--high $p_T$ range, capturing to a large extent the NNLO corrections in the NLO shift. 
This is more pronounced in the sm-prescription suggesting that the central scale choice of $\mu_0=\tfrac{1}{2}E_{T,H}$ is probing regimes in the scale variation that are close a FAC point.

\subsection{Inclusive jet production}
\label{sec:jet_inclusive}

\begin{figure}
    \centering
    \includegraphics[width=\linewidth]{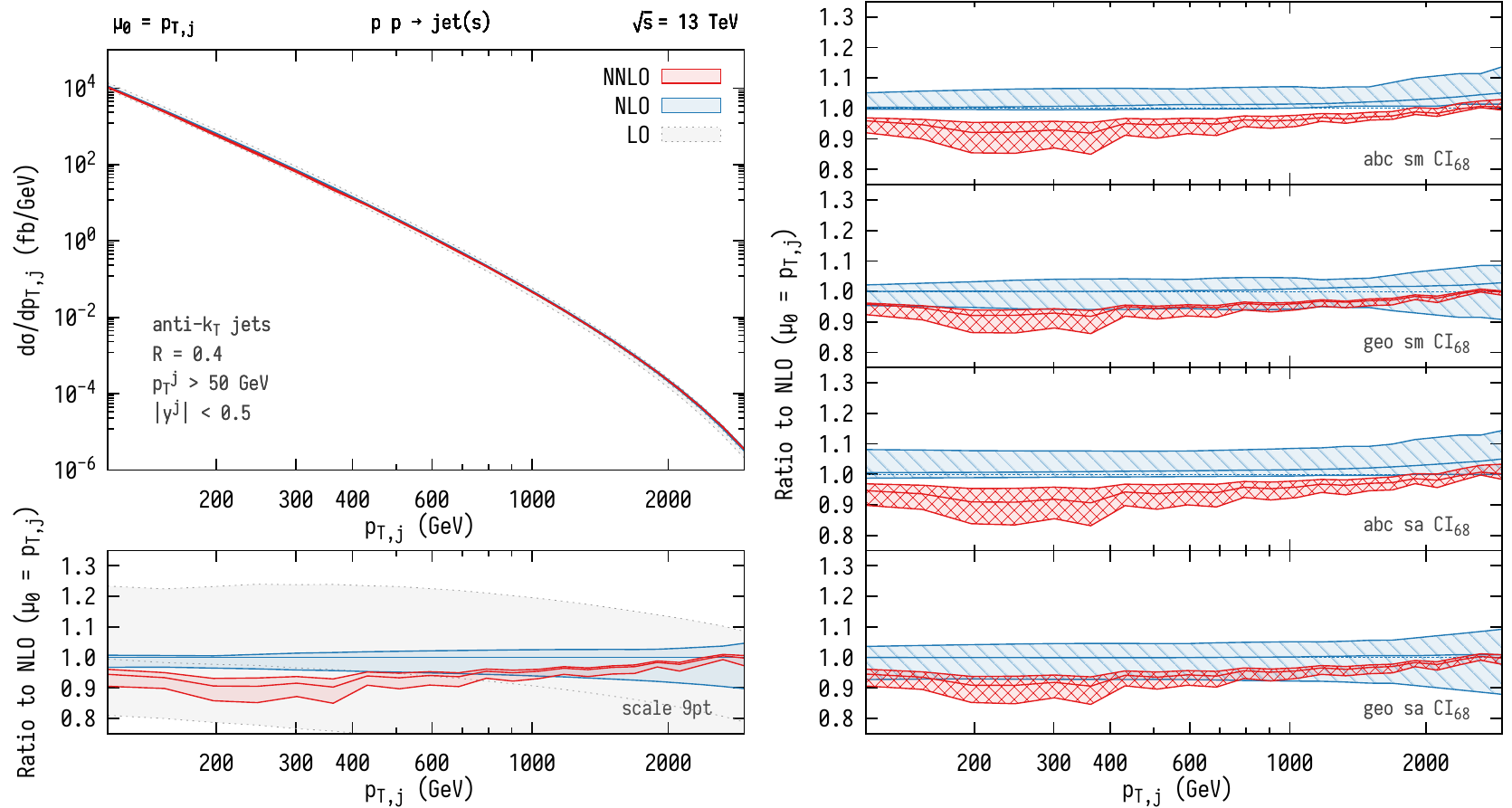}
    \caption{%
        Inclusive jet production up to NNLO with $\mu_0=p_{T,j}$ as the central scale choice. Left: absolute $d\sigma/dp_{T,j}$ distributions and 9-point scale variation envelopes normalised to the NLO prediction at $\mu_0$. Right: 68\% CIs for the $abc$ and geometric models using the scale-marginalisation~(top) and scale-averaging~(bottom) prescriptions (also normalised to NLO).
    }
    \label{fig:jet_pt}
\end{figure}

\begin{figure}
    \centering
        \includegraphics[width=\linewidth]{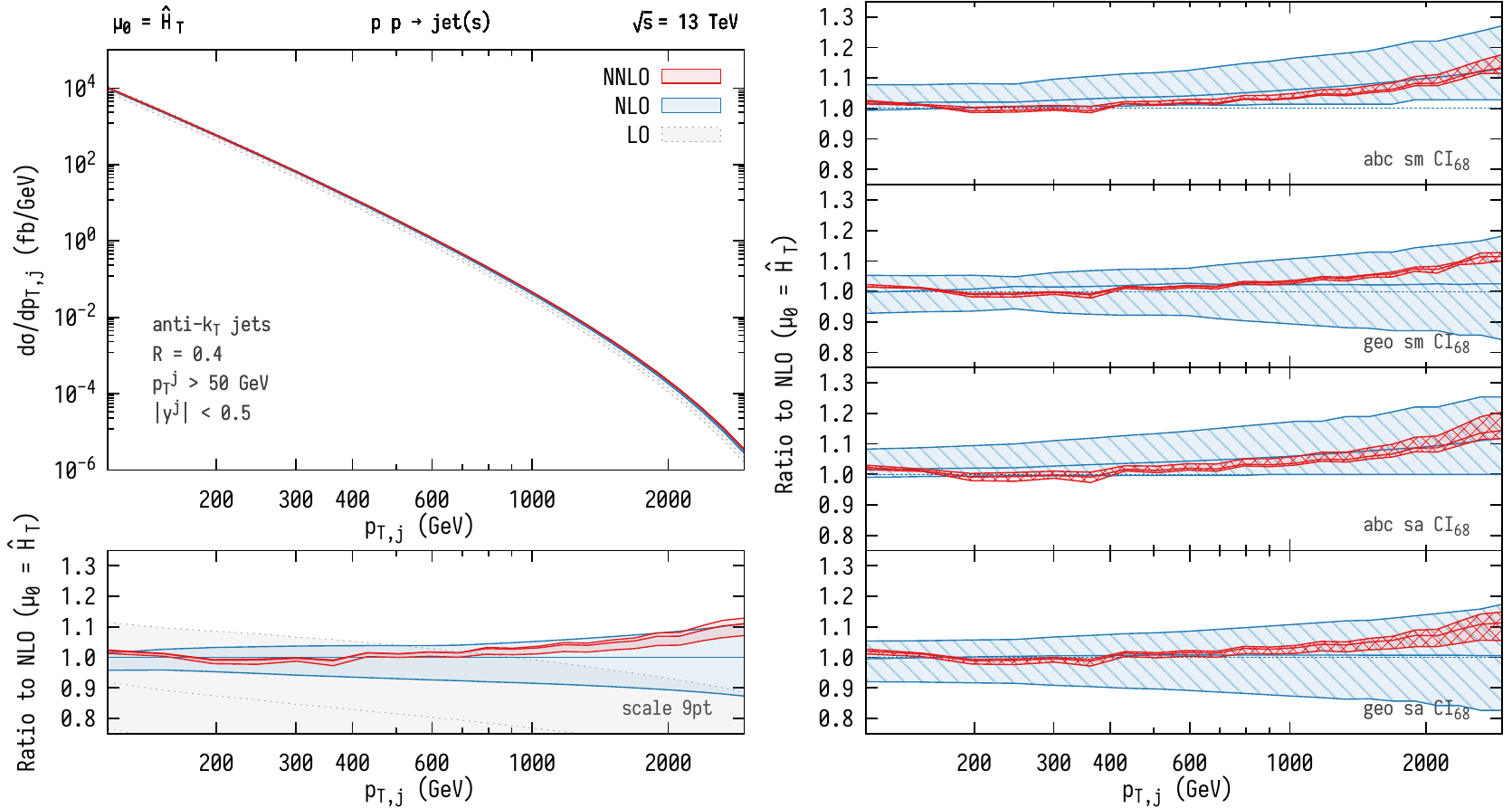}
    \caption{%
        Inclusive jet production up to NNLO with $\mu_0=\hat{H}_T$ as the central scale choice. Left: absolute $d\sigma/dp_{T,j}$ distributions and 9-point scale variation envelopes normalised to the NLO prediction at $\mu_0$. Right: 68\% CIs for the $abc$ and geometric models using the scale-marginalisation~(top) and scale-averaging~(bottom) prescriptions (also normalised to NLO).
    }
    \label{fig:jet_ht}
\end{figure}

Jet production is the most prominent process at hadron colliders and an essential tool for the study of QCD.
The inclusive transverse momentum distribution  $d\sigma/dp_{T,j}$ is particularly interesting in the context of theory uncertainties, due to its sensitivity on the specific choice of the central scale that further persists at NNLO due to instabilities in the sub-leading jet distribution that impacts the inclusive spectrum~\cite{Currie:2018xkj}.
In order to gain insights into this sensitivity and how the different prescriptions for assigning MHO uncertainties compare in these scenarios, we  consider two different central scale choices: the jet transverse momentum $p_{T,j}$ and the scalar sum of the transverse momenta of the partons $\hat{H}_T = \sum_{i\in\text{partons}} p_{T,i}$, which are shown in figures~\ref{fig:jet_pt} and \ref{fig:jet_ht}, respectively.

We begin our discussion with the scale choice $\mu_0=p_{T,j}$, which makes the NLO corrections artificially small. From the results presented in figure~\ref{fig:jet_pt}, we see that in the low-$p_T$ region all approaches give rise to sizeable NNLO CIs.
This is most clearly visible for the $abc$-model, where no reduction in the size of the CI band is observed as we move from NLO to NNLO. Neither of the bands overlap, because the $abc$-model expects a positive NNLO correction, which is not true for this central scale choice.
The geometric model, instead, is able to assign more conservative MHO uncertainties at NLO that allow for some overlap between the last two orders.
Overall, we can conclude that the Bayesian models considered here are not able to ``un-do'' an inappropriate scale choice through either the sm- or sa-prescriptions.

More specifically, 
no prescription based on a simple re-scaling from a central scale choice will be able to adapt between different \emph{dynamical} scales that encapsulate a different kinematic dependence.
This becomes most apparent in observables that are sensitive to infrared physics, because it is the cancellation between real and virtual corrections that strongly impacts such a sensitivity. 
For two dynamical scale choices that have a different dependence on the kinematics, the difference in the scales assigned to the real and virtual corrections can vary substantially. 
Such a feature can never be captured by a global scaling, as done in the sa- and sm-prescriptions considered here, that varies the scales for the real and virtual contributions in the same manner.

Using $\hat{H}_T$ as the central scale choice as advocated in ref.~\cite{Currie:2018xkj}, we observe in figure~\ref{fig:jet_ht} a stable progression of the series throughout with the next order, typically lying within the uncertainty estimate of the previous order.
The 68\% CIs are slightly larger for the Bayesian approaches at NLO, while at NNLO all uncertainty estimates are mutually compatible.
This further supports this scale as an appropriate choice given its robustness with respect to the different prescriptions in assigning theory uncertainties.
Finally, it is interesting to note that the $abc$-model is able to adapt to the sizeable positive corrections in the tail of the distribution, thus giving a noticeable positive shift at NLO that reduces the overall impact of NNLO corrections.

\section{Discussion}
\label{sec:discussion}

\begin{figure}
    \centering
    \includegraphics[width=0.7\linewidth]{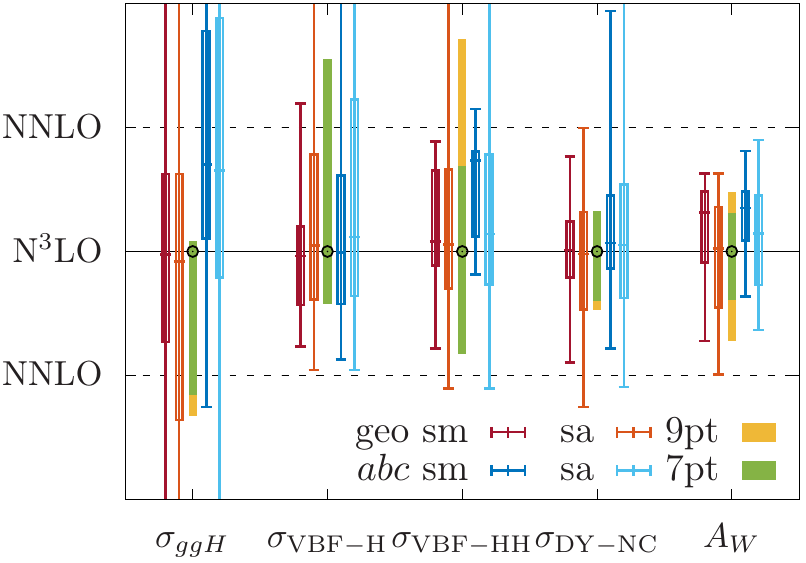}
    \caption{Comparison of MHO estimates at N${}^3$LO between different models and scale prescriptions for the gluon-fusion Higgs cross-section, single and di-Higgs production in vector boson fusion, neutral current Drell-Yan production and the lepton charge asymmetry. See the corresponding sections for each process for further details.}
    \label{fig:pull_plot}
\end{figure}

The estimation of MHOs is a vital aspect of collider phenomenology. Traditional approaches are based on scale variation, an ad hoc prescription to obtain uncertainty intervals for MHOs without a clear probabilistic underpinning.
In this paper, we have analysed Bayesian approaches to estimate the MHO uncertainties from the progression of the perturbative expansion of a physical observable. Building on the recent work in ref.~\cite{Bonvini:2020xeo}, we have focused in particular on the question of how the dependence on the unphysical perturbative scales $\mu_F$ and $\mu_R$ can be incorporated into Bayesian inference. Our goal was to scrutinise various Bayesian models and to assess their common features by studying their performance on a wide and representative selection of hadron collider observables known at high orders in perturbation theory, namely NNLO and/or N$^3$LO. In figure~\ref{fig:pull_plot} we show a summary plot of 68\% and 95\% CIs for MHOs in our Bayesian approach and the traditional scale variation intervals  for selected collider processes known up to N$^3$LO.
In the following we summarise our main findings and conclusions, pointing out directions for future research.

\paragraph{Biases related to the choice of the scale prescription.} 
An important question is how the dependence on perturbative scales can (or should) be included into Bayesian MHO estimation.
In ref.~\cite{Bonvini:2020xeo} it was proposed to treat the perturbative scales on par with other hidden parameters of a Bayesian model, which we refer to as the scale-marginalisation (sm) prescription. In section~\ref{sec:scale_marginalisation} we have introduced an alternative prescription, called scale-averaging (sa), which does not require the perturbative scales to be assigned a statistical interpretation. We do not see any compelling physics argument to prefer one prescription over the other.

The two prescriptions do not just differ in their philosophy of how the perturbative scales are interpreted, but we have demonstrated that they may lead to substantially different estimates for MHOs. In particular, in section~\ref{sec:scale_discussion} we have shown that (under some additional reasonable assumptions for Bayesian models) each of the two prescriptions has intrinsic biases towards specific scales: the sm-prescription tends to favour FAC points as the `best' value of the perturbative scale, while the sa-prescription is biased towards the values of the observable at the PMS points. These biases are clearly visible in the examples in section~\ref{sec:hadronic_examples} (even for models for which the assumptions entering the proofs are not satisfied). Care is thus needed when applying Bayesian techniques to incorporate scale information: the choice of the prescription may bias the predictions in favour of a specific scale choice. It would be interesting to see if one can construct other Bayesian models that incorporate scale information in an unbiased way.

\paragraph{The choice of the Bayesian model and the priors.} In order to estimate MHOs in a Bayesian approach, one has to specify a model for the progression of the perturbative series. The geometric model studied in ref.~\cite{Bonvini:2020xeo} neglected the sign of perturbative corrections, which enforces the posterior distributions for MHOs (at fixed scale) to be symmetric around zero. To allow for asymmetric MHO probability distributions we introduced the $abc$-model in section~\ref{sec:no_scale} (see appendix B of ref.~\cite{Bonvini:2020xeo} for alternatives). In section~\ref{sec:qft_examples} we benchmarked the $abc$-model on a number of perturbative QFT quantities known to high orders, for which scale dependence is not relevant. Our results show that the specific choice of the model and its priors can strongly influence the size and position of MHO CIs, especially when the number of input data is small.

From the physics perspective, there is no preference between Bayesian models. 
While from an aesthetic point of view, one may prefer simple models, a judicious but ad hoc choice has to be made. It might be therefore desirable to develop new models in the future suited for specific classes of perturbative series, e.g., see our QED examples.

\paragraph{The choice of the central scale and the range of scales considered.} 
The prior on the perturbative scale in eq.~\eqref{eq:scale_prior} for the sm-prescription (see ref.~\cite{Bonvini:2020xeo}) (or the weighting function in eq.~\eqref{eq:weight} for the sa-prescription) depends on a choice of the central scale $\mu_0$ and the size $F$ of the interval for the scale integration. Our analysis in section~\ref{sec:hadronic_examples} shows that the CIs have a substantial dependence on the choice of $\mu_0$ and $F$. For example, once the range $\mu_F,\mu_R\in[\mu_0/F, \mu_0 F]$ is wide enough to include an FAC point, the CIs in the sm-prescription may become relatively $F$-independent, but mostly because the probability distributions become strongly peaked and biased towards to the FAC point. The sa-prescription, instead, will typically increase with $F$, thereby rendering the probability distributions $F$-dependent. 

The study of the inclusive jet cross section at NNLO in section~\ref{sec:jet_inclusive} shows that the convergence of the CIs in the Bayesian approach (at least for the prescriptions and models studied here) is not any less sensitive to the central scale choice as the traditional scale-variation intervals. It would be interesting to see if there are other prescriptions or choices for the prior $P_0(\mu)$ and the weighting function $w(\mu)$ that lead to probability distributions and CIs that are truly independent of the choice of the central scale $\mu_0$ and interval size $F$.
The main challenge in this regard arises from the limitations of a naive re-scaling of the central scale by constant factors that, in general, will not able to capture the difference in the kinematic dependence among possible choices for a  dynamical scale.

\paragraph{Relation to MHO estimates using scale-variation.} 
Let us point out a very practical connection between 
the existing theoretical results with the 9-point scale variation and scale prescriptions in the Bayesian approach. The scale dependence is typically mild enough so that a low-order Gauss-Legendre quadrature rule can be used to approximate the scale integrals in the sm- and sa-prescriptions. For the specific choice of integration interval $F\approx 2.45$, the traditional scale variation points are placed at the quadrature nodes. Therefore the Bayesian CIs can be efficiently computed from already existing predictions.

Furthermore, for the observables studied in section~\ref{sec:hadronic_examples},
we have demonstrated  that the traditional 7- or 9-point scale variation intervals are often similar in size to the Bayesian 68\% and 95\% CIs. Similar behaviour has been observed before  \cite{Cacciari:2011ze, Bagnaschi:2014wea,Bonvini:2020xeo}.
This is not a coincidence, and in section~\ref{sec:MHO-connection-7pt} we have shown that (under certain assumptions)
 the 95\% CIs in the sa-prescription must include the scale-variation intervals. We stress that scale-variation intervals themselves are devoid of any statistical interpretation.
 
Finally, we observe 
that the 68\% CIs at different perturbative orders in the Bayesian approach share some qualitative features known from scale-variation. For example, neither the scale variation envelopes nor the 68\% CIs overlap  for Drell-Yan production at N$^3$LO and NNLO or di-photon production at NNLO and NLO (however some models lead to barely overlapping 68\% CIs, e.g., see figure~\ref{fig:yy_myy}). For processes where scale-variation intervals indicate good perturbative convergence,  Bayesian models often also show a good, or even better, nesting of 
68\% CIs at different orders. The similarities of the two approaches are not coincidental, as both prescriptions share assumptions on the convergence of the perturbative series, namely that we are at low enough order so the perturbative series exhibits a convergent behaviour.
Despite these similarities, scale-variation intervals and Bayesian CIs are very different objects from a statistical point of view. In particular, within the Bayesian approach, one can choose in a quantitative manner how aggressive or conservative estimates of MHO should be for a given process and for a given choice of model.

\paragraph{Conclusions.} Based on our comprehensive analysis, we conclude that Bayesian approaches provide solid alternatives to the traditional scale variation as a means to estimate MHOs, but there are several caveats that one needs to keep in mind (and that possibly need to be improved in the future, e.g., by developing novel Bayesian models). The choice of the Bayesian setup, in particular the scale prescription and the priors, may have a substantial impact on the inference. In a certain sense, the Bayesian approach is not any less subjective or ad hoc than customary scale variation. Therefore care is needed not to introduce any biases by the model choice.

A significant advantage of Bayesian techniques over scale variation is that the CIs computed from Bayesian approach have a clear probabilistic interpretation (at least within the context of a given model). This opens the way to apply statistical techniques, e.g., to systematically combine MHO uncertainties with uncertainties from different sources or to include correlations between observables or bins for differential distributions.
It would also be interesting to explore how far one can treat the choice of the model itself in a Bayesian way, in order to minimise bias from the model choice, e.g., using ideas from \emph{Bayesian Model Averaging (BMA)} (see, e.g., ref.~\cite{bma}, and references therein). 
We leave these studies for future work.

The source code to the computer program \texttt{MiHO} that was used to obtain the results shown in the paper is made publicly available at \url{https://github.com/aykhuss/miho}.

\acknowledgments

RS is supported by the United States Department of Energy under
Grant Contract DE-SC0012704.

\appendix

\section{The position of the peak of the distribution}
\label{app:peak}

In this appendix we present a (heuristic) proof of the position of the peak of the distribution $\cP(\Sigma|\bfSigma_n)$. We work under the assumptions discussed in section~\ref{sec:scale_discussion}, which we recall here for convenience:
\begin{enumerate}
\item The model $(P,P_0)$ is symmetric.
\item We work with the prior in eq.~\eqref{eq:scale_prior} or the weighting function in eq.~\eqref{eq:weight}.
\item The distributions $P(\Sigma^{(n+1)}(\mu)|\bfSigma_n(\mu))$ have a single peak in $\Sigma^{(n+1)}(\mu)$ for each fixed value of $\mu$, and the peak is reached for some $\mu\in  [\mu_0/F,\mu_0F]$.
\item The peak is more and more pronounced as $m$ increases, i.e., as more information on the progression of the series becomes known. 
\item There is a single FAC or PMS point in the interval $[\mu_0/F,\mu_0F]$.
\end{enumerate}

\subsection{The position of the peak for the scale-marginalisation prescription}
Using the prior on the scale in eq.~\eqref{eq:scale_prior} and the approximation in eq.~\eqref{eq:perturbative_P}, eq.~\eqref{eq:P_scale-marginalization_approx} takes the form:
\beq\label{eq:P_scale-invariant_2}
 \cP_{\text{sm}}(\Sigma|\bfSigma_n)  = \frac{\int_{L_-}^{L_+} dL\,P(\Sigma-\Sigma_n(e^L)|\bfSigma_{n}(e^L))\,P(\bfSigma_{n}(e^L))}{\int_{L_-}^{L_+}  dL\,\cPFS(\bfSigma_n(e^L))}\,.
\eeq
In the previous equation we have changed variables from $\mu$ to $L=\log\mu$, and we defined $L_{\pm}=\log(\mu_0F^{\pm 1})$. In order to determine the position of the peak of $\cP_{\text{sm}}(\Sigma|\bfSigma_n)$ as a function of $\Sigma$, we analyse where the bulk of the integral in the numerator receives its contributions. Bayes' formula implies:
\beq\label{eq:peak_1}
\cPFS(\bfSigma_n(\mu)) = \cPFS(\Sigma_0(\mu))\,\prod_{k=1}^n\cPFS(\Sigma^{(k)}(\mu)|\bfSigma_{k-1}(\mu))\,.
\eeq
Since $\cPFS(\Sigma^{(k)}(\mu)|\bfSigma_{k-1}(\mu))$ is assumed to be a peaked function, and the peak becomes more pronounced as $k$ increases, the product in the right-hand side in eq.~\eqref{eq:peak_1} is dominated by the peak from $k=n$. Since we assume $\cPFS$ to be symmetric, the position of the peak is the point where $\Sigma^{(n)}(\mu)=0$, and by assumption there is a $\mu_{\text{FAC}} \in[\mu_0/F,\mu_0F]$ such that $\Sigma^{(n)}(\mu_{\text{FAC}})=0$, i.e., $\mu_{\text{FAC}}$ is a FAC point at N$^n$LO.
Hence, $\cPFS(\bfSigma_n(\mu))$ is peaked at the FAC point $\mu=\mu_{\text{FAC}}$, and we can approximate the integral by
\beq
 \cP_{\text{sm}}(\Sigma|\bfSigma_n)  \approx P\big(\Sigma - \Sigma_n(\mu_{\text{FAC}})|\bfSigma_n(\mu_{\text{FAC}})\big) = P(\Sigma - \Sigma_{n-1}(\mu_{\text{FAC}})|\bfSigma_n(\mu_{\text{FAC}}))\,.
 \eeq
Since $P$ is symmetric, $P(\Sigma - \Sigma_{n-1}(\mu_{\text{FAC}})|\bfSigma_n(\mu_{\text{FAC}}))$ reaches its peak when $\Sigma - \Sigma_{n-1}(\mu_{\text{FAC}})=0$, i.e., the position of the peak is located at the FAC point at N$^n$LO:
 \beq\label{eq:bonvini_peak}
 \Sigma_n^{\text{peak}}\approx \Sigma_{n-1}(\mu_{\text{FAC}}) = \Sigma_{n}(\mu_{\text{FAC}})\,.
 \eeq

\subsection{The position of the peak for the scale-averaging prescription}
We use the same notations and conventions as in the previous section.
Using the weighting function in eq.~\eqref{eq:weight} and the approximation in eq.~\eqref{eq:perturbative_P}, eq.~\eqref{eq:P_mixture-distribution} takes the form:
\beq\label{eq:P_mixture-distribution_2}
 \cP_{\text{sa}}(\Sigma|\bfSigma_n)  = \frac{1}{2\log F}\int_{L_-}^{L_+} dL\,P(\Sigma-\Sigma_n(e^L)|\bfSigma_{n}(e^L))\,.
\eeq
We proceed in a similar way as in the previous section, and we want to understand where the bulk of the integral comes from. The probability in the integrand is peaked by assumption. To understand the impact on the position of the peak of $\cP_{\text{sa}}(\Sigma|\bfSigma_n)$, let us in first approximation write the probability distribution in the integrand as a $\delta$-function with support at $\Sigma-\Sigma_n(e^L)=0$:
\beq\label{eq:P_mixture-distribution_2}
 \cP_{\text{sa}}(\Sigma|\bfSigma_n)  \approx \frac{1}{2\log F}\int_{L_-}^{L_+} dL\,\delta(\Sigma-\Sigma_{n}(e^L))\,.
\eeq
By assumption, the $\delta$-function has support in the range $[L_-,L_+]$, and so there is $L_\Sigma = \log\mu_\Sigma\in[L_-,L_+]$ such that $\Sigma-\Sigma_{n}(\mu_\Sigma)=0$ (note that $\mu_{\Sigma}$ depends on $\Sigma$!). This gives:
\beq
 \cP_{\text{sa}}(\Sigma|\bfSigma_n)  \approx \frac{1}{\Sigma'_n(\mu_\Sigma)}\,.
 \eeq
 We see from this argument that $\cP_{\text{sa}}(\Sigma|\bfSigma_n)$ reaches is maximum when $\Sigma'_n(\mu_\Sigma)\approx 0$, and so $\Sigma^\text{peak}_n=\Sigma_n(\mu_\Sigma)=\Sigma_n(\mu_{\text{PMS}})$ is at the PMS point at N$^n$LO.

\section{Analytic solution of the $abc$-model}\label{app:abc_model}
In this appendix we show the analytical solution of  the $abc$ model for $\xi=1$. The general case $\xi\neq 1$ can be solved using similar techniques. Recall that for $\xi=1$, after integration over $c$ we have
\begin{align}
P_{abc}(\bfdelta_n)&=
\int \frac{da\,db\, P_0(a)}{|a|^{n(n+1)/2}} \frac{ \epsilon\,\eta^{\epsilon}}{2^{n+2}(n+2+\epsilon) } \max\left(\eta, \max_{0\leq k\leq n}\left|\frac{\delta_k}{a^k}-b\right|, |b|\right)^{-(n+2+\epsilon)}.
\end{align}
We perform the integral over $b$ first. We split factors $\tfrac{\delta_i}{a^i}$ into two sets according to
\begin{align}
  \Delta^{(+)}(a) &=
  \max \Bigl( \Bigl\{  \tfrac{\delta_i}{a^i} \Big\vert \tfrac{\delta_i}{a^i} \geq 0 \Bigr\} \cup \{ 0 \} \Bigr) \geq 0\,,
  \\
  \Delta^{(-)}(a) &=
  \min \Bigl( \Bigl\{  \tfrac{\delta_i}{a^i} \Big\vert \tfrac{\delta_i}{a^i} \leq 0 \Bigr\} \cup \{ 0 \} \Bigr) \leq 0\,,
\end{align}
and we define $b_\lor \equiv \tfrac{\Delta^{(+)}+\Delta^{(-)}}{2}$. We  distinguish two cases:

\renewcommand{\labelenumi}{\textbf{(\Roman{enumi})}}
\begin{enumerate}
  \item $\eta \leq |b_\lor|$: 
 The integration region splits into two regions
  \begin{align}
    \int_{-\infty}^{+\infty} db \; \max \{\ldots\}^{-(n+2+\epsilon)}
    &=
    \int_{-\infty}^{b_\lor} db \; \left[ \Delta^{(+)} - b \right]^{-(n+2+\epsilon)}
    \nonumber\\&\quad
    + \int_{b_\lor}^{\infty} db \; \left[ b - \Delta^{(-)} \right]^{-(n+2+\epsilon)}
    \nonumber\\&=
    \frac{2^{n+2+\epsilon}}{n+1+\epsilon} \left(\Delta^{(+)}-\Delta^{(-)}\right)^{-(n+1+\epsilon)} \,.
  \end{align}
  \item $\eta > |b_\lor|$: 
  In this case, we have to consider three regions
  \begin{align}
    \int_{-\infty}^{+\infty} d b \; \max \{\ldots\}^{-(n+2+\epsilon)}
    &=
    \int_{-\infty}^{-\eta+\Delta^{(+)}} d b \; \left[ \Delta^{(+)} - b \right]^{-(n+2+\epsilon)}
    \nonumber\\&\quad
    + \int_{-\eta+\Delta^{(+)}}^{\eta+\Delta^{(-)}} d b \eta
    \nonumber\\&\quad
    + \int_{\eta+\Delta^{(-)}}^{\infty} d b \; \left[ b - \Delta^{(-)} \right]^{-(n+2+\epsilon)}
    \nonumber\\&=
    \eta  (\Delta^{(-)}-\Delta^{(+)}+2 \eta )+\frac{2 \eta^{-\epsilon-n-1} }{n+\epsilon+1}\,.
  \end{align}
\end{enumerate}
Finally, we obtain the probability distribution 
\begin{align}
P_{abc}(\bfdelta_n)
  &=
  \int_{-1}^{1} da\frac{(1-|a|)^{\omega}}{|a|^{n(n+1)/2}}\frac{(1+\omega) \epsilon\,\eta^{\epsilon}}{2^{n+3}(n+2+\epsilon) }
  \nonumber\\ &\quad\times
  \begin{cases}
     \frac{2^{n+2+\epsilon}}{n+1+\epsilon} \left(\Delta^{(+)}-\Delta^{(-)}\right)^{-(n+1+\epsilon)}\,,
  & \text{for $\eta \leq |b_\lor|$,}
  \\
    \eta  (\Delta^{(-)}-\Delta^{(+)}+2 \eta )+\frac{2 \eta^{-\epsilon-n-1} }{n+\epsilon+1}\,,
  & \text{for $\eta > |b_\lor|$ \,.}
  \end{cases}
\end{align}
The result for $\xi \neq 1$ takes a similar functional form, but one has to consider additional cases, as shown in figure \ref{fig:bc_ranges}. 
In practice, we perform the integral over $a$ numerical in our implementation of the $abc$ model in the code \texttt{MiHO}. 
Nonetheless, it is instructive to discuss the analytical solution for a specific  choice of $\omega$. We note that the $\Delta^{(\pm)}$ depend on $a$ and are piecewise power functions of $a$;  $\Delta^{(\pm)} \sim a^{-i}$ for some power $i$. As such, the $a$ integration domain is divided into disjoint intervals $0<\ldots<a_k<a_{k+1}<\ldots<1$ on which the maximum function can be evaluated  (see ref.~\cite{Bonvini:2020xeo} for how to find the $a_k$) and the most complicated integral we encounter has the form
\begin{align}
  \int_{a_k}^{a_{k+1}} d a \;
  (1-|a|)^\omega  a^{l} \left( \frac{\alpha}{a^i} + \frac{\beta}{a^j} \right)^{\nu}\,,
\end{align}
with $\alpha,\beta>0$, $l\in\mathbb{Z}$, $i,j\in\mathbb{N}$. Restricting ourselves to integer $\omega$, the most general integral that we need is of the form 
\begin{align}
    I_{pq}(\mu,\nu;a_k,a_{k+1},\alpha,\beta)&\equiv
  \int_{a_k}^{a_{k+1}} d a \;
  a^{\mu} \left( \frac{\alpha}{a^p} + \frac{\beta}{a^q} \right)^{\nu}\\
  &= \frac{\alpha^\nu}{p-q}\left(\frac{\alpha}{\beta}\right)^{\frac{\mu-\nu p+1}{p-q}}F\left(\frac{\mu -\nu p +1}{p-q}-1,\nu;\frac{\beta}{\alpha} a_k^{p-q},\frac{\beta}{\alpha}a_{k+1}^{p-q}\right),
\end{align}
with $F$ related to Gaussian hypergeometric function 
\begin{align}
    F(\rho,\sigma;A,B) \equiv \frac{B^{1+\rho}}{1+\rho}\, {}_2F_1\left(2+\rho + \sigma, 1+\rho;2+\rho;\frac{B}{1+B}\right) - (B \leftrightarrow A)\,.
\end{align}

\section{Supplementary plots for sections 3 and 4}
\label{app:plots}

\begin{figure}[ht]
 \centering
    \includegraphics[width=\linewidth]{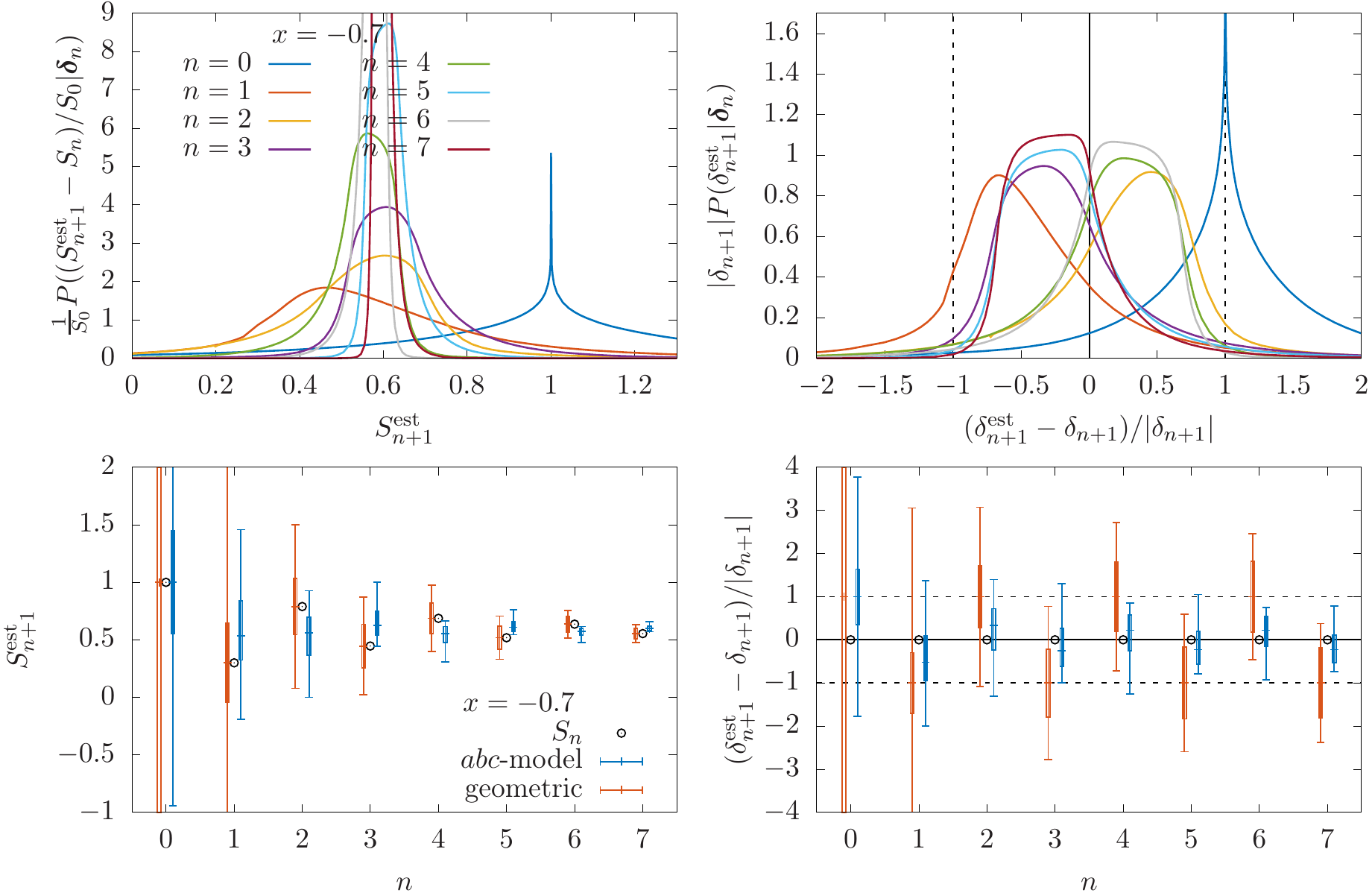}
        \caption{\label{fig:alt_geom_progression}
     Top left panel: The probability distribution from the $abc$ model for $S_{n+1}^\text{est}$  for different values of $n$ for the geometric series with $x=-0.7$. Top right panel: The same as the left panel, but we show the probability for the scaled deviation from the known correction $(S_{n+1}^\text{est}-S_{n+1})/|S_{n+1}-S_{n}|$.
    Bottom left panel: The median (plus), 68\% CI (errorbox) and 95\% CI (errorbar)  for the posterior of $S_{n+1}^\text{est}$, computed from the $abc$ (blue) and geometric (red) models using information on the previous orders. The exact values of $S_n$ are shown as black circles. Bottom right panel: The same as the left panel, but the exact $S_n$ value is subtracted from $S_{n+1}^\text{est}$ and the difference is normalised by $|S_{n+1}-S_{n}|$.
      }
    \end{figure}

For completeness we include the plots for probability distributions and CIs for the alternating geometric series, figure~\ref{fig:alt_geom_progression}. Compare to figure~\ref{fig:geom_progression} and see section~\ref{sec:sensitivity_to_priors} for a detailed discussion.

\begin{figure}[ht]
 \centering
    \includegraphics[width=\linewidth]{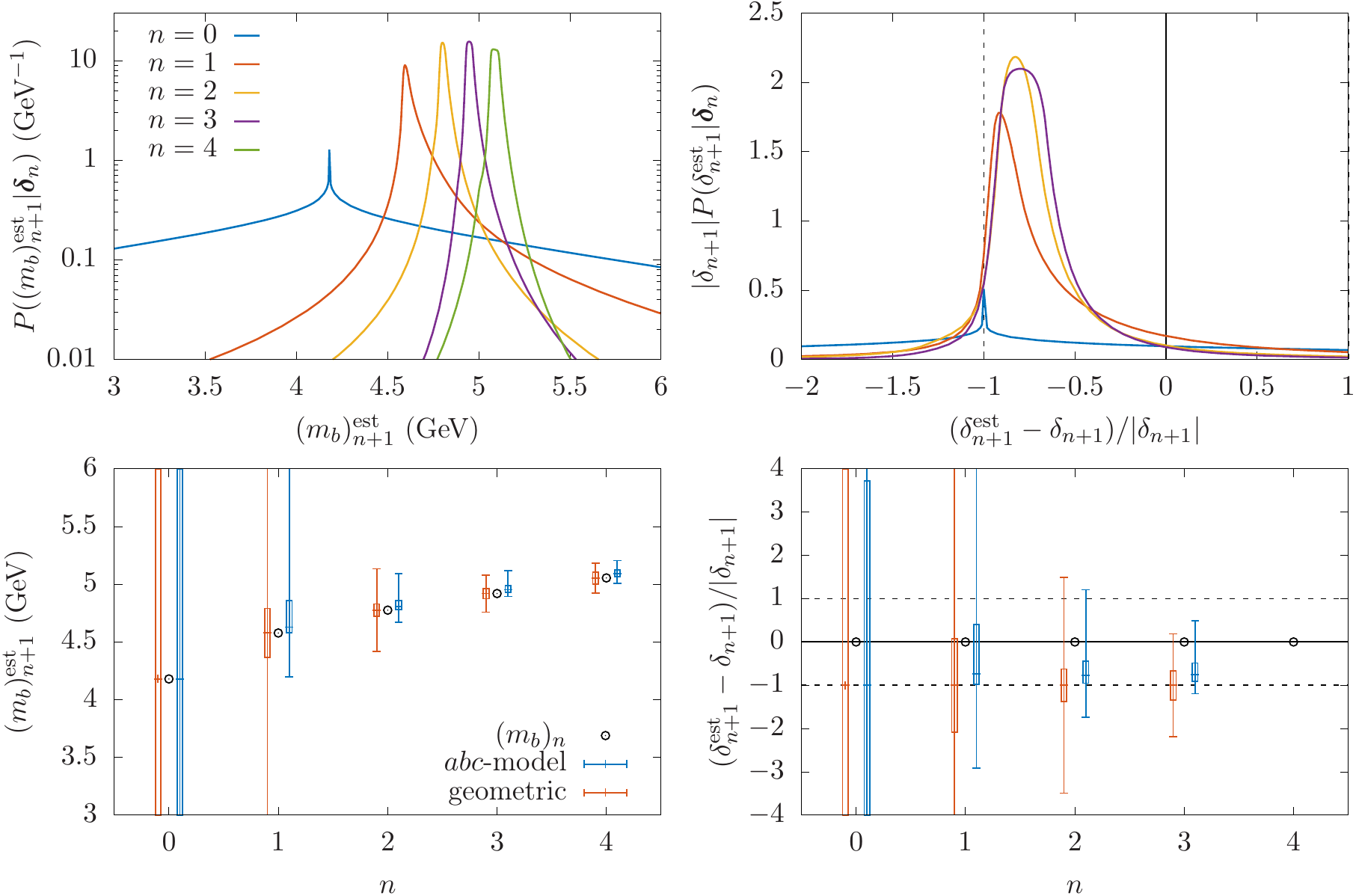}
    \caption{\label{fig:mb_progression}Top left panel: The probability distributions from $abc$-model for the on-shell bottom quark mass $(m_b)^\text{est}_{n+1}$ evaluated at $\mu_R=m_b$ and for different values of $n$. Top right panel: The same distributions
    normalised to the exact N${}^{n+1}$LO  correction.
    Bottom left panel: the median (plus), 68\% CI (errorbox) and 95\% CI (errorbar) for the posterior of $(m_b)_{n+1}^\text{est}$ , computed from the $abc$ (blue) and geometric (red) models using information on the previous orders. The exact values of $(m_b)_n$ are shown as black circles. Bottom right panel: CIs scaled to the exact N${}^{n+1}$LO correction.
    }
    \end{figure}

\begin{figure}[ht]
 \centering
    \includegraphics[width=\linewidth]{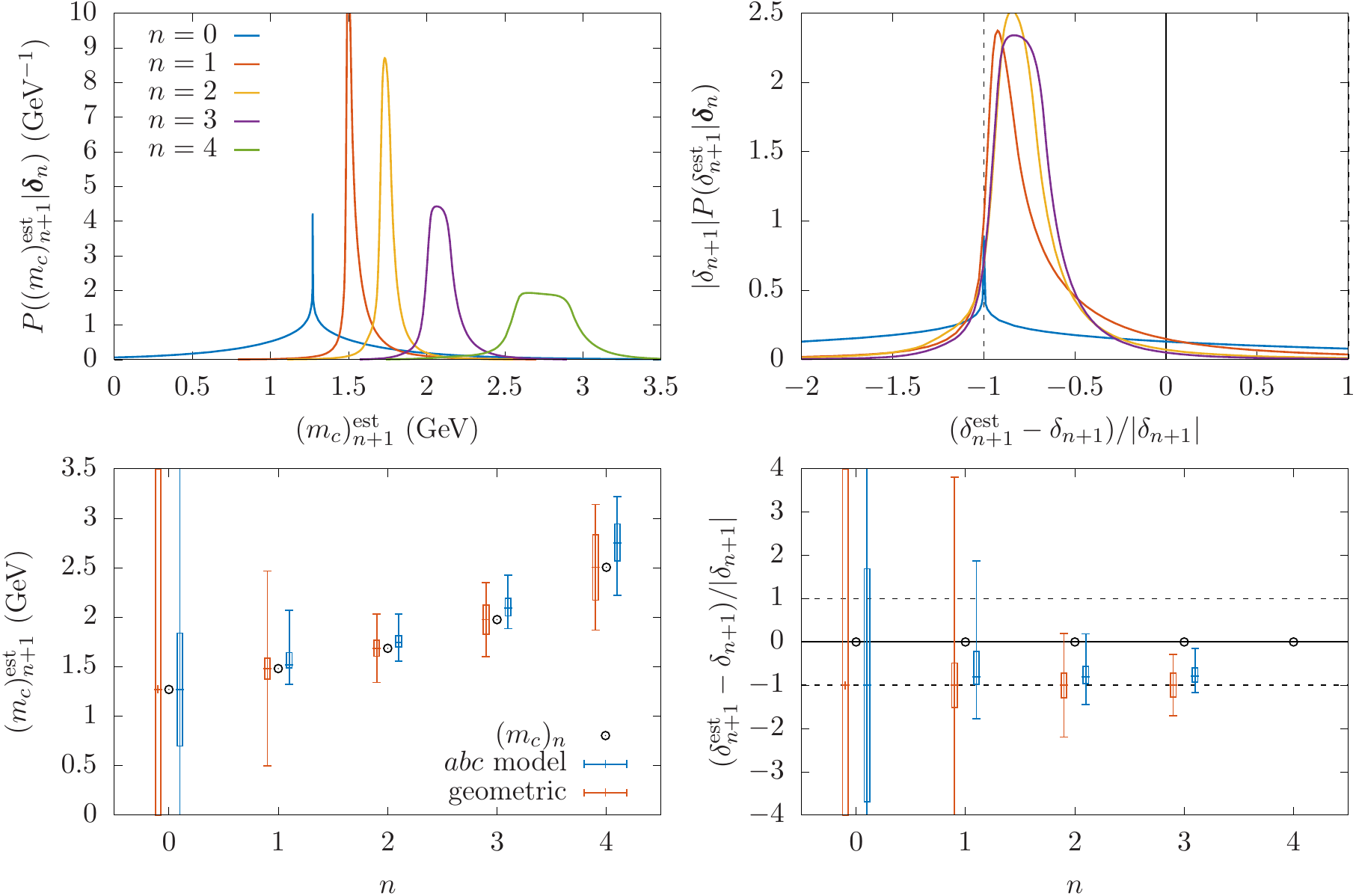}
    \caption{\label{fig:mc_progression} 
    The probability distributions and CIs for charm quark mass. See caption of figure~\ref{fig:mb_progression}.
    }
    \end{figure}

    We also document the plots for probability distributions and CIs for the on-shell bottom and charm quark masses in figures~\ref{fig:mb_progression} and \ref{fig:mc_progression}. Compare to figure~\ref{fig:mt_progression} and see section~\ref{sec:quark_masses} for a detailed discussion.

    
    \clearpage

\bibliographystyle{JHEP}
\bibliography{master}% Produces the bibliography via BibTeX.

\end{document}